\documentclass[11pt]{article}
\usepackage{jcapmod}
\usepackage{graphicx}
\usepackage{amsfonts, amssymb,amsmath}
\usepackage{hyperref}
\usepackage{slashed}
\usepackage{braket}
\usepackage[normalem]{ulem}
\usepackage{comment}
\usepackage{siunitx}
\usepackage{multirow}
\usepackage{subcaption}
\usepackage{placeins}
\usepackage{afterpage}

\definecolor{ForestGreen}{rgb}{0.13, 0.55, 0.13}
\definecolor{airforceblue}{rgb}{0.36, 0.54, 0.66}
\definecolor{orange}{rgb}{1.0, 0.5, 0.0}

\newcommand{\fnl}{f_{\rm NL}^{\rm loc}}
\newcommand{\bx}{\mathbf{x}}

\newcommand{\SIGpc}{{\rm Gpc/h}}

\begin{document}
\begin{titlepage}
	\baselineskip=15.5pt \thispagestyle{empty}
	
	\bigskip\
	
	\vspace{1cm}
	\begin{center}
		{\fontsize{15.5}{10}\selectfont \sffamily \bfseries The Persistence of Large Scale Structures I: Primordial non-Gaussianity}
	\end{center}

	%\vspace{0.2cm}
	\begin{center}
		{\fontsize{12}{30}\selectfont Matteo Biagetti,$^{1,2}$ Alex Cole,$^{1,3,4,5}$ and Gary Shiu$^5$} 
	\end{center}
	
	\begin{center}
		
		\vskip8pt
		
		\vskip8pt
		\textsl{ $^1$ Institute for Theoretical Physics, University of Amsterdam, 1098 XH Amsterdam, NL }\\
		\textsl{ $^2$ Institute for Fundamental Physics of the Universe, via Beirut 2, 34151, Trieste, Italy}\\
		\textsl{$^3$ GRAPPA, University of Amsterdam, 1098 XH Amsterdam, NL}\\
        \textsl{ $^4$ Kavli Institute for Theoretical Physics, University of California, Santa Barbara, CA 93106, USA}\\
        \textsl{ $^5$ Department of Physics, University of Wisconsin-Madison, Madison WI 53706, USA}
			\vskip8pt
	\end{center}

	\vspace{1.2cm}
	\hrule \vspace{0.3cm}
	\noindent {\sffamily \bfseries Abstract}\\[0.1cm]
We develop an analysis pipeline for characterizing the topology of large scale structure and extracting cosmological constraints based on \emph{persistent homology}. Persistent homology is a technique from topological data analysis that quantifies the multiscale topology of a data set, in our context unifying the contributions of clusters, filament loops, and cosmic voids to cosmological constraints. We describe how this method captures the imprint of primordial local non-Gaussianity on the late-time distribution of dark matter halos, using a set of N-body simulations as a proxy for real data analysis. For our best single statistic, running the pipeline on several cubic volumes of size $40~(\si{Gpc/h})^{3}$, we detect $\fnl=10$ at $97.5\%$ confidence on $\sim 85\%$ of the volumes. Additionally we
test our ability to resolve degeneracies between
the topological signature of $\fnl$ and variation of $\sigma_8$ and argue that correctly identifying nonzero $\fnl$ in this case is possible via an optimal template method.  Our method relies on information living at $\mathcal{O}(10)$ Mpc/h, a complementary scale with respect to commonly used methods such as the scale-dependent bias in the halo/galaxy power spectrum. Therefore, while still requiring a large volume, our method does not require sampling long-wavelength modes to constrain primordial non-Gaussianity. Moreover, our statistics are interpretable: we are able to reproduce previous results in certain limits and we make new predictions for unexplored observables, such as filament loops formed by dark matter halos in a simulation box.
\vspace{0.5cm}  \hrule
\vskip 1cm

\end{titlepage}

\thispagestyle{empty}
\setcounter{page}{2}
\setcounter{tocdepth}{2}
\tableofcontents

\def\thefootnote{\arabic{footnote}}
\setcounter{footnote}{0}

\baselineskip= 15pt

\newpage

\section{Introduction}
In the last decade, the study of the large scale structures of the universe has undergone exponential progress thanks to increasingly powerful experiments. Galaxy surveys mapping the entire sky, observing millions of bright galaxies up to high redshifts, have transformed cosmology into a data-driven, precision science. Still, it is an observational science rather than an experimental one: there is only a single universe in which we live and measurements of its properties cannot be repeated. To make up for this lack of repeatability,
we must take care to exploit as much information in cosmological observables as possible.
Commonly used statistical analyses of the spatial distribution of galaxies and voids, or of the temperature anisotropies of the cosmic microwave background, employ low-order correlation functions (mainly the two-point correlation function) in order to 
constrain cosmological parameters and initial conditions. This approach has led to the construction of a consistent history of the universe from its origins to today and a standard model of cosmology, which relies on a handful of very well-constrained parameters (see for instance \cite{Aghanim:2018eyx,Alam:2020sor} for recent results). As experimental efforts in mapping the galaxies in the universe are being designed with unprecedented accuracy, the standard model is put to test in search for inconsistencies. Alongside the refinement of conventional methods based on the analytical and numerical prediction of galaxy correlation functions to match the sensitivity of these experiments, new methods are coming into the game in order to exploit the huge wealth of information available.

In this paper, we explore the statistics of the persistent homology of large-scale structure in the universe. Persistent homology \cite{frosini1992measuring,B94, robins1999towards,edelsbrunner2000topological} is a tool from Topological Data Analysis (TDA) that computes the ``shape'' of discrete data via its multiscale topology. The power of persistent homology lies in its ability to track \emph{individual} topological features as they are created and destroyed by successive coarse-graining transformations. By now, persistent homology has been fruitfully applied in a wide variety of fields, including sensor networks \cite{desilva2007}, image processing \cite{carlsson2008local}, genomics \cite{chan2013topology}, protein structure \cite{Gameiro2015}, neuroscience \cite{2018arXiv180605167S}, and string theory \cite{Cole:2018emh}, to name only a few. 
In the context of cosmology, persistent homology makes contact with previously developed morphological statistics such as the Minkowski Functionals \cite{mecke1991euler,Mecke:1994ax}. Among the Minkowski Functionals, the genus \cite{gott1986sponge,Bardeen:1985tr} is topological in nature. As we elaborate in Sec.\ \ref{sec:PHreview}, persistent homology can be viewed as a refinement of the genus that contains strictly more topological information. Methods of persistent topology have appeared in recent cosmology literature. In the context of the cosmic microwave background, the topological signature of local non-Gaussianity has been investigated to a preliminary extent \cite{2012ApJ...755..122C,Cole:2017kve,Feldbrugge:2019tal} (see also \cite{Pranav:2018lox}). In \cite{Elbers:2018fus} persistent homology methods have been applied to study the time evolution of the reionization bubble network. For large scale structure, general structure identifiers like \texttt{DisPerSE} \cite{Sousbie:2010fp} and \texttt{SCHU} \cite{Xu:2018xnz} have been developed. Previous ``watershed''-based algorithms for identifying cosmic voids like \texttt{ZOBOV} \cite{Neyrinck:2007gy} can be regarded as spiritually related to persistent homology (see Section \ref{sec:PHreview} for more details). Additionally, qualitative properties of toy cosmological models have been investigated with these methods \cite{vandeWeygaert:2011ip,vandeWeygaert:2013kma,Pranav:2016gwr}. The consistency of Sloan Digital Sky Survey data as summarized by persistent homology was investigated in \cite{KIMURA2017722}. However, the capacity of persistence-based methods to directly provide \emph{quantitative} constraints on cosmology (and the attendant systematics that must be taken into account) is in our view an underdeveloped subject. (For recent progress in this direction using cosmic shear, see \cite{Heydenreich:2020hrr}.) We therefore work in this paper to fill this gap, presenting a pipeline for quantitatively constraining cosmological parameters.

The output of a galaxy survey, as a list of positions in 3-dimensional space, is a natural input for persistent homology. In this context, the relevant cosmological objects are clusters, filament loops, and cosmic voids. By tracking these features through various coarse-graining scales we exploit information that is not accessible to low-order correlation functions. This higher-order information includes the sizes of individual filament loops and voids and the patterns by which sub-loops and sub-voids are embedded within their parent structures. This information is complementary to the usual information accessed by low-order correlation functions. On the other hand, the higher-order nature of these observables puts them outside the regime of standard perturbative methods
commonly used for low-order correlation functions (see \cite{Bernardeau:2001qr} for a review and \cite{DAmico:2019fhj,Ivanov:2019pdj} for recent methods applied to data). 
For this reason, we must resort to a combination of analytic and numerical tools in order to gather as much information as possible, both on the expected behavior of topological features and the expected systematics.

Consequently, we test our methods on N-body simulations. In particular, we use halo catalogs extracted from dark matter simulations with varying cosmologies as a proxy for observed galaxy positions in the sky. Even though the relation between galaxy and halo distribution is non-trivial, halos provide a good framework to test a pipeline for the statistical analysis that can then be applied to surveys. In our case, it also helps to start with a controlled setting where we know the input cosmologies.

While our final goal is to use persistent homology to simultaneously constrain a wide range of cosmological parameters, as a starting point for building our pipeline, we investigate primordial non-Gaussianity. One of the greatest appeals of studying large scale structures is learning about initial conditions. Primordial non-Gaussianity is a key aspect of these initial conditions, and provides an important link between high energy physics and cosmology. Whether the mechanism that generates the seeds of density perturbations is cosmic inflation or other means, the detailed patterns in the distribution of galaxies or in the cosmic microwave background encode the unknown high energy physics at work in the early universe. Observations of the cosmic microwave background \cite{Akrami:2018odb} have so far provided the strongest constraints on the two-point function of primordial perturbations, which  however only provides coarse-grained information about the free theory of the fields generating these perturbations, while not answering fundamental questions about the field content and interactions during this early phase. Precision measurements of the three and higher-point correlations of the temperature anisotropies of the cosmic microwave background or of observed galaxies can potentially go further in unveiling the interactions governing the early universe. In the context of inflation, the primordial bispectrum (three-point function of curvature perturbations) for single field slow-roll \cite{Maldacena:2002vr} and general single field inflation \cite{Chen:2006nt} is indeed parametrized by the interactions of the inflaton field. Thus, given an inflationary Lagrangian, one can work out the corresponding galaxy distribution and their higher order correlations.
The most competitive constraints on the primordial bispectrum have also been performed up to now by CMB experiments \cite{Akrami:2018odb}, determining for instance that local-type primordial non-Gaussianity of order $\mathcal O(\fnl) = 10$ (see Sec.\ \ref{sec:png} for a definition) is excluded with $95\%$ confidence. On the other hand, large scale structure observations seem to hold the best figures in the coming decade \cite{Meerburg:2019qqi}.

Motivated by these figures, we employ a dataset of large scale N-body simulations with local type primordial non-Gaussianity in the initial conditions of amplitude $\fnl=10$ and use it to build a pipeline for quantitatively constraining local non-Gaussianity. Along the way, we study the signature of primordial non-Gaussianity in our topological observables. Some of these observables make contact with and agree with previous analytic results including the effect of local non-Gaussianity on the void size function \cite{Kamionkowski:2008sr}. Intriguingly, we find a similar effect on the filament loop radius function, which as far as we know has not yet been derived from first principles (see \cite{Bond:1995yt,Shen:2005wd,Cadiou:2020xmo} for analytical work on the Gaussian case). 

Indeed, we find that in our setup the filament loops are the most sensitive to $\fnl=10$. Our best single statistic is $B_1$ (in the particular case of an $\alpha$DTM$\ell$-filtration, see Sec.\ \ref{sec:PHreview}), which counts the creation of filament loops as a function of coarse-graining scale. Using small-box simulations with stronger non-Gaussianity, we build templates for primordial non-Gaussian signatures and apply them to simulations with $\fnl=10$. By running the pipeline on several  cubic volumes of size $40~(\si{Gpc/h})^{3}$, we ``detect'' $\fnl=10$ at $97.5\%$ confidence on $\sim 85\%$ of the volumes. The template method is not specific to the case of primordial non-Gaussianity, and should be applicable to other cosmological parameters.

The paper is organized as follows. In Section \ref{sec:PHreview} we give a short review of  persistent homology, paying particular attention to aspects of the topic relevant to our pipeline. In Section \ref{sec:phhalo} we specialize persistent homology to large scale structures, setting the stage for data analysis by anticipating possible systematics we can encounter and motivating our choice of statistics. We also compute and discuss the signature of primordial non-Gaussianity on topological features. Section \ref{sec:pipe} details the pipeline and the main results of the analysis. We conclude in Section \ref{sec:conclusions}.

\section{Lightning review of persistent homology}\label{sec:PHreview}
In this section, we describe relevant aspects of persistent homology as a tool for quantitative characterization of the morphology of discrete data sets. The aim of this section is to give the reader a rather operational understanding of persistent homology. For more details on the many mathematically beautiful aspects of the subject, we refer the reader to \cite{zomorodian2005topology,edelsbrunner2010computational,oudot2015persistence}.

\subsection{The basic idea}\label{sec:basics}
Suppose we are given a collection of points, hereafter a \emph{point cloud}, embedded in a metric space. For our purposes, these will be the positions of dark matter halos in an N-body simulation. We would like to characterize the structure of the point cloud. In particular, we would like to assign a notion of topology to the distribution. There are several reasons for doing this. First, from the perspective of large scale structure, it has long been understood that the statistical distribution of dark matter tracers and voids are discerning probes of fundamental physics (see e.g. the latest results from the SDSS survey on galaxy-galaxy \cite{Ivanov:2019pdj,DAmico:2019fhj,Ivanov:2019hqk} and void-galaxy correlation functions \cite{Nadathur:2020vld}). As we will see, these objects are naturally defined in terms of topology. Topological features probe the tail of the initial density perturbation and are enhanced by gravitational evolution, so we expect them to be especially worthwhile to study in late-time observables. Moreover, from a data science perspective, we would like our analysis pipeline to be robust, behaving well under the addition of noise and observational effects. Classical topology studies precisely those features of a space that cannot be created or destroyed by continuous deformations. As we will see, extending this notion to discrete data sets will earn us corresponding robustness results. 

\begin{figure}
    \centering
    \includegraphics[width=0.5\textwidth]{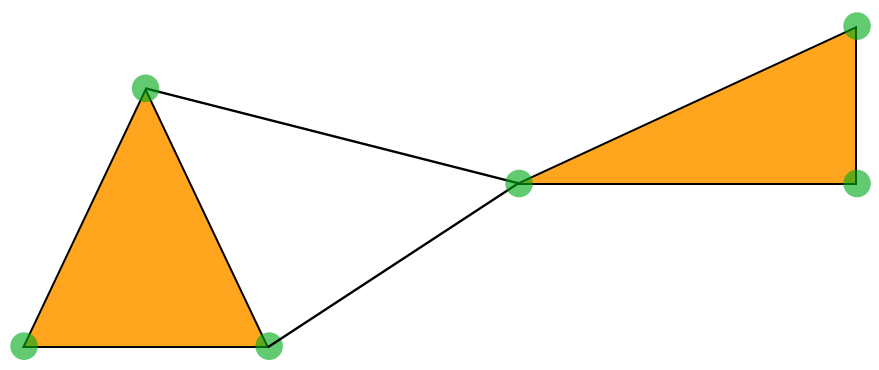}\includegraphics[width=0.5\textwidth]{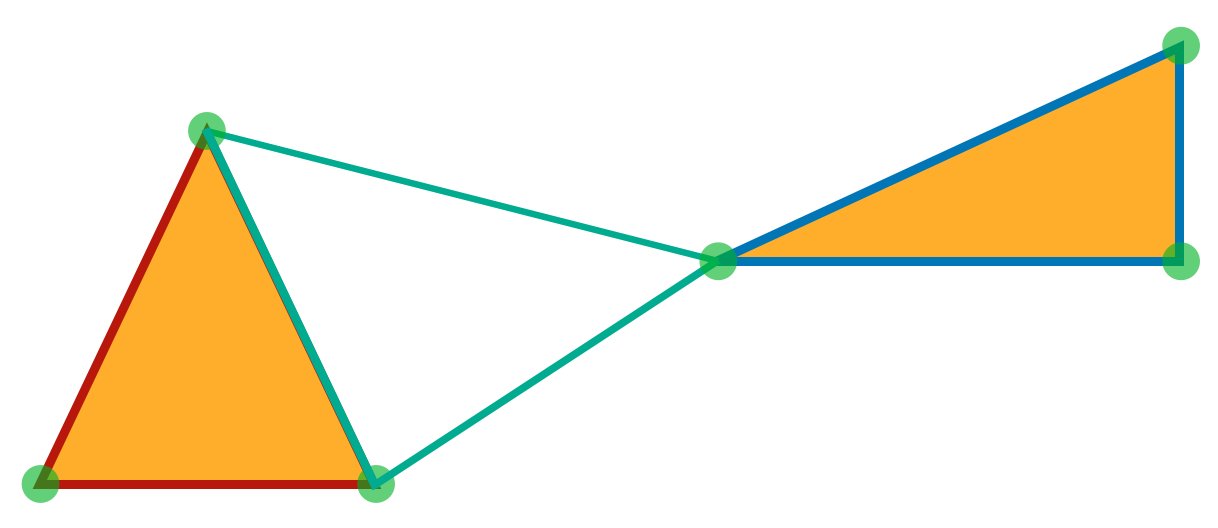}
    \caption{Left: an example of a simplicial complex. In this work, we use families of simplicial complexes to represent the topology of the distribution of DM halos over a range of scales. Right: topological 1-cycles in the simplicial complex are highlighted. The green 1-cycle is topologically nontrivial. It cannot be ``deformed away.'' It is equivalent to (i.e.\ belongs to the same homology class as) the 1-cycle involving two green edges and two red edges. 
    %The blue 1-cycle is topologically trivial.
    }
    \label{fig:complex_example}
\end{figure}

To describe global aspects of our point cloud, we need to invoke some linking between different points. This can be achieved by embedding our data in
a \emph{simplicial complex}. Simplicial complexes are built from \emph{simplices}. For our purposes, relevant simplices include vertices (0-simplices), edges (1-simplices), triangles (2-simplices), and tetrahedra (3-simplices). It is often useful to represent a simplex by its collection of vertices, e.g.\ $[x_0 x_1]$ represents the edge connecting vertices $x_0$ and $x_1$. A simplicial complex is a collection of simplices that is closed under intersection of simplices and under taking faces (e.g.\ if an edge is present, the two vertices making up its boundary are also present). See Fig.\ \ref{fig:complex_example} for a graphical depiction of a simplicial complex.

The topology of a simplicial complex can be characterized via its \emph{homology}.  For a simplicial complex embedded in a 3-dimensional space, there are 3 homology groups $H_p$, $p=0,1,2$. The ranks of these groups are called the Betti numbers $b_p=|H_p|$. The elements of $H_p$ are equivalence classes of objects called $p$-cycles. An element of $H_0$ represents a connected component of the simplicial complex. Analogously, elements of $H_1$ correspond to ``loops'' and elements of $H_2$ are ``voids.'' There is one nontrivial 1-cycle in the simplicial complex in Figure \ref{fig:complex_example}. Moreover, the Betti numbers can be computed efficiently. For example, given a finite simplicial complex in $S^d$, $d\leq 3$, the incremental algorithm of Delfinado and Edelsbrunner computes the Betti numbers in time $O(N_{\rm simplices})$ \cite{delfinado1995incremental}.

\begin{figure}
\begin{subfigure}{.33\textwidth}
\centering
\includegraphics[width=5cm]{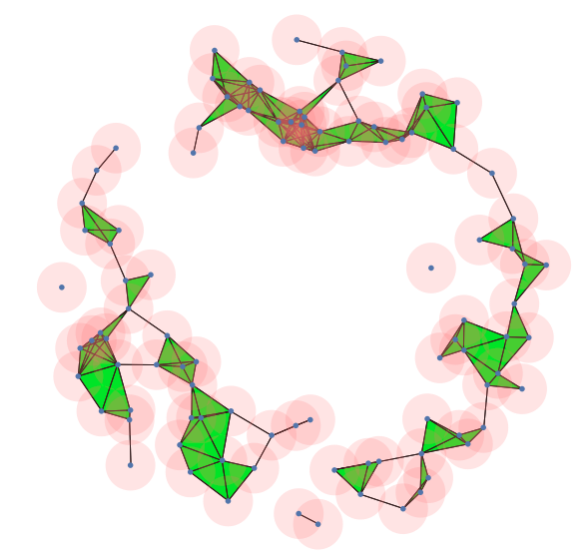}
\end{subfigure}\hfill
\begin{subfigure}{.33\textwidth}
\centering
\includegraphics[width=5cm]{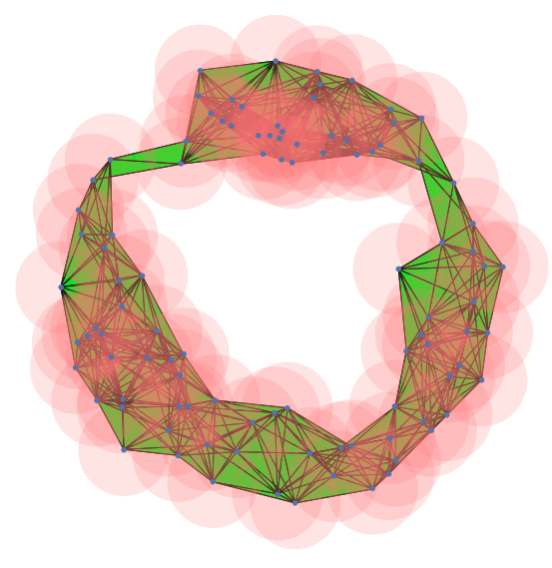}
\end{subfigure}\hfill
\begin{subfigure}{.33\textwidth}
\centering
\includegraphics[width=5cm]{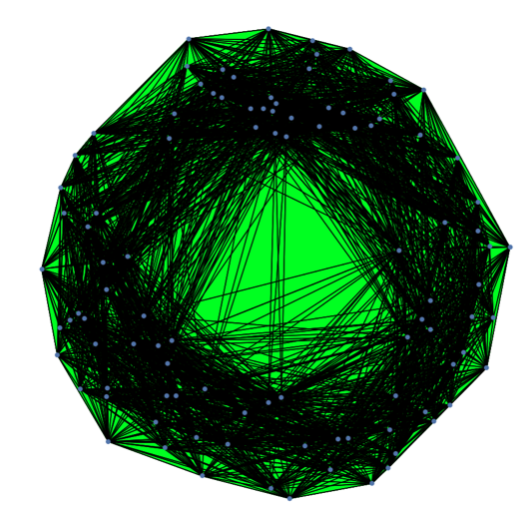}
\end{subfigure}
\caption{The Vietoris-Rips filtration for an annulus at three coarse-graining scales. Early in the filtration (left), small-scale structure of the distribution is probed. At medium scales (middle), a strong 1-cycle is born, which persists until large scales (right), at which point the topology becomes trivial.}
    \label{fig:VR}
\end{figure}

A particular data set can be represented by a single simplicial complex, but such a procedure is ambiguous and unstable: there are many choices of simplicial representations, and these different choices can lead to different topological features.
The main idea of \emph{persistent} homology is to represent the data via a growing family of simplicial complexes. The family, called a \emph{filtration}, often represents the data set at different length scales. For example, the Vietoris-Rips (VR) filtration is defined as follows. One starts with a point cloud $X$ embedded in a metric space with distance function $d(x,y)$. The filtration is parameterized by a coarsening scale $r>0$. In generic cases we will denote the filtration parameter by $\nu$. The vertex set of each complex in the VR filtration is the point cloud itself, $X$. For each $r>0$, an edge between vertices $x$ and $y$ is included if $d(x,y)<2r$. Higher-dimensional simplices are included if all necessary faces are present. One then has the interpretation that pairs of vertices are connected by an edge if balls of radius $r$ drawn around the vertices overlap. Thus the VR filtration can be used to compute the multiscale topology of a data set. This is depicted in Fig.\ \ref{fig:VR}, where we show a VR filtration at three coarse-graining scales on a point cloud sampled from an annulus.  While the VR filtration is sufficient for simple data sets, it is rather inefficient. Note that as $r\to \infty$ there is an edge between every pair of vertices, a triangle between every triple of vertices, and so on. This is impractical for large data sets. We will use more sophisticated filtrations, defined in Section \ref{sec:filtrations}. For later reference, we note that persistence can also be phrased in terms of sublevel filtrations of distance functions. To be more concrete, given a point cloud $X\subset \mathbb{R}^n$, consider the distance function $d_X:\mathbb{R}^n\to \mathbb{R}_{\geq 0}$ defined by $d_X(x)=\min_{y\in X}||y-x||$. The function $d_X$ computes the distance from a point $x$ to the closest $y\in X$. It is then clear that \emph{sublevel sets} $d_X^{-1}(-\infty,\nu]$ of the distance function give a coarse-graining of the set $X$ with coarsening scale $\nu$. Moreover, $\nu_1<\nu_2$ implies that $d_X^{-1}(-\infty,\nu_1]\subseteq d_X^{-1}(-\infty,\nu_2]$. In other words, the sublevel sets also define a filtration for which we can compute persistence. (Note that to use the simplicial machinery described earlier, we must choose some suitable triangulation of the domain.)

\subsection{Persistence diagrams and derived statistics}
\begin{figure}
    \centering
    \includegraphics[width=0.5\textwidth]{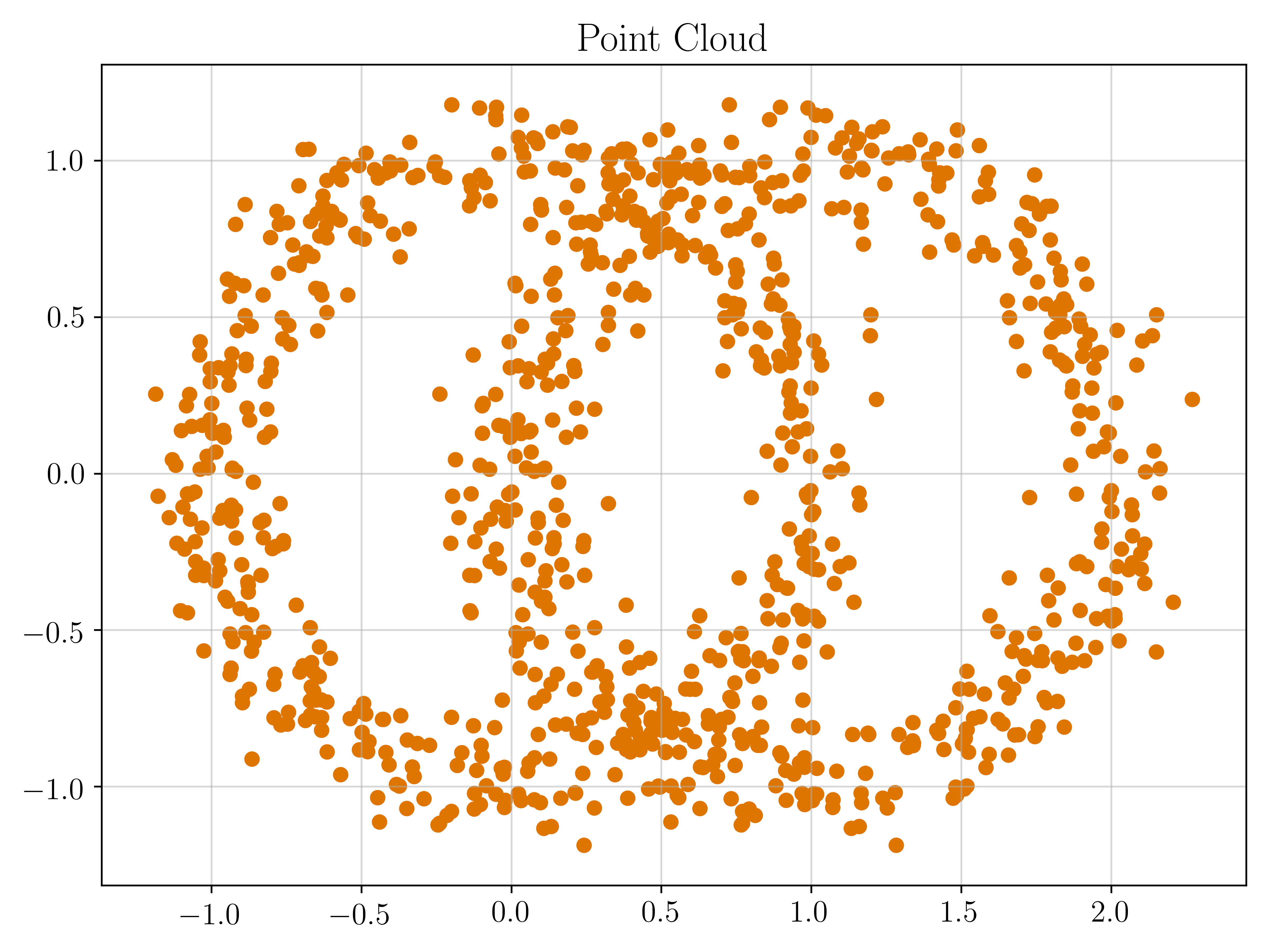}\includegraphics[width=0.5\textwidth]{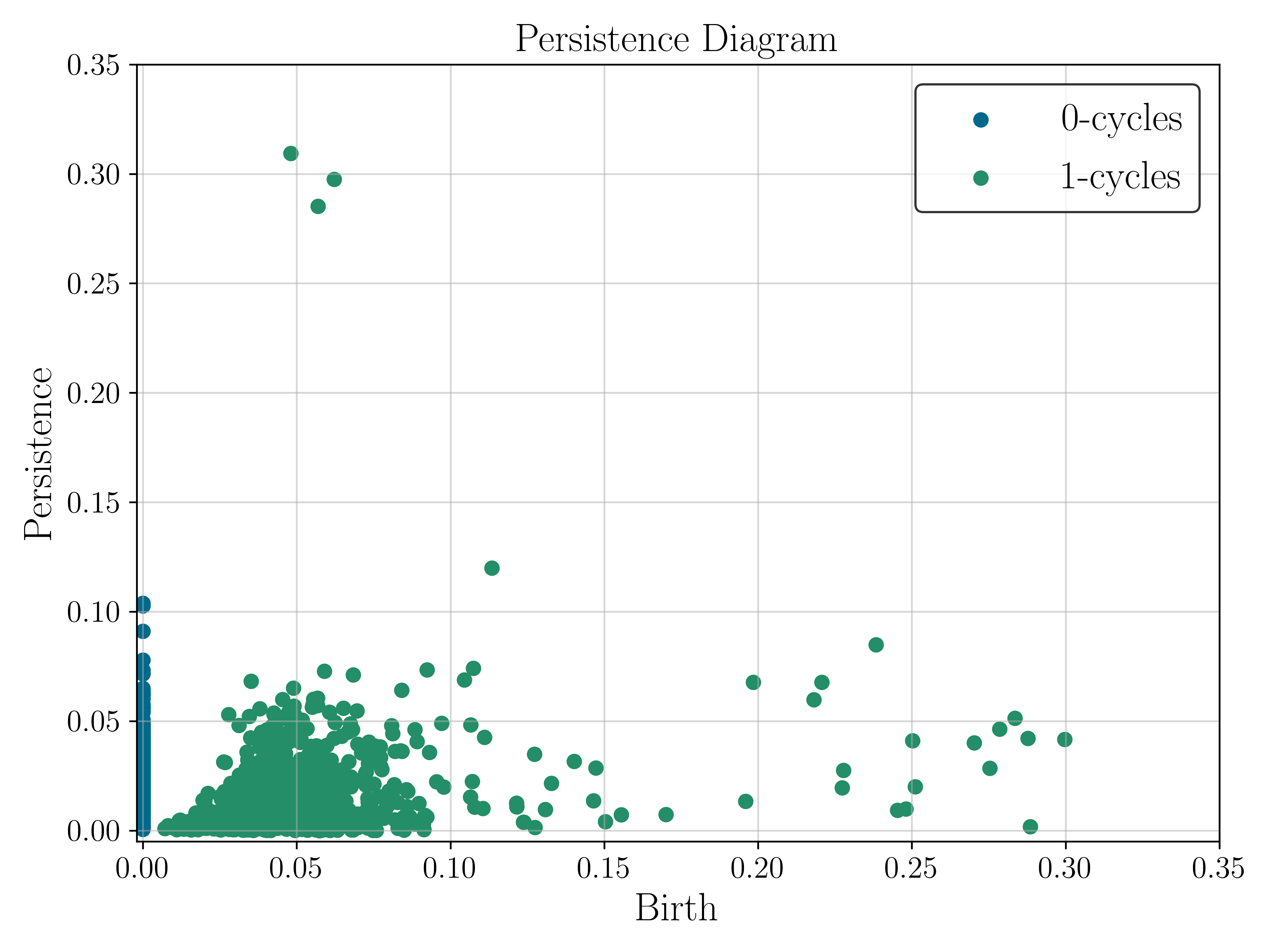}
    \caption{Two overlapping rings (left) and the persistence diagram of a corresponding filtration (right). By eye, we identify three strong independent 1-cycles, which manifest as three 1-cycles at large persistence in the persistence diagram.}
    \label{fig:Figure8}
\end{figure}

When a $p$-simplex is added to a filtration, it either creates a new $p$-cycle or destroys an existing $(p-1)$-cycle. More precisely, it creates or destroys corresponding homology elements, which are equivalence classes of cycles.\footnote{In the case of destruction, a more precise statement is that the simplex either trivializes a particular homology class or makes two previously distinct homology classes equivalent. In the latter case, one adopts the \emph{elder rule}, under which the ``younger'' homology class is destroyed.} The simplices, however, are unique. A persistent homology computation then amounts to computing the \emph{persistence pairs} $(\sigma,\tau)$ that create and destroy homology classes arising throughout the filtration. 
We can then say that a particular homology element with persistence pair $(\sigma,\tau)$ is born at $\nu(\sigma)$ and dies at $\nu(\tau)$. (There can also be homology elements that do not die. These will not be relevant for our purposes.) We refer to the difference $\nu(\tau)-\nu(\sigma)$ as the \emph{persistence} of the feature. The persistence pairs can be computed in worst-case scaling $O(N_{\rm simplices}^3)$. The output of such a computation can be represented pictorially by a \emph{persistence diagram}, which is a scatter plot of the birth and death times of individual homology generators. We will in fact generally plot persistence diagrams in the coordinates $(\nu_{\rm birth},\nu_{\rm persist}\equiv \nu_{\rm death}-\nu_{\rm birth})$. An example is shown in Fig.\ \ref{fig:Figure8}, where we show the persistence diagram for a filtration defined by a point cloud sampled from two overlapping annuli. The immediate intuition is that long-lived topological features correspond to ``true'' features in the data, while short-lived features may be caused by noise. Such long-lived features are represented in Figure \ref{fig:Figure8} by the three dots at high persistence. This intuition is formalized in various \emph{stability theorems} (see e.g.\ \cite{cohen2007stability}), which state that under a small perturbation to the input data, the change in the resulting persistence diagram is accordingly small under a natural metric on the space of persistence diagrams. Intuitively, small perturbations will not create or destroy long-lived topological features. See e.g.\ \cite{edelsbrunner2010computational} for more details. On the other hand, we will later observe that the distinction between long-lived and short-lived features is not as clear-cut as in Figure \ref{fig:Figure8} when applying these methods to a halo catalog (cfr. Figure \ref{fig:alpha1000sub-boxValid}) and that short-lived features can themselves provide useful information for the purpose of statistical inference.

While a persistence diagram is generally suitable for visualization, it is not an ideal object for performing statistical analyses. For example, we may have some statistical process $\mathcal{P}(z)$ depending on parameter $z$ whose individual realizations can be analyzed via persistent homology, and we want to measure some realizations and constrain $z$. For this purpose, it will be useful to further process persistence diagrams into statistics that live in vector spaces, admitting $\chi^2$ tests and so on.

The derived statistics we will consider are objects called persistence images (PIs) as well as various one-dimensional curves constructed by slicing a persistence diagram. PIs are essentially processed histograms of the persistence diagrams \cite{adams2017persistence}. More precisely, each cycle in the persistence diagram is smoothed via a kernel. The kernel is weighted by a factor that scales with $\nu_{\rm persist}\equiv\nu_{\rm death}-\nu_{\rm birth}$ and vanishes at $\nu_{\rm persist}=0$, for example $\log(1+\nu_{\rm persist})$. This weighting preserves stability properties inherited from the persistence diagrams. The resulting density function is then binned as a two-dimensional histogram and often downsampled for further smoothness and a tractable low-dimensional statistic.

The utility of the PIs is that they closely represent the interplay between birth and persistence evident in a persistence diagram. The main difficulty of the PIs is the need to choose various smoothing parameters, e.g.\ the widths of kernels and spacing between histogram bins. 
On the other hand, from a persistence diagram, one can construct one-dimensional curves that necessarily miss some of the two-dimensional information in the persistence diagram, but do not require choosing smoothing parameters. 

\begin{figure}
    \centering
    \includegraphics[width=0.5\textwidth]{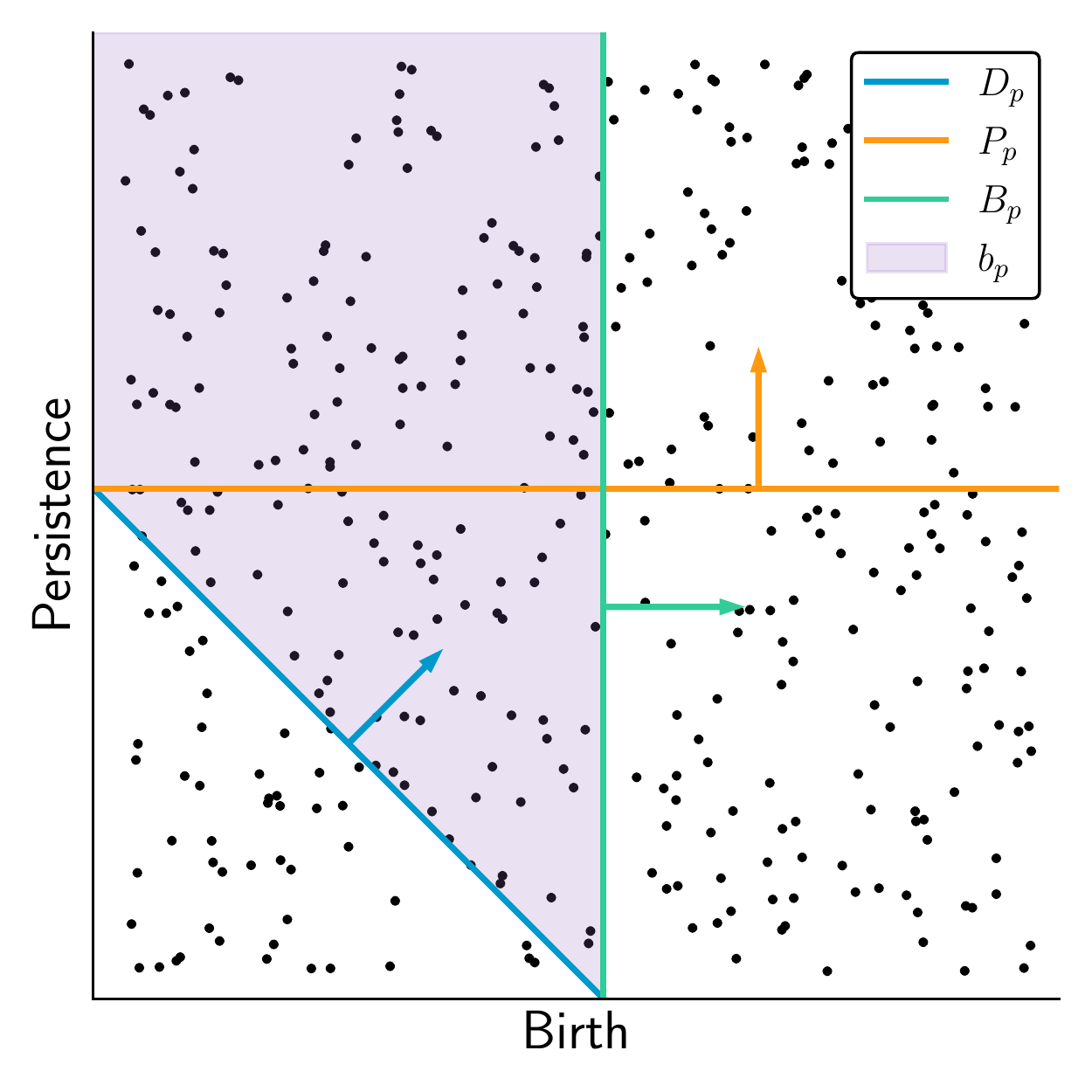}
    \caption{Construction of various one-dimensional curves from persistence diagram. At a given threshold $\nu$, each curve counts the number of cycles in a particular region of the diagram, indicated by arrows. The green curve $B_p(\nu)$ counts the number of features that are born after $\nu$. The orange curve $P_p(\nu)$ counts the number of features with persistence greater than $\nu$. The blue curve $D_p(\nu)$ counts the number of features that die after threshold $\nu$. Note that the Betti number $b_p(\nu)$ counts the number of ``living'' features, in other words $b_p(\nu)=-B_p(\nu)+D_p(\nu)$. Graphically, this corresponds to features in the shaded purple region.}
    \label{fig:curveConstruct}
\end{figure}

We construct four classes of curves from a given persistence diagram. These count the number of cycles in a given subregion of the diagram. We will consider the curves:
\begin{align}
    B_p(\nu)&=\textrm{count}_p(\nu_{\rm birth}>\nu)\\
    P_p(\nu)&=\textrm{count}_p(\nu_{\rm persist}>\nu)\\
    D_p(\nu)&=\textrm{count}_p(\nu_{\rm death}>\nu)
\end{align}
where $\textrm{count}_p$ refers to counting the cycles in a particular region of the $p^{\rm th}$ persistence diagram. The regions we select have the interpretation of counting (i) the number of cycles born after $\nu$ (ii) the number of cycles with persistence greater than $\nu$ (iii) the number of cycles that die after $\nu$.  The construction of these curves is shown in Fig.\ \ref{fig:curveConstruct}. Each curve interpolates between the total number of $p$-cycles in a given diagram and $0$. {We also consider the Betti numbers as functions of $\nu$ (previously used in a similar context by \cite{vandeWeygaert:2013kma,Pranav:2016gwr}) which can be constructed from the above curves as $b_p(\nu)=-B_p(\nu)+D_p(\nu)$.} {For the statistical analysis, we normalize each curve to interpolate between 1 and 0.} In other words, we are using the empirical distribution functions of $\nu_{\rm birth},\nu_{\rm persist}$, and $\nu_{\rm death}$ for topological features of each dimension as statistics. {We also choose to normalize the Betti numbers, as this appears to remove unwanted variance from these statistics, see Section\ \ref{sec:curveNorm}.}

To connect to the Minkowski Functionals, we note that the genus can be computed via the Euler-Poincar\'e formula
\begin{equation}
    \chi(\nu)=\sum_p (-1)^p b_p(\nu)
\end{equation}
It is clear from this formula that the individual Betti numbers probe topology at a finer resolution than the genus. Moreover, there is more information in persistent homology than the Betti numbers. The power of persistent homology lies in its ability to track \emph{individual} features. The Betti numbers, which count ``living'' features at different scales, aggregate information from the persistence diagram in a certain manner. The other topological curves $B_p,D_p,P_p$ represent complementary information to the Betti numbers within persistent homology.
\subsection{Relevant filtrations }\label{sec:filtrations}
As introduced above, the key point of persistent homology is the use of filtrations, i.e.\ families of complexes. The choice of proper filtration for a given dataset depends on the problem at hand; for this reason we test three different filtrations, whose features we summarize below.
%\footnote{Note that other filtrations, such as the sublevel filtration described earlier, have also been used for cosmology, see \cite{Neyrinck:2007gy,Xu:2018xnz}.}
We will show full results for the third filtration, which best captures the primordial signatures we are after, while summarizing results from the other filtrations in Appendix \ref{app:table}.

\begin{figure}
    \centering
    \includegraphics{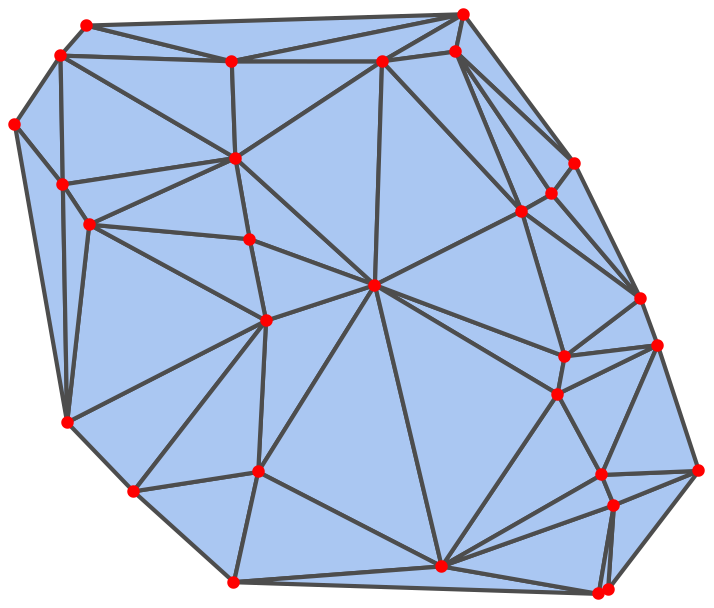}
    \caption{The $\alpha$-filtration consists of subcomplexes of the Delaunay complex. Here an example of a Delaunay complex in two dimensions is shown. One can probe multiscale features of the set of points by adding simplices to the filtration corresponding to their size.}
    \label{fig:Delaunay}
\end{figure}

\paragraph{$\alpha$-filtration.} The $\alpha$-filtration \cite{edelsbrunner1994three} takes its inspiration from Delaunay complexes. The Delaunay complex can be defined in the following way. Given a set of points $X\subset \mathbb{R}^d$, the \emph{Voronoi cell} corresponding to point $x\in X$ is defined by $V_x=\{y\in \mathbb{R}^d ~|~ ||x-y||\leq ||z-y||~\forall z\in X\}$. In other words, the Voronoi cell of point $x$ includes all points in $\mathbb{R}^d$ for which $x$ is the nearest (or tied as the nearest) point among all $z\in X$. The collection of Voronoi cells tessellates the domain $\mathbb{R}^d$ and constitutes the \emph{Voronoi diagram of $X$}. The \emph{Delaunay complex} is defined via the Voronoi diagram by $\{\sigma\subseteq X~|~\cap_{x\in \sigma} V_x\neq \emptyset\}$. The Delaunay complex gives a triangulation of the ambient space by filling in the convex hulls of abstract simplices. An example of this geometric realization is shown in Fig.\ \ref{fig:Delaunay}. The set of simplices of the $\alpha$-filtration is the same as that of the Delaunay complex, with the ``filtration time'' a particular simplex is added corresponding to its size, see \cite{cgal:dy-as3-19b} for details. Thus we are still able to describe the multiscale topological nature of our data, benefitting from efficiency aspects of the Delaunay complex. The worst-case number of simplices for $N_{\rm halo}$ points in 3 dimensions scales as $\mathcal{O}\left(N_{\rm halo}^2\right)$ \cite{otter2017roadmap}.

One perhaps undesirable aspect of the $\alpha$-filtration is that the topological objects it identifies, in particular voids, die at smaller scales than one would expect from traditional void-finding algorithms. This is in fact true for any distance-based filtration that does not account for outliers. For example, a single halo at the center of an otherwise pristine, i.e.\ empty, void will decrease the coarsening scale at which the feature dies by a factor of two. In other words, our procedure is not stable against the addition of outliers.\footnote{We distinguish between the addition of outliers (which introduces new points in underdense regions) and the addition of noise (which perturbs the positions of existing points).}

We can try to remedy this situation by modifying the $\alpha$-filtration. In this paper we will perform two modifications of the $\alpha$-filtration. Both keep the same set of simplices (those from the full Delaunay complex) but modify the scales at which these simplices are added to the filtration.

\paragraph{$\alpha$DTM-filtration.} In what we call the $\alpha$DTM-filtration, we assign the simplices a filtration time based on the Distance-to-Measure (DTM) function. Given a set of point $X$, the empirical DTM function is given by \cite{chazal2011geometric}
\begin{equation}
    {\rm DTM}(x)=\frac{1}{k}\left(\sum_{X_i\in N_k(x)}||x-X_i||^p\right)^{1/p}
\end{equation}
where $N_k(x)$ is the list of the $k$ nearest neighbors in $X$ to $x$ and $p$ is a free parameter. Most commonly $p$ is taken to be 2. One can build intuition by considering the case $k=1$, in which the DTM function gives the distance to the nearest point in $X$, in other words the distance function $d_X$ described at the end of Section \ref{sec:basics}. Increasing $k$ ``smooths'' this distance function so that it takes small values where it is near many points and large values near outliers. Then performing a sublevel filtration using the DTM function, outliers will be added relatively late in the filtration, leading to a smaller effect on the persistent homology.

Note that sublevel filtrations of the DTM function are used by the void-finding algorithm SCHU \cite{Xu:2018xnz}. There the DTM function is evaluated on a grid. 
%In many contexts (including ours), evaluating on a grid with high resolution and sufficiently large $k$ involves prohibitive computational expense.
Depending on the quantity of interest, evaluating on a grid with high resolution and sufficiently large $k$ requires an accurate choice of the grid size to optimize the computational expense.
One way to get around this is to evaluate the DTM function at certain relevant points of the space. This amounts to choosing a smart tessellation of the ambient space. In our context, we will evaluate the DTM function on the Delaunay complex.

In the $\alpha$DTM-filtration, we assign a vertex from the full Delaunay complex a filtration time that is the DTM value evaluated at the vertex's location. Higher-dimensional simplices in the Delaunay complex are assigned filtration time $\max_{\rm faces}{\rm DTM}({\rm face})$, which ensures a well-defined (i.e.\ monotonically growing) filtration. This is called a lower star filtration. {Note that this amounts to performing a sublevel filtration on the Delaunay complex with sublevel function given by the DTM function. This is analogous to the procedure performed by the void-finder \texttt{ZOBOV} \cite{Neyrinck:2007gy}, which instead uses the inverse volume of a Voronoi cell as the sublevel function. Recall that the Voronoi decomposition is dual to the Delaunay triangulation.} 

\paragraph{$\alpha$DTM$\ell$-filtration.} One perhaps unsatisfying aspect of the $\alpha$DTM-filtration is that it obscures information regarding the scales of certain features. We thus introduce another filtration that interpolates between the $\alpha$-filtration and its $\alpha$DTM variant. We call this third filtration the $\alpha$DTM$\ell$-filtration. This construction was introduced in \cite{chazal2017robust}. Given a point in the our point cloud $x\in X$ we consider the radius function
\begin{equation}
    r_x(\nu)=\left(\nu^p-f(x)^p\right)^{1/p}\quad \textrm{if }\nu\geq f(x)
\end{equation}
and $-\infty$ otherwise, and by definition a ball of radius $r=-\infty$ is the empty set. In general the function $f(x)$ is arbitrary; we will take $f(x)=\textrm{DTM}(x)$. The role of $f(x)$ is to delay the contribution of a particular point to the entire filtration. In other words, $f(x)$ parameterizes the ``unimportance'' or ``nuisance'' of the point $x$. Since we want to minimize the contribution of outliers to the filtration, $f(x)=\textrm{DTM}(x)$ is a good choice. Here $p$ controls the mixing between $f(x)$ and $\nu$. We will take $p=2$\footnote{While leaving a rigorous tuning of this parameter to future work, the choice $p=2$ is motivated by the fact that it lives at a trade off between large $p$, where large DTM vertices are added late, but their balls soon catch up with the rest, and small $p$, where large DTM vertices take longer to catch up.}.  For very large $p$, $r_x(\nu)$ is $-\infty$ until $\nu=f(x)$, at which point it jumps to $\nu$. Beyond the vertices, we add an edge $[x_1 x_2]$ that is present in the Delaunay complex once the relevant balls overlap, i.e.\ $B_{r_{x_1}(\nu)}\cap B_{r_{x_2}(\nu)}\neq \emptyset$. Triangles and tetrahedra are then added when all necessary faces are present. In this construction, topological features will suffer less from the presence of outliers that in the $\alpha$-filtration.
\section{Persistent homology in halo catalogs}\label{sec:phhalo}
In the previous section, we have defined techniques and tools from TDA that can be applied to a wide range of datasets. We now to specialize these techniques to our dataset, which consists of several halo catalogs produced from N-body simulations of the universe on large scales. In other words, our point clouds are sets of halo positions in 3-dimensional cubic boxes of cosmological size.
Using halos in real space as a proxy for galaxies, we implicitly take a number of unrealistic approximations, since i) halos typically do not contain a single galaxy in them, therefore there is not a one-to-one statistical correspondence between halos and galaxies, ii) galaxy formation is a complicated process, the modeling of which requires taking into account baryon effects, iii) galaxies are observed in redshift space. Additionally, in observational contexts one generally deals with masked regions where the data cannot be trusted. For topological observables, this can be accomplished in the framework of relative homology \cite{Pranav:2018lox,Heydenreich:2020hrr}. Because of these approximations, the results of our analysis should be considered as a preliminary assessment of applicability to real data. The important aspect of building the pipeline on the simplified case with halos is that we keep a degree of control over the relevant physics.

In this section we introduce a few preliminaries on applying persistent homology to large scale structure that are preparatory to defining a statistical analysis pipeline in the next section. We also study qualitative features of the imprint of primordial non-Gaussianity on our statistics and we find that our results agree with several previously known signatures of non-Gaussianity in large scale structure.

\subsection{Dataset}
We use the \textsc{Eos Dataset}, a suite of N-body simulations created to investigate the imprint of primordial non-Gaussianity in large-scale structures at low redshift\footnote{Information on the full dataset is available at \url{https://mbiagetti.gitlab.io/cosmos/nbody/eos/}. The reference paper investigated the imprint of primordial non-Gaussianity on the halo power spectrum \cite{Biagetti:2016ywx}.}. We summarize the simulations used in this work in Table \ref{tab:eos}. The reference cosmology is flat $\Lambda$CDM with $\Omega_m= 0.3$, $h= 0.7$, $n_s= 0.967$ and $\sigma_8=0.85$.  The  transfer  function  was  obtained from the Boltzmann code CLASS \cite{Blas:2011rf}. The initial particle displacements were implemented using the publicly available code 2LPTic \cite{Scoccimarro:1997gr,Crocce:2006ve} for  realizations  with  Gaussian  initial  conditions and  \cite{Scoccimarro:2011pz} for non-Gaussian initial conditions of the local type. The starting redshift of the simulations is $z_{in}= 99$. The simulations were evolved using the public code Gadget2 \cite{Springel:2005mi}. The halo catalog is produced using Rockstar \cite{Behroozi:2011ju}, for which we use a linking length of $\lambda= 0.28$. We look at halos with a minimum of $50$ particles, which corresponds to a minimum mass of $M_{\rm min} = 9.2\times 10^{12} M_\odot$.

\begin{table}
\begin{center}
\begin{small}
\begin{tabular}{|c||c|c||c|c|c|c|}\hline
    \bf{ID} &$\mathbf{\sigma_8}$&$\mathbf{\fnl}$& \bf{realizations}&$\mathbf{N_p^{1/3}}$&$\mathbf{L_{\rm box}}$ (Mpc/h)&$\mathbf{m_p} (10^{10} M_\odot)$ \\\hline\hline
    \textsf{G85L} & $0.85$&$0$&$15$&$1536$&$2000$&$18.3$\\\hline
    \textsf{NG10L} &$0.85$&$10$&$15$&$1536$&$2000$&$18.3$\\\hline
    \textsf{NG250L} &$0.85$&$250$&$10$&$1536$&$2000$&$18.3$\\\hline
    \textsf{G87L} &$0.87$&$0$&$3$&$1536$&$2000$&$18.3$\\\hline
    \textsf{G85S} & $0.85$&$0$&$5$&$1024$&$1000$&$7.8$\\\hline
%    \textsf{G83L} & $0.83$&$0$&$3$&$1536$&$2000$&$18.3$\\\hline
%    \textsf{G83S} &$0.83$&$0$&$5$&$1024$&$1000$&$7.8$\\\hline
    \textsf{G87S} &$0.87$&$0$&$5$&$1024$&$1000$&$7.8$\\\hline
%    \textsf{NG250L} &$0.85$&$250$&$10$&$1536$&$2000$&$18.3$\\\hline
    \textsf{NG250S} &$0.85$&$250$&$5$&$1024$&$1000$&$7.8$\\\hline
%    \textsf{NGm250L} & $0.85$&$-250$&$10$&$1536$&$2000$&$18.3$\\\hline
%    \textsf{NGm250S} &$0.85$&$-250$&$5$&$1024$&$1000$&$7.8$\\\hline
\end{tabular}

\caption{ A summary of the simulations in the \textsc{Eos Dataset} used for this work. There are two sets with different resolution. The \textsf{L} sets are large box sets of $8 (\si{Gpc/h})^3$ volume and $N_p=1536^3$ particles with mass $m_p = 1.83 \times 10^{10} M_\odot$.  The second set \textsf{S} has simulations with $1 (\si{Gpc/h})^3$ volume and $N_p=1024^3$ particles with  mass $m_p = 18.3 \times 10^{10} M_\odot$ . Snapshots have been saved at redshift $z=0,1$ and $2$.}
\label{tab:eos}
\end{small}
\end{center}
\end{table}

\subsection{The sizes of topological features}\label{sec:size}

As a first step in the analysis, we build intuition on the typical scales of topological features in the simulations and connect them to scales we expect to observe in large scale structure. Another objective of this section is to show that dividing the simulation box into sub-boxes does not bias our results, if the sub-boxes are large enough. The advantage of dividing into sub-boxes is that we are able to speed up our persistence calculations.

\begin{figure}
    \centering
    \includegraphics[width=0.325\textwidth]{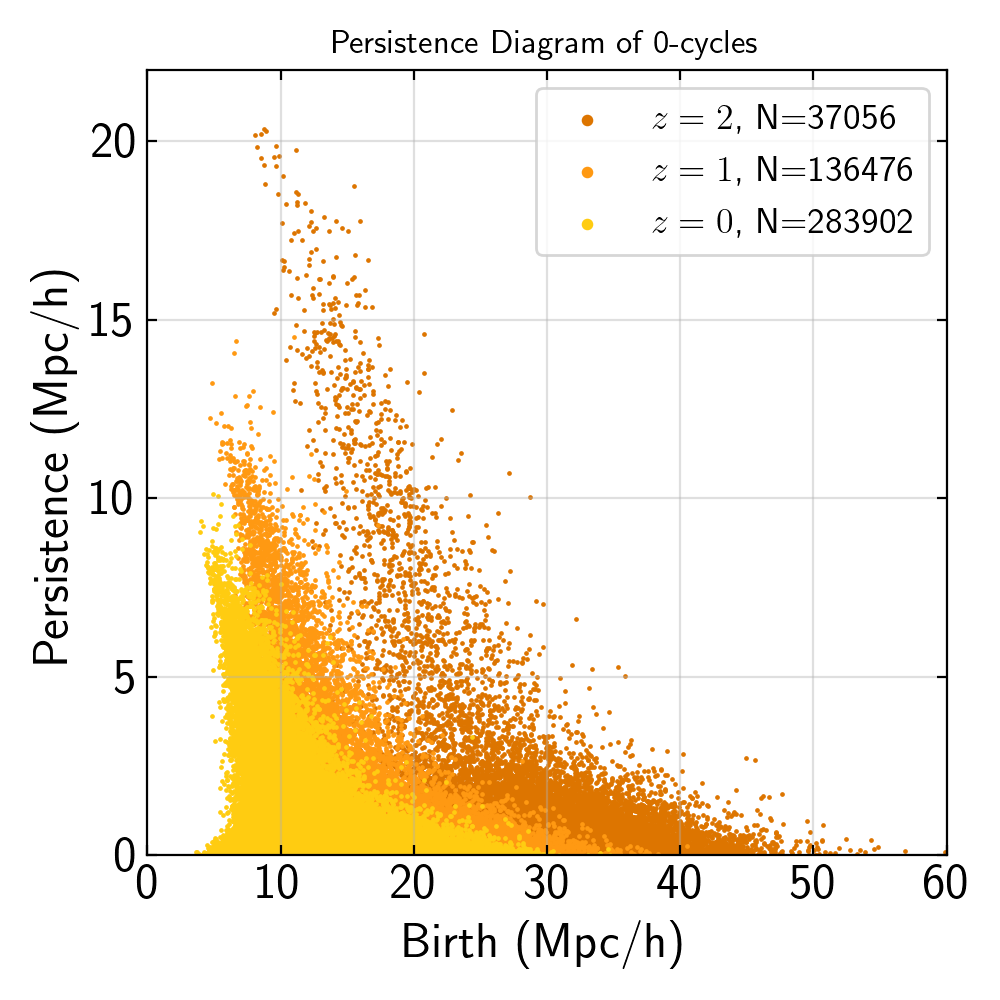}
    \includegraphics[width=0.325\textwidth]{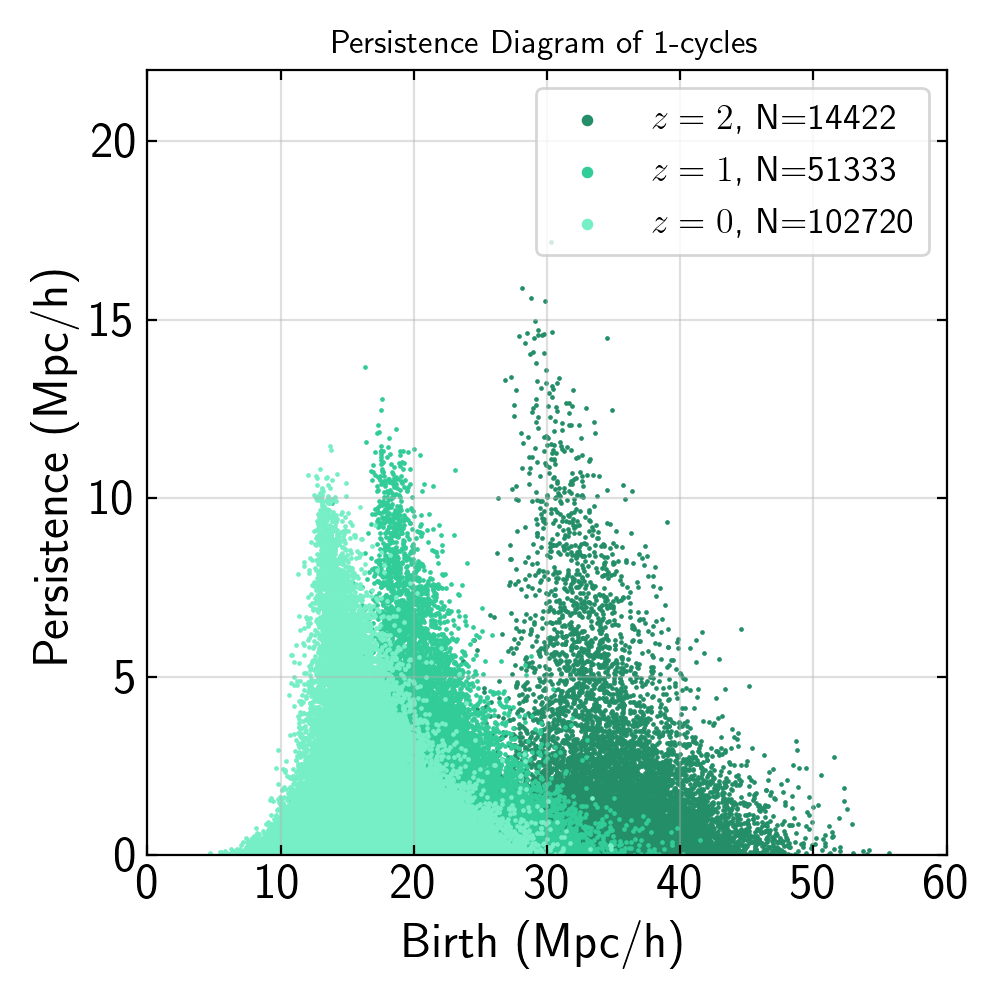}
    \includegraphics[width=0.325\textwidth]{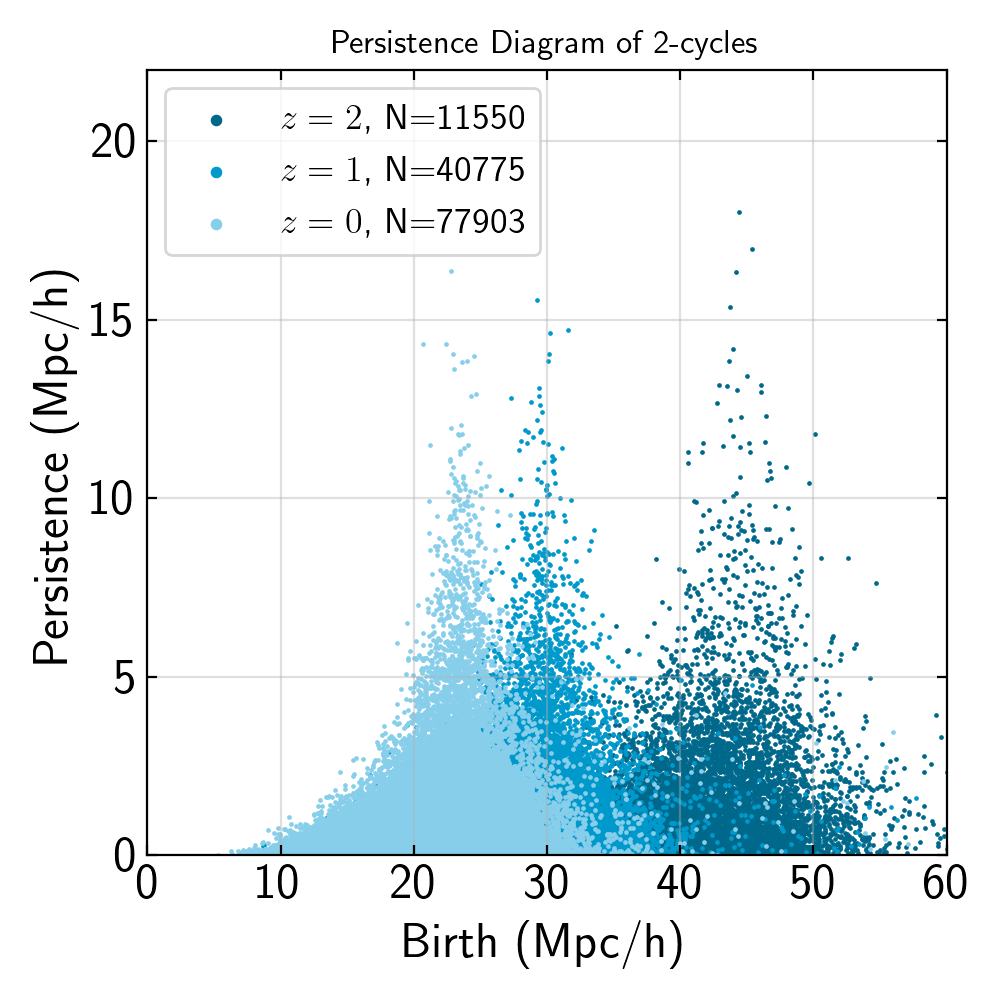}
    \caption{The persistence diagram for $0$-cycles (left), $1$-cycles (middle) and $2$-cycles (right) for an $\alpha$DTM$\ell$-filtration in a 1 $(\si{Gpc/h})^3$ sub-box of \textsf{G85L} at redshift $z=0,1$ and $2$. The total number of  independent cycles at each redshift is given.}
    \label{fig:alpha1000sub-boxValid}
\end{figure}

As explained in the previous section, the main tool for summarizing the persistent topology of a dataset is the persistence diagram. 
Let us take therefore as an example the persistence diagram for an $\alpha$DTM$\ell$-filtration in a sub-box with side $1$ $(\si{Gpc/h})^3$ of realization $\#1$ of the set with Gaussian initial conditions, \textsf{G85L}, shown in Figure \ref{fig:alpha1000sub-boxValid}.  In Figure \ref{fig:alpha1000sub-boxValid}, each dot represents $0$-, $1$- and $2$- cycles at redshifts $z=0, 1$ and $2$. 

\paragraph{0-cycles.} Each homology class of 0-cycles corresponds to a cluster of halos. In the $\alpha$DTM$\ell$-filtration, a 0-cycle is added to the filtration at the DTM value of the halo, in other words $\nu_{\rm birth}=\textrm{DTM}(x_{\rm halo})$. Therefore halos in overdense regions (i.e.\ those with small $\textrm{DTM}(x_{\rm halo})$) will be added to the filtration first.
As we increase the coarsening scale, halos are connected by edges and correspondingly no longer independent as 0-cycles. We therefore observe the deaths of homology classes as clusters merge. Recall that in the $\alpha\textrm{DTM}\ell$-filtration the balls around outliers grow more slowly. 
The scale at which a cluster dies, $\nu_{\rm death}$, represents the scale at which it is included into another cluster\footnote{For small $\nu$, the DTM value plays a larger role in the interpretation of $\nu$, due to the definition of the $\alpha$DTM$\ell$-filtration. For large $\nu$ (including in general $\nu_{\rm death}$), we can think of $\nu$ as providing the length scale of the feature. In general, if the length scale is not clean because of DTM, we checked also the alpha filtration and found consistent scales.}. The largest cluster is the one in which all halos in the catalog belong to the same cluster.
For this halo catalog, the largest cluster is achieved for a connecting scale of around  $30-50\, \si{Mpc/h}$ depending on redshift. 
\paragraph{1-cycles.} 
In our dataset, $1$-cycles are loops formed by connecting halos separated by some maximum filtration scale $\nu_{\rm birth}$. 
These loops trace dark matter filamentary structures, which host most of the halos, and consequently galaxies, in the universe \cite{Bond:1995yt,Colberg:2004cd,Hahn:2006mk,Tempel:2013wha}. The persistence diagram for $1$-cycles, Figure \ref{fig:alpha1000sub-boxValid} (middle panel), exhibits a peak at each redshift, indicating the typical scale at which the loops with longest persistence are formed. 
We also observe a maximum scale of death, varying with redshift, which generally indicates the shortest ``semi-minor axis'' (imagining a planar ellipse) of the largest loops. A 1-cycle death also occurs when two filament loops join, i.e.\ a sub-loop becomes topologically equivalent to its parent loop. The largest death scales correspond to the sizes of the parent loops. While the interpretation of these features is clear, predicting them analytically from initial conditions remains a challenge \cite{Bond:1995yt,Shen:2005wd,Cadiou:2020xmo}. 
\paragraph{2-cycles.}  The diagram for $2$-cycles represents the distribution of voids across scales of the simulation. 
These can be directly related to voids in a halo catalog (see \cite{Xu:2018xnz} for a void-finding algorithm based on persistent homology), although there can be differences with the results of conventional void-finders due to differing definitions of voids. The radius of a void generally corresponds to the filtration scale at its death, therefore the largest voids are roughly found at the peak in Figure \ref{fig:alpha1000sub-boxValid} (right panel) for each redshift. Note that again a void death can also correspond to the combination of a parent void and sub-void. The largest scales of death correspond to the radius of the full void. This scale at redshift $z=0$ is $ \sim 45 ~ \si{Mpc/h}$, which is somewhat smaller than the result from other void finding algorithms on the same type of catalog \cite{Chan:2018piq}.

A common denominator among these three types of features is that the typical scale of the largest feature is $\mathcal O(10)$ Mpc/h, which is around a hundred times smaller than a side of the sub-box in which they are calculated. This statement is however oversimplified: because these are features living in a 3-dimensional space, we might prefer to compute the effective volume of these features, which is not necessarily spherical. For example, among the filament loops, we only probe the maximum size of the ``semi-minor axis,'' but a loop might have a very ``squashed'' shape and therefore a much larger ``semi-major axis.'' 
Given this uncertainty, we might worry that dividing the simulation into sub-boxes creates some topological features and destroys others \footnote{In addition, there might be features that are larger than the size of the sub-boxes themselves. These features would have to be of scale of order $\mathcal O(1000)$ Mpc/h. From Figure \ref{fig:alpha1000sub-boxValid} it seems that even features of $\mathcal O(100)$ Mpc/h are suppressed, therefore it is reasonable to assume that very large features are very rare and do not impact the statistics we are interested in.}. These effects are localized at the boundary of a given sub-box. To minimize these effects, we should find the right compromise between having large sub-boxes and computational speed. For the \textsc{Eos Dataset}, the total volume of a single realization is $8 (\si{Gpc/h})^3$. We verified that dividing the box into the fewest number of sub-boxes, i.e.\ $8$ sub-boxes of $1\, \si{Gpc/h}$ per side gives no significant differences with respect to the full box, while providing computational speedups. The worst-case number of simplices in the $\alpha$-filtration is $N_{\rm halo}^{O(2)}$ giving a worst case running time for the persistence calculation (after the filtration is built) of $O(N_{\rm halo}^{6})\propto \textrm{Vol}^{6}$. We will therefore run the persistence calculation on sub-boxes with volume $(1\, \si{Gpc/h})^3$ throughout the analysis and subsequently recombine sub-boxes into complete realizations when performing our statistical analyses.

\paragraph{Conservative cuts of the persistence diagram.} 
Recall the intuition that long-lived topological features represent ``true'' aspects of the data. This intuition is formalized in the aforementioned stability theorems, which rely crucially on the fact that long-lived features cannot be created or destroyed by small perturbations.
For this reason, one might want to ignore low-persistence regions of a persistence diagram when performing inference. In an observational context, this would account for errors in the determination of galaxy positions.
For our current purposes, another reason to ignore certain regions of
a persistence diagram is the accuracy of our simulation code: while studies of the dark matter power spectrum show that different codes agree within one percent accuracy at scales of a few megaparsecs \cite{Schneider:2015yka}, including baryons presents several challenges (see e.g. \cite{Martizzi:2013aja,Cui:2014aga,Bocquet:2015pva} for studies of the impact of baryons on the halo mass function). 
In this case, not only the scale of persistence, but also the birth scale is affected; we should not trust features that are
created and destroyed
at scales where we cannot trust the dynamics. We decide to cut our persistence diagram by excluding all features dying at scales smaller than $5$ Mpc/h. Note that this corresponds to the radii of the individual balls in our coarse-graining process, so that halos joined at $\nu=5$ Mpc/h are in fact separated by a distance $10$ Mpc/h. This ensures that the topological signatures we rely on to detect non-Gaussianity are not based on simulated dynamics that we cannot trust. We then generate our statistics from these modified persistence diagrams as previously described. We have checked that this cut affects our final results to within $5\%$.
\subsection{The number density of topological features}
Our statistical method relies on comparing persistence diagrams (or more precisely, statistics derived from them) of different datasets.
We expect the total number of halos in a sub-box to vary not only as a function of cosmology, but also across different sub-boxes and realizations of the same cosmology, due to cosmic variance.
Even only considering variations across different cosmologies, we expect degeneracies. These variations are generically not under control when dealing with real data due to e.g.\ galaxy bias, and they can have a significant impact on the persistence diagram. We therefore choose to avoid this systematic by subsampling our halo catalog. Additionally, subsampling is necessary for our asymptotic template method, see Sec.\ \ref{sec:degen}.

\paragraph{Subsampling procedure.}\label{sec:subsamp}
We subsample the point cloud at the level of sub-boxes.
Given a set of simulations we want to use, we find the sub-box that has the minimum number of halos, $N_{\rm min}$, in it and subsample all other sub-boxes, including those from different cosmologies if present in the set, to $N_{\rm min}$ by randomly removing halos from the point cloud, before performing our computation (see \cite{Paranjape:2020wuc} for a discussion of the effect of subsampling on Voronoi cells). To remove possible variance due to this subsampling, we perform the computation for several subsamples of a given sub-box and average the result. We have checked that, for the simulations and subsamples considered, our results are stable under the inclusion of additional subsamples for $N\geq 3$ subsamples.\footnote{Subsampling was also used by \cite{KIMURA2017722}, which studied the internal consistency of data from the SDSS survey. }

\paragraph{Curve normalization.}
\label{sec:curveNorm}
The topological curves we calculate count the number of homology classes in particular regions of the persistence diagrams. As such, they probe the total number of topological features. Even for fixed cosmology and subsampled simulations, the number of features can vary between simulations. In fact, for the cosmologies we are comparing (e.g.\ $\fnl=10$ vs.\ $\fnl=0$), this variance can compete with variation due to different underlying cosmology. As such, we find it advantageous to normalize the topological curves to achieve a maximum value of unity. For $B_p(\nu),P_p(\nu),$ and $D_p(\nu)$, the curves subsequently have the interpretation of empirical distribution functions of births, persistences, and deaths. We also normalize the Betti curves $b_p(\nu)$, which do not have the same interpretation as the other curves, but also benefit from having their variance across simulations reduced. 

\paragraph{Redshift dependence.} As we already remarked, the number of topological features is necessarily dependent on the number density of the catalog. The number density of halos increases at low redshifts, so, naively, one would think that the best option is to look at nearby halos/galaxies. However, when constraining parameters related to the initial conditions, gravitational non-linearities can erase the primordial information and they do so on larger and larger scales at lower and lower redshifts. A simple example of this process is the non-linear collapse of overdense regions of the universe, which virialize forming compact objects such as dark matter halos: the dynamics of dark matter particles inside the halo involves stochastic processes that result in dissipation and drift \cite{Ma:2003cq}. There is therefore a tension between having a sufficient number of tracers and primordial signature remaining unburied after late-time physics. Figure \ref{fig:alpha1000sub-boxValid} gives an idea of how our topological features change with redshift. The main difference appears to be the halo number density, which decreases as the redshift increases. As such, clusters, filament loops and voids are born and die at larger scales. Additionally, there are fewer topological features in general at larger redshifts. 
In the next section, we find that the best constraints come from redshift $z=1$. Our interpretation is that at $z=2$ there are not enough topological features to extract information about $\fnl$, while at $z=0$ nonlinearities dominate over the primordial signal.

\subsection{Primordial non-Gaussianity and persistent homology}\label{sec:png}
In this section, we review the local ansatz for primordial non-Gaussianity and its well-known signature in the halo power spectrum. We then examine the topological signature of local non-Gaussianity as computed by persistent homology, investigating the change in persistence images as non-Gaussianity is introduced. We argue that the persistence images (and some of the topological curves) summarize certain well-known effects of non-Gaussianity, and in addition predict signatures in new observables that merit follow-up studies.
\subsubsection{Local non-Gaussianity and scale-dependent bias}
While our pipeline can be used to constrain any cosmological parameter \cite{bcsinprogress}, here we investigate primordial non-Gaussianity of the local type, which is modeled as a Taylor expansion of the primordial fluctuations around a Gaussian field at position $\bx$,
\begin{equation}\label{eq:locpng}
    \zeta(\bx) = \zeta_{\rm G}(\bx) + \frac 35\fnl\left(\zeta^2_{\rm G}(\bx) - \langle \zeta_{\rm G}^2\rangle \right) +\mathcal O\left(\zeta_{\rm G}^3\right) ,
\end{equation}
with $\zeta$ the comoving curvature perturbation, $\zeta_{\rm G}$ a Gaussian random field, and $\fnl$ the non-linearity parameter which defines the amplitude of non-Gaussianity. We neglect cubic and higher-order contributions in this work. This type of primordial non-Gaussianity is generically realized in multi-field models of inflation \cite{Salopek:1990jq,Bartolo:2001cw,Bernardeau:2002jf,Bernardeau:2002jy,Rigopoulos:2005ae,Rigopoulos:2005xx,Seery:2005gb,Vernizzi:2006ve}, and the corresponding bispectrum peaks  in  the  squeezed  limit,  where  single-field  models necessarily  produce  very  small non-Gaussianity as implied by consistency relations \cite{Maldacena:2002vr,Creminelli:2004yq}. The most stringent bounds on $\fnl$ are currently provided by the Planck collaboration, with $|\fnl| \lesssim 10$ at $95\%$ confidence level \cite{Akrami:2018odb}.

\begin{figure}
    \centering
    \includegraphics[width=0.49\textwidth]{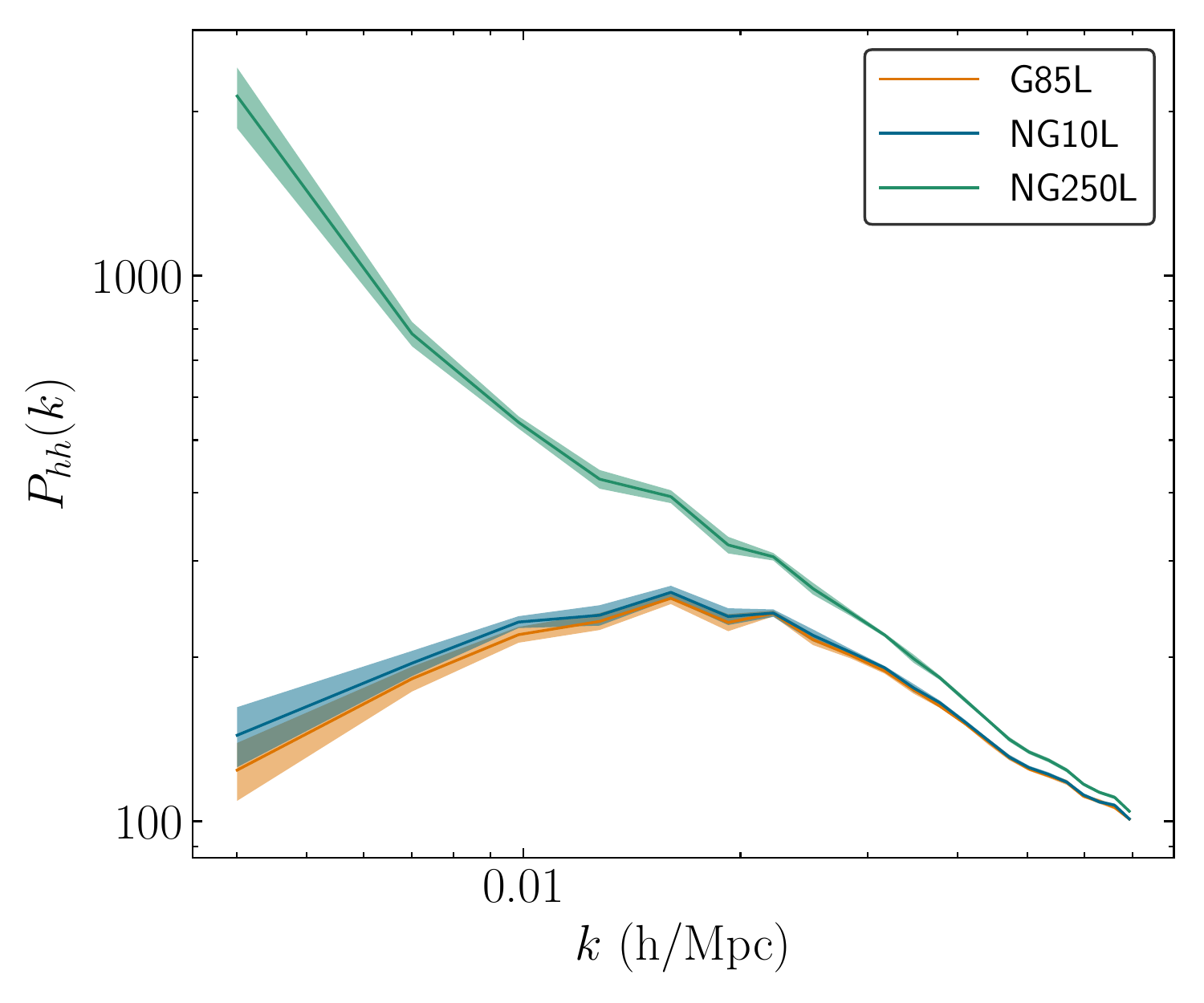}
    \includegraphics[width=0.49\textwidth]{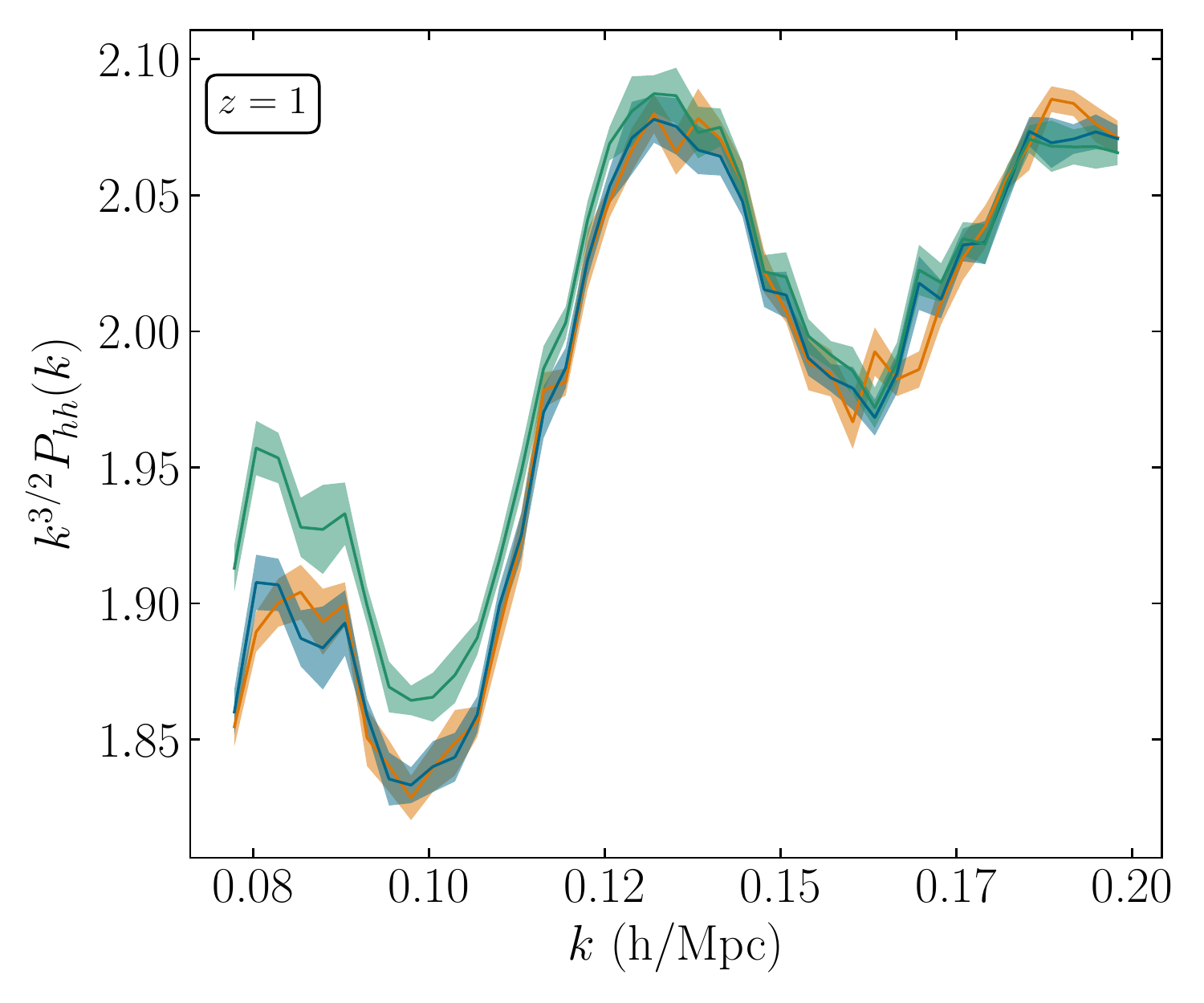}
    \caption{Halo power spectrum as measured from the \textsc{Eos} subsets with Gaussian and non-Gaussian initial conditions, \textsf{G85L}, \textsf{NG10L} and \textsf{NG250L}, corresponding to $\fnl=0$, $\fnl=10$ and $\fnl=250$ respectively, as a function of scale at redshift $z=1$ at large (left panel) and small (right panel) scales. The small scale halo power spectrum (right) has been multiplied by $k^{3/2}$ for easier visualization. Shaded areas indicate the standard deviation from the mean over $12$ realizations.}
    \label{fig:halopspng}
\end{figure}

In large-scale structure, a coupling of long and short modes of the type Eq.\ \eqref{eq:locpng} generates a distinctive signature in the large-scale clustering of dark matter tracers, such as galaxies, which is observable in the power spectrum 
\begin{equation}\label{eq:pngloc}
    P_g (k) = \left(b_g + \frac{12}{5}\frac{\fnl}{\mathcal M(k)} b_\zeta\right)^2 P_m (k),
\end{equation}
where $b_g$ is the linear galaxy bias, $\mathcal M(k) \propto k^2$ at large scales is the transfer function relating the primordial curvature perturbation $\zeta$ to the linear dark matter density field $\delta$, and $b_\zeta$ is the bias associated to primordial non-Gaussianity (see \cite{Dalal:2007cu,Matarrese:2008nc,Slosar:2008hx} and \cite{Biagetti:2019bnp} for a review). While in the absence of primordial non-Gaussianity, the galaxy power spectrum at large scales should relate to the matter power spectrum via a constant, the second term in Eq. \eqref{eq:pngloc} enhances (suppresses) power at large scales for positive (negative) $\fnl$ with a factor scaling as $1/k^2$ and proportional to a bias amplitude $b_\zeta$. In Figure \ref{fig:halopspng} we show the halo power spectrum as measured from the \textsc{Eos Dataset} for the simulations \textsf{G85L}, \textsf{NG10L} and \textsf{NG250L}, corresponding to $\fnl=0$, $\fnl=10$ and $\fnl=250$ respectively, at redshift $z=1$.
\subsubsection{Local non-Gaussianity and persistent homology}
The signature of primordial non-Gaussianity in 
persistent homology
has been investigated in the literature to a certain extent \cite{Cole:2017kve,Feldbrugge:2019tal}, but to the best of our knowledge, not in the context of large-scale structure\footnote{Topology in the form of the genus has been used to characterize the signature of primordial non-Gaussianity in large scale structure \cite{Matsubara:1994wn,Matsubara:1995dv,Matsubara:1995ns,Seto:1996gc,Gay:2011wz}. As remarked previously, persistent homology provides strictly more topological information than the genus.}.
Because of the high-order nature of the correlations encoded in the persistent homology calculation, it is challenging to predict the complete behavior of a persistence diagram under the addition of non-Gaussianity. On the other hand, the quantities we compute numerically are easily interpretable as standard topological features. The interpretability of novel approaches to cosmological data analysis is important to keep in mind as machine learning (in particular deep learning) approaches become ubiquitous. 
This interpretability allows us to make cuts on our statistics that ensure they are using information generated by the simulations at trustable scales (cfr. \ref{sec:size}). Another advantage of this interpretability is the possibility of discovering new signatures of physics that can later be derived from first principles. 

\paragraph{Persistence Images.} We therefore turn to investigate the  signature of local primordial non-Gaussianity in the multiscale topology of large scale structure.
In this section, we examine the persistence images (PIs) computed from our simulations and how they are affected by local non-Gaussianity. This allows us to gather some intuition regarding the signature of local non-Gaussianity in topological observables and predict which topological curves might be most efficient in detecting $\fnl$. We use this information in the next section, where we describe how to quantitatively constrain $\fnl$.
\begin{figure}[h]
\centering
\includegraphics[width=0.33\textwidth]{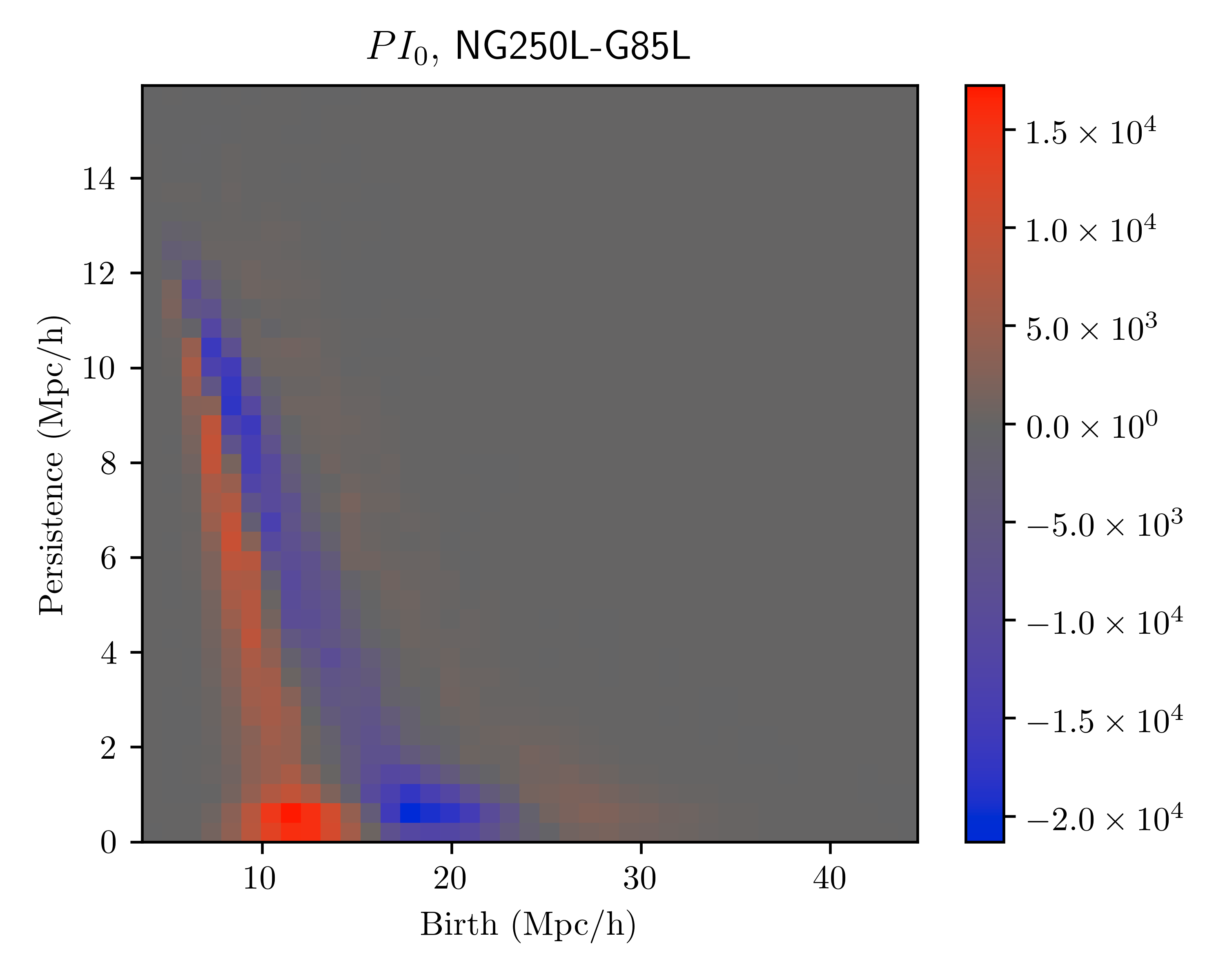}\includegraphics[width=0.33\textwidth]{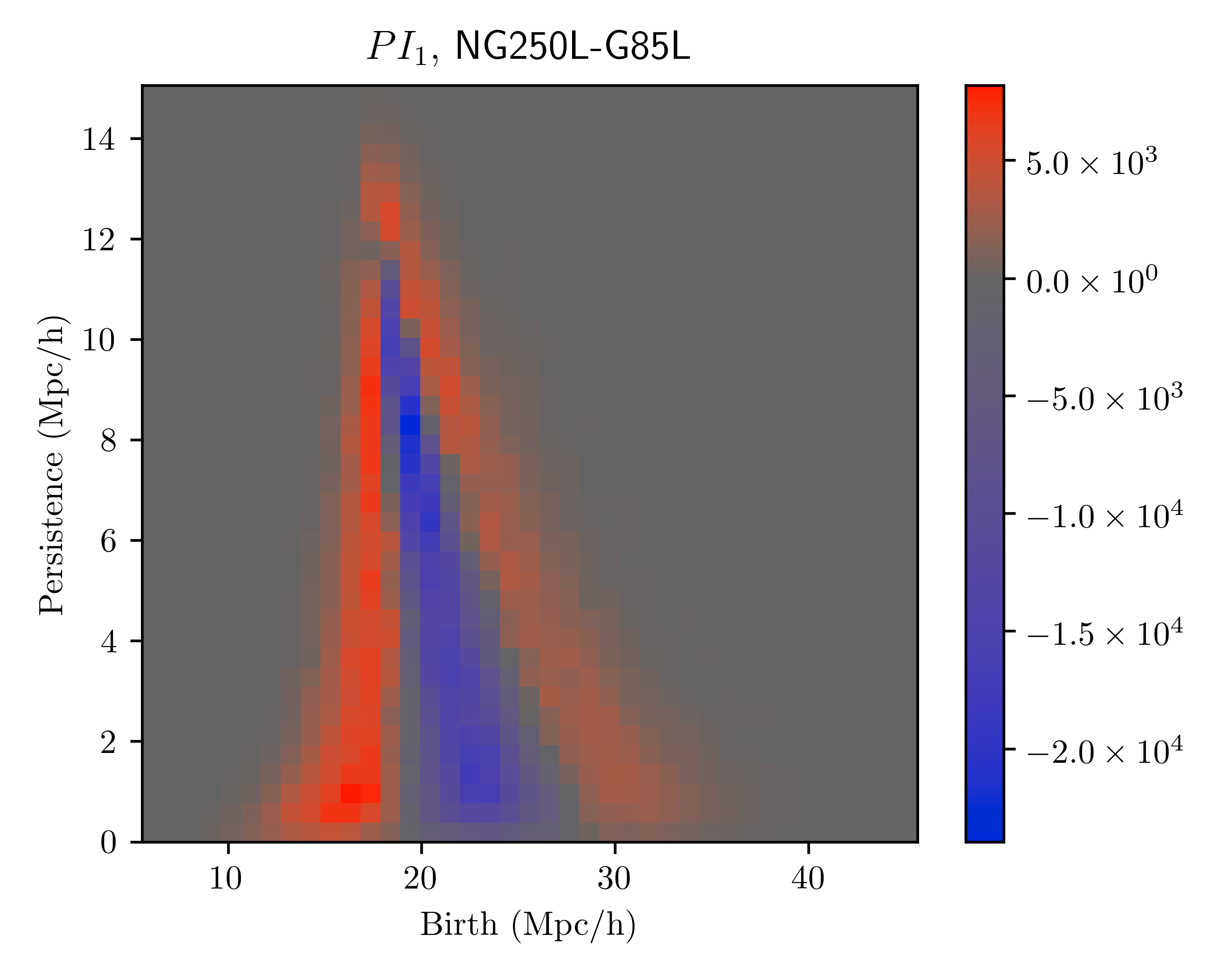}\includegraphics[width=0.33\textwidth]{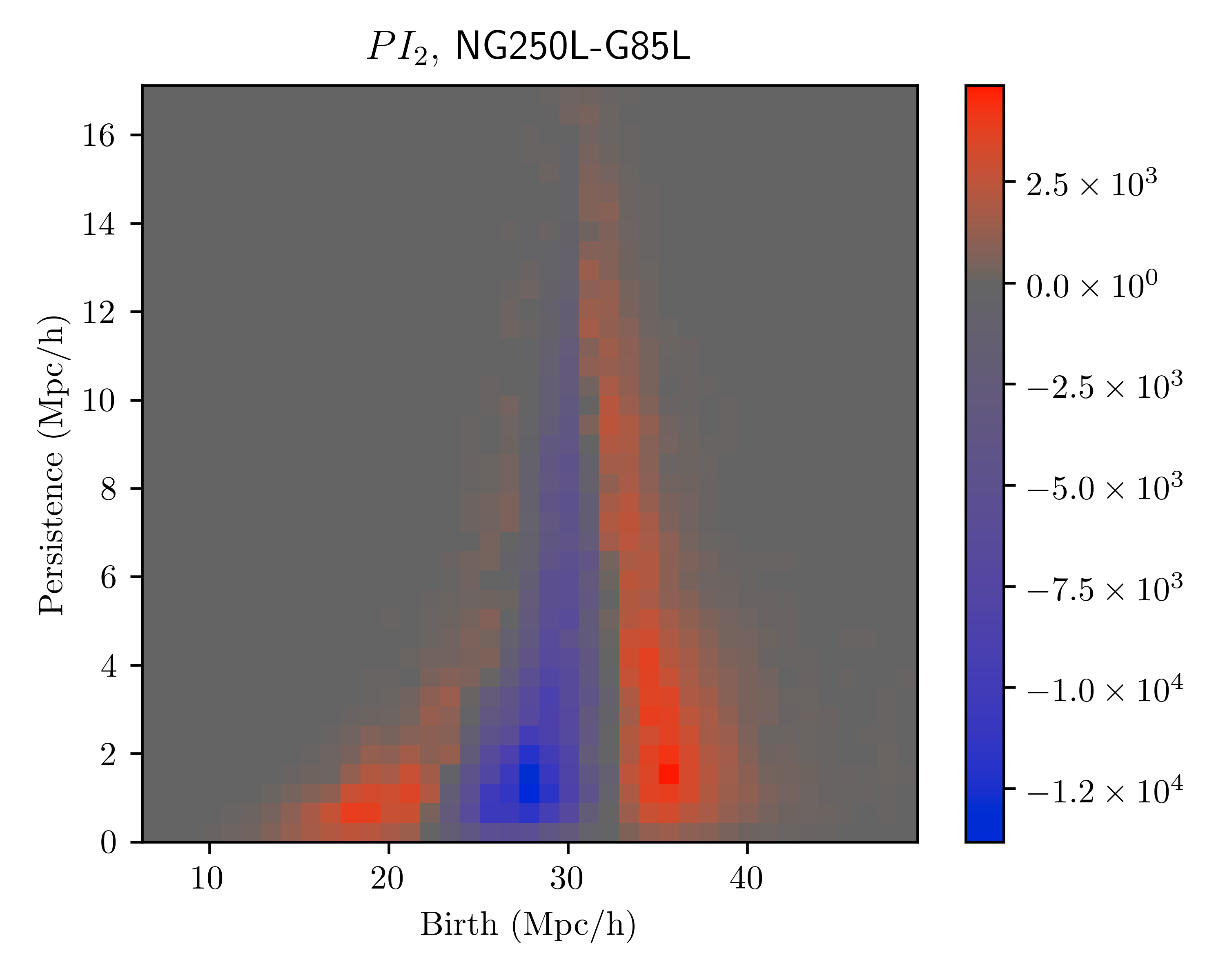}
\includegraphics[width=0.33\textwidth]{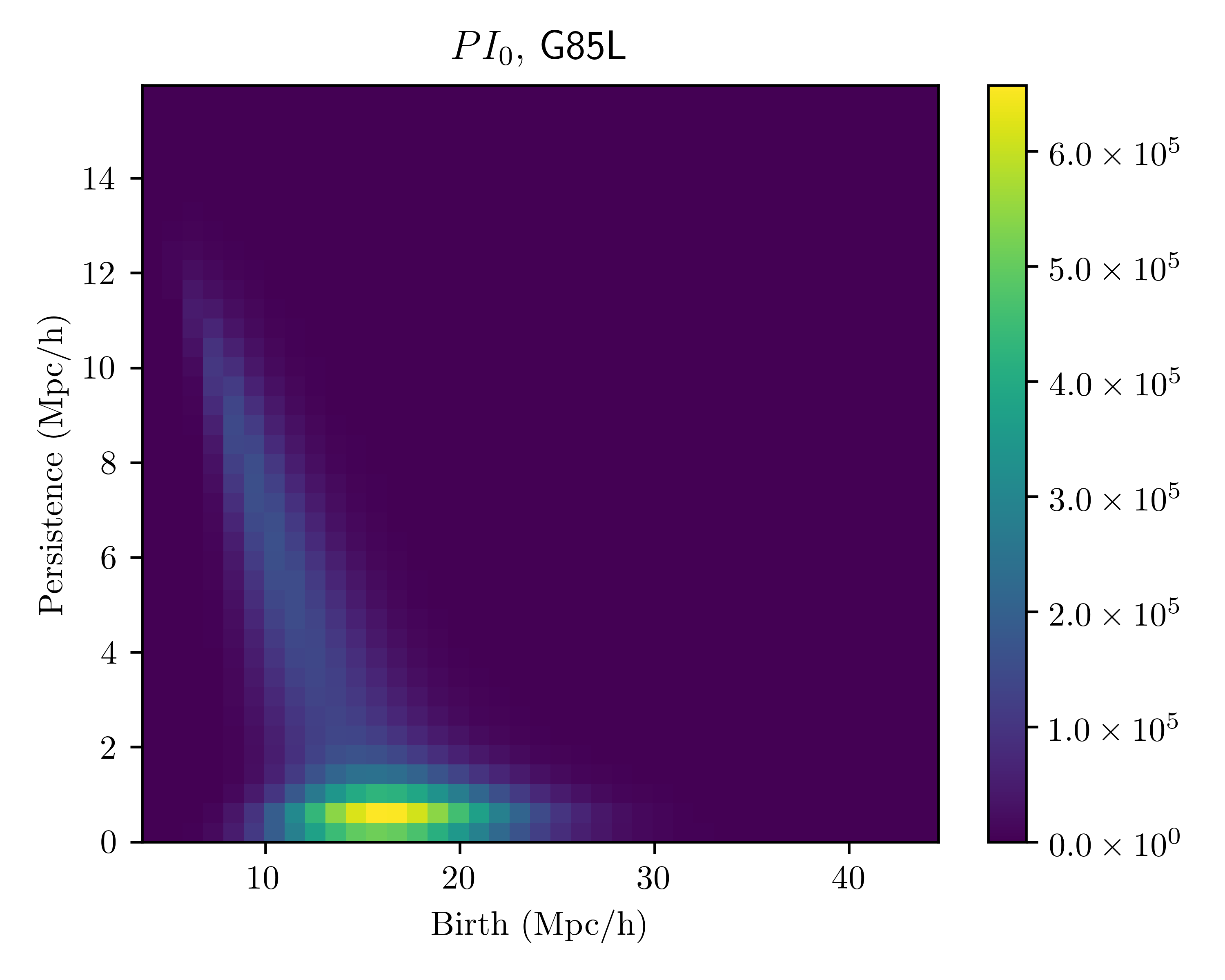}\includegraphics[width=0.33\textwidth]{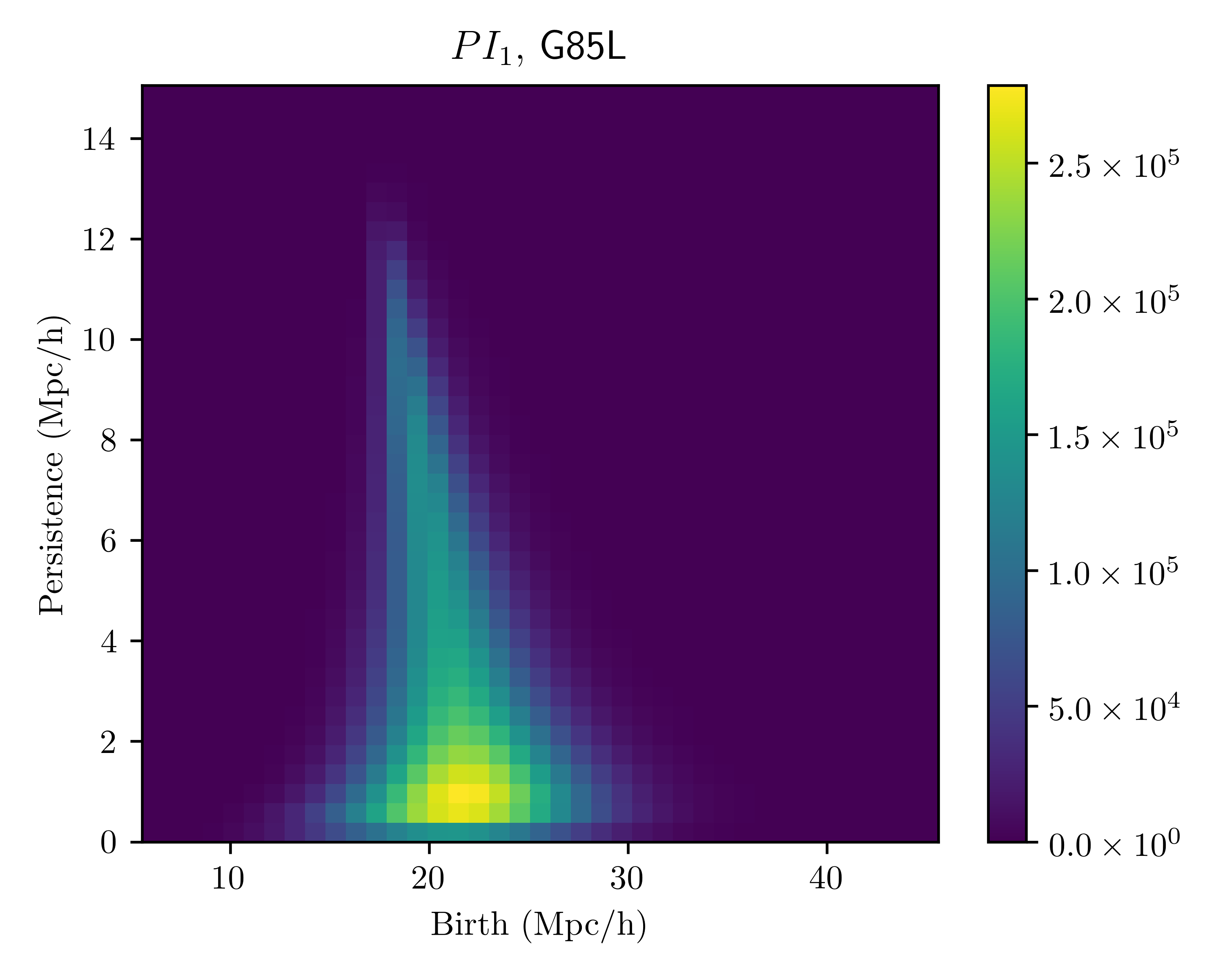}\includegraphics[width=0.33\textwidth]{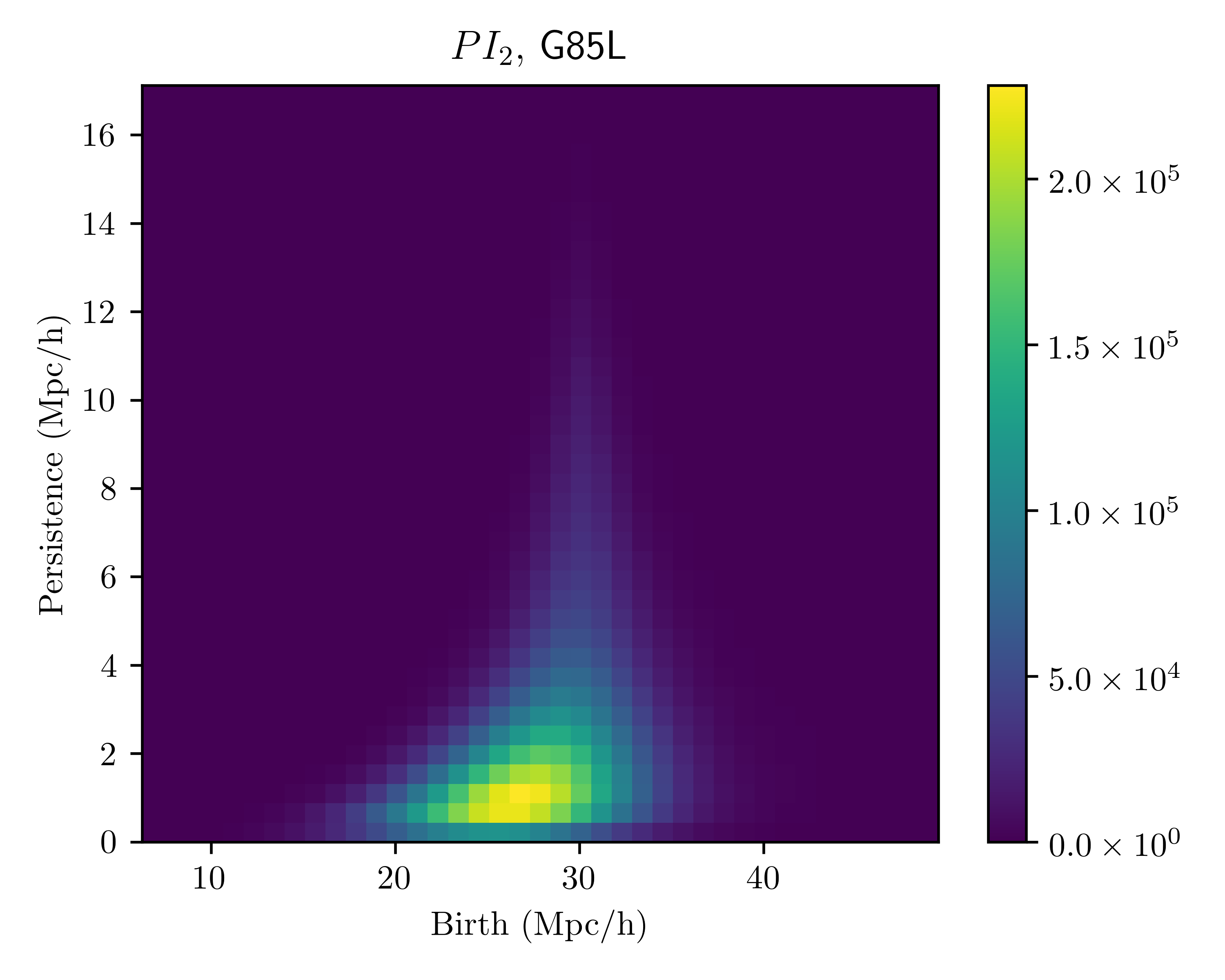}
\caption{The effect of local non-Gaussianity on PIs for the $\alpha$DTM$\ell$-filtration. In the top row we subtract the PIs averaged over \textsf{G85L} from the PIs averaged over \textsf{NG250L}. In the bottom row we show the PIs averaged over \textsf{G85L}. 
}\label{fig:PIintuition}
\end{figure}
We can attempt to interpret the physics of our signature by examining the PIs, see Figure \ref{fig:PIintuition}. Although we are ultimately interested in more realistic values of $\fnl$, we study the signature generated by $\fnl=250$, 
as the signal is large and results in clean patterns when averaged over our simulation volume. Generating clean images at e.g.\ $\fnl=10$ appears to require more simulation volume. 

We show the difference between PIs with non-Gaussian initial conditions and Gaussian initial conditions in Fig.\ \ref{fig:PIintuition}. The first feature that sticks out is a similarity between the signatures of nonzero $\fnl$ in all the topological features. At small birth scales, there is a region in which the non-Gaussian simulations exhibit more topological features. At intermediate birth scales, the Gaussian simulations have more features. At large birth scales, the non-Gaussian simulations once again dominate. Most of these regions tend to be oriented more vertically (i.e.\ along the direction of persistence) than horizontally (along the direction of birth).
The  PIs lead us to expect the topological curves $B_p(\nu)$ and $b_p(\nu)$ to outperform $P_p(\nu)$ in detecting $\fnl$. Slicing a persistence diagram across constant persistence equivalently sums over all births, giving $P_p(\nu)$ a smaller signal to detect. For $0$-cycles (left panel), deaths also look promising given the inclination of the PIs along the line of death (cfr. Figure \ref{fig:curveConstruct}). A decent strategy for detecting some particular physics is to study the signature in PIs before deciding how to slice the diagram. 

In addition to suggesting smart 1-dimensional statistics to use,
the multiband structure in the PIs is reminiscent of well-known effects of local non-Gaussianity. In particular, the halo mass function and void abundance both exhibit a ``turnover'' behavior. Focusing on the voids, positive (negative) $\fnl$ is predicted to give more (fewer) voids of small radius, and fewer (more) voids of large radius \cite{Kamionkowski:2008sr}. This behavior should manifest in the statistic $D_2$, taking the death of a 2-cycle to correspond to the size of the corresponding void. This behavior can be seen in $PI_2$. It is shown more explicitly in Fig.\ \ref{fig:templateAsymp}, which also features a range at large radius in which the non-Gaussian case exhibits more voids, perhaps due to details of our void definition or the fact that deaths count the combination of sub-voids and parent voids instead of just the void radius. We plan to investigate this feature, which is not computed by conventional void-finders, in future work.
Intriguingly, a similar structure is evident for the distribution of 1-cycle deaths $D_1$. This motivates new analytical work to derive the distribution of these features from first principles.

This structure extends even to the 0-cycles. 
In the $\alpha$DTM$\ell$-filtration, 0-cycles are born at the DTM value of the corresponding halo. 
Again, we observe a similar multiband structure. This encodes a change in the distribution of DTM values of the halos via their births. The meaning of the multiband structure in this case is the following: overdense regions become more overdense while some underdense regions become more underdense. For halos already belonging to overdense regions, the DTM value decreases (as their neighborhood becomes more overdense), while for some outliers the DTM value increases (as their neighborhood becomes more underdense). This agrees with the heuristic observation that positive $\fnl$ affects structure as if it were more gravitational evolution in overdense regions and less in underdense regions \cite{Dalal:2007cu,Kang:2007gs,Grossi:2008fm,Pillepich:2008ka}.

All in all, the interpretability of our statistics allows us to make contact with previously derived quantities, while motivating future study to derive other aspects of our results from first principles.

\section{Analysis pipeline}\label{sec:pipe}

Having determined how to prepare our dataset for the analysis, we describe the analysis pipeline. We would like to simulate as closely as possible an analysis of real data. The techniques we will use require comparing datasets. In galaxy surveys, one typically runs sets of simulations, so called mock catalogs, that resemble the survey's geometry and characteristics, with a fixed fiducial cosmology. These mock catalogs are used to compute expected systematics and uncertainties in the survey. In our setup, we reproduce this procedure by defining ``survey data'' and ``mock data'' from the simulation boxes. We then attempt to extract cosmological information from the survey data by comparing it to the mock data, which has a known fiducial cosmology. Our pipeline is composed of several steps:

\begin{itemize}
    \item \textsf{Step 1}: Run persistence calculation on all sub-boxes.
    \item \textsf{Step 2}: Find anomalies between survey and mock dataset.
    \item \textsf{Step 3}: Generate a template for each cosmological parameter of interest.
    \item \textsf{Step 4}: Compare datasets using templates and constrain cosmological parameters.
    \item \textsf{Step 5}: Account for degeneracies between different cosmological parameters
\end{itemize}

The pipeline constitutes a general method for comparing datasets using persistent homology. First anomalies between the mock dataset and the survey dataset are identified. Then a template-based method is used to associate such an anomaly with some particular physics. A notion of template optimality is then used to understand degeneracies between different cosmological parameters. 
Let us now go through all the steps.

\subsection{\textsf{Step 1}: Persistence calculation}
Given a sub-box of a simulation, we construct three filtrations: the $\alpha$-filtration, the $\alpha$DTM-filtration, and the $\alpha$DTM$\ell$-filtration. We show only the results for the $\alpha$DTM$\ell$-filtration in the main body, with the results for the other filtrations summarized in Appendix \ref{app:table}. We compute these filtrations 
for subsampled sub-boxes, as described in the previous section.
For each filtration, we compute the persistence diagrams and process them into the previously described statistics, as summarized in Figure \ref{fig:curveConstruct}. We end up with data vectors $\vec S^X_{\textsf{ ID}_i}$ whose components are number counts of topological features at increasing filtration parameter $\nu$, for each simulation set as defined in Table \ref{tab:eos}, with $i=1,...,N_r$ an index scanning through $N_r$ realizations and $X=b_0,\,b_1,\,b_2,\,B_1,\,B_2,\,P_0,\,P_1,\,P_2,\,D_1,\,D_2$ scanning through statistics. Upon properly normalizing the number counts as indicated in Section \ref{sec:subsamp}, these vectors, except for the Betti curves, represent empirical distribution functions. 

\begin{figure}
    \centering
    \includegraphics[width=0.49\textwidth]{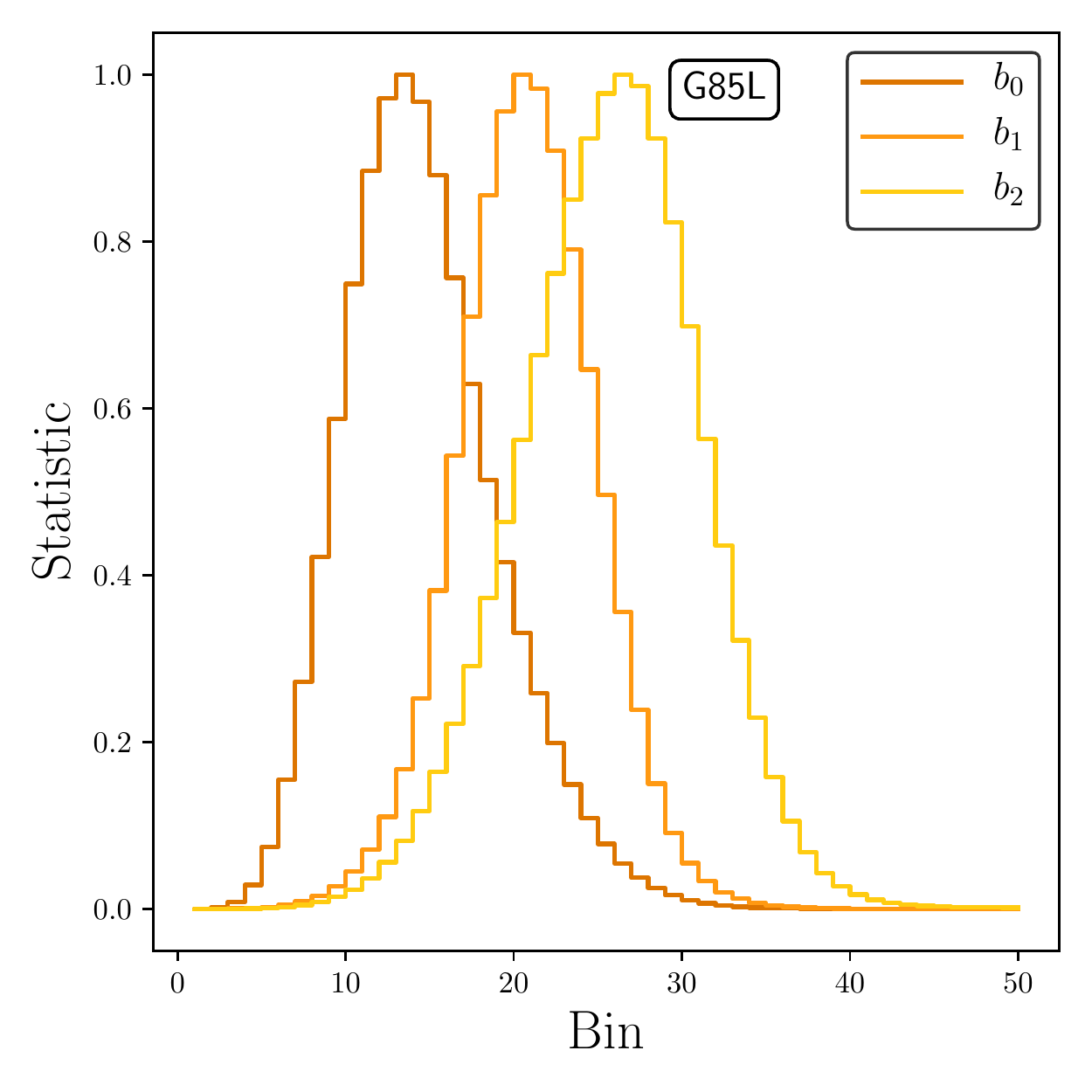}
    \includegraphics[width=0.49\textwidth]{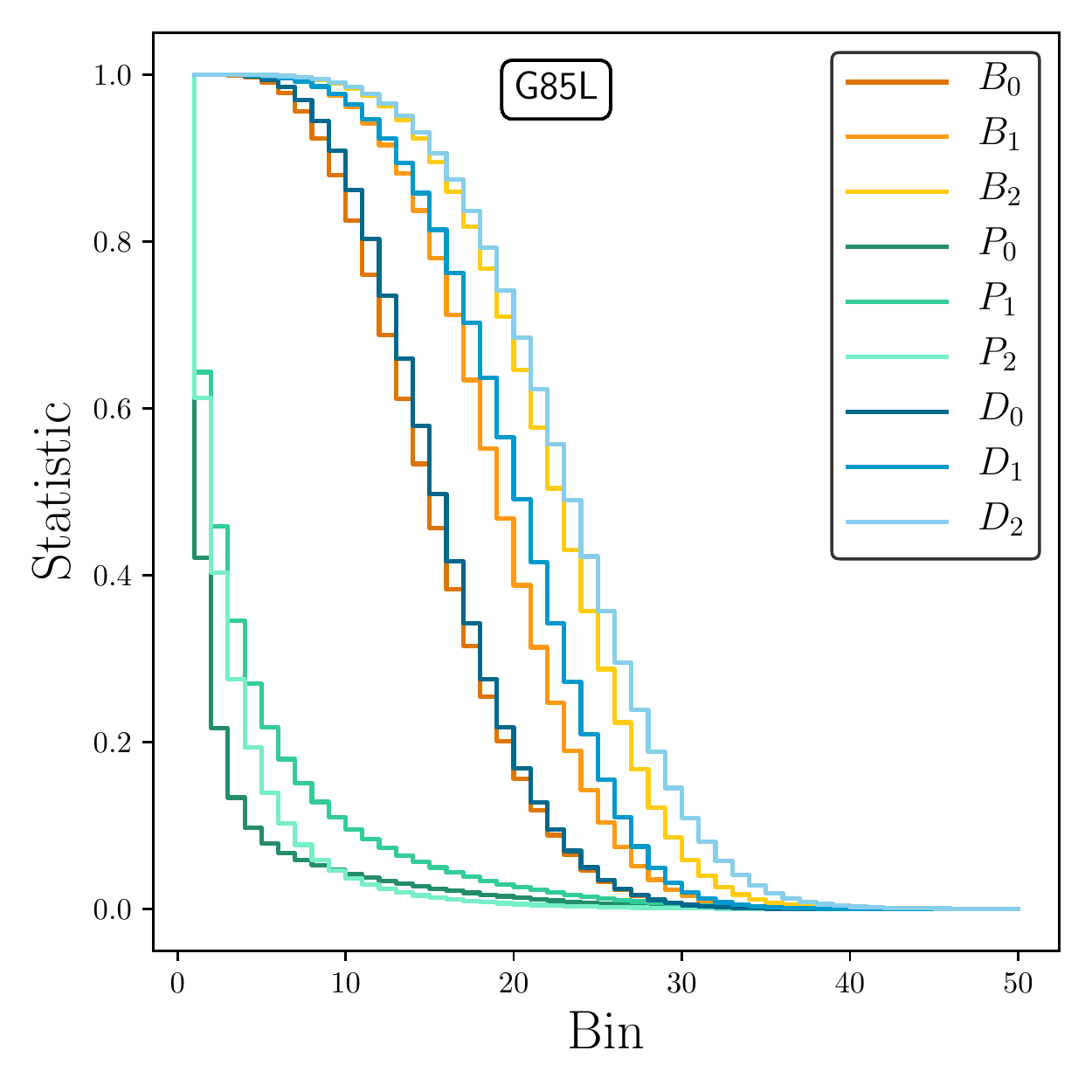}
    \caption{Step plot of one-dimensional curves with binning $N=50$, averaged over $12$ realizations of \textsf{G85L} at redshift $z=1$. On the left panel, we show Betti curves for $0$-,$1$- and $2$-cycles, on the right panel we show all one-dimensional curves as defined in the previous section (cfr. Figure \ref{fig:curveConstruct}).}
    \label{fig:curves}
\end{figure}

The number of components in each vector is determined by a binning of the filtration parameter $\nu$: the finest trustable binning would coincide with the N-body resolution. Since this is a very large and fine grained data-vector, we divide each 1-dimensional curve into $N$ linearly spaced bins, where we take the sum of the total counts inside the bin as value for the bin. We have tested several values for $N$ ranging from $50$ to $1000$ and verified that results do not change appreciably in this range. We take $N=1000$ for our final analysis, while we show $N=50$ in Figure \ref{fig:curves} for ease of visualization, also taking the mean over $15$ realizations of \textsf{G85L} at redshift $z=1$\footnote{In Figure \ref{fig:curves} the bin sizes for different statistics $X_p$ correspond to different binning scales}. When subsampling the sub-box, each vector is the result of averaging over $3$ different subsamplings of the simulation sub-boxes. 
When combining multiple realizations to enlarge the effective volume of our data (see next section for a definition of volume splits), we average each statistic over those realizations.
For the purpose of deriving strong constraints, one could imagine concatenating the different statistics into one large vector. We leave each statistic separate for the sake of identifying the sensitivities of various curves to different physics.

We use the library \texttt{GUDHI} \cite{gudhi:urm} for performing our persistent homology calculations. \texttt{GUDHI} computes its Delaunay triangulations using \texttt{CGAL} \cite{cgal:dy-as3-19b}. We compute nearest neighbors for the DTM function using the \texttt{scikit-learn} package \cite{scikit-learn}. The code performing these calculations to produce persistence diagrams and images, along with explanatory notebooks and examples, is publicly available\footnote{\url{https://gitlab.com/mbiagetti/persistent_homology_lss}}.%\textrm{https://gitlab.com/mbiagetti/persistent\textunderscore homology\textunderscore lss}}.

\subsection{\textsf{Step 2}: Anomaly detection}
\label{sec:anomaly}

First, let us set some terminology. We will distinguish subsets of \textsc{Eos} simulations between \textsf{fiducial boxes} and \textsf{survey boxes}. The former represent mock catalogs used in galaxy surveys, for which a fiducial cosmology is fixed as input. The latter have unknown cosmology, which we want to constrain\footnote{Of course, we do know the cosmology of the simulations we run. It would be interesting in this context to implement a blinded analysis as a further test of our method.}. While our survey data will be composed of multiple realizations of an N-body simulation, galaxy catalogs come from a single realization of the universe, our universe. 
Regardless, if a galaxy catalog is large enough, we can imagine it as being composed of a sum of subvolumes. Conversely, we should be able to patch together multiple independent subvolumes to approximate a galaxy catalog of larger volume. This is consistent as long as we are computing statistics that do not depend on scales larger than a subvolume. From Sec.\ \ref{sec:size}, we see that this is indeed the case for our topological statistics. 
\subsubsection{Dataset splitting}
\begin{figure}
    \centering
    \includegraphics[width=0.9\textwidth]{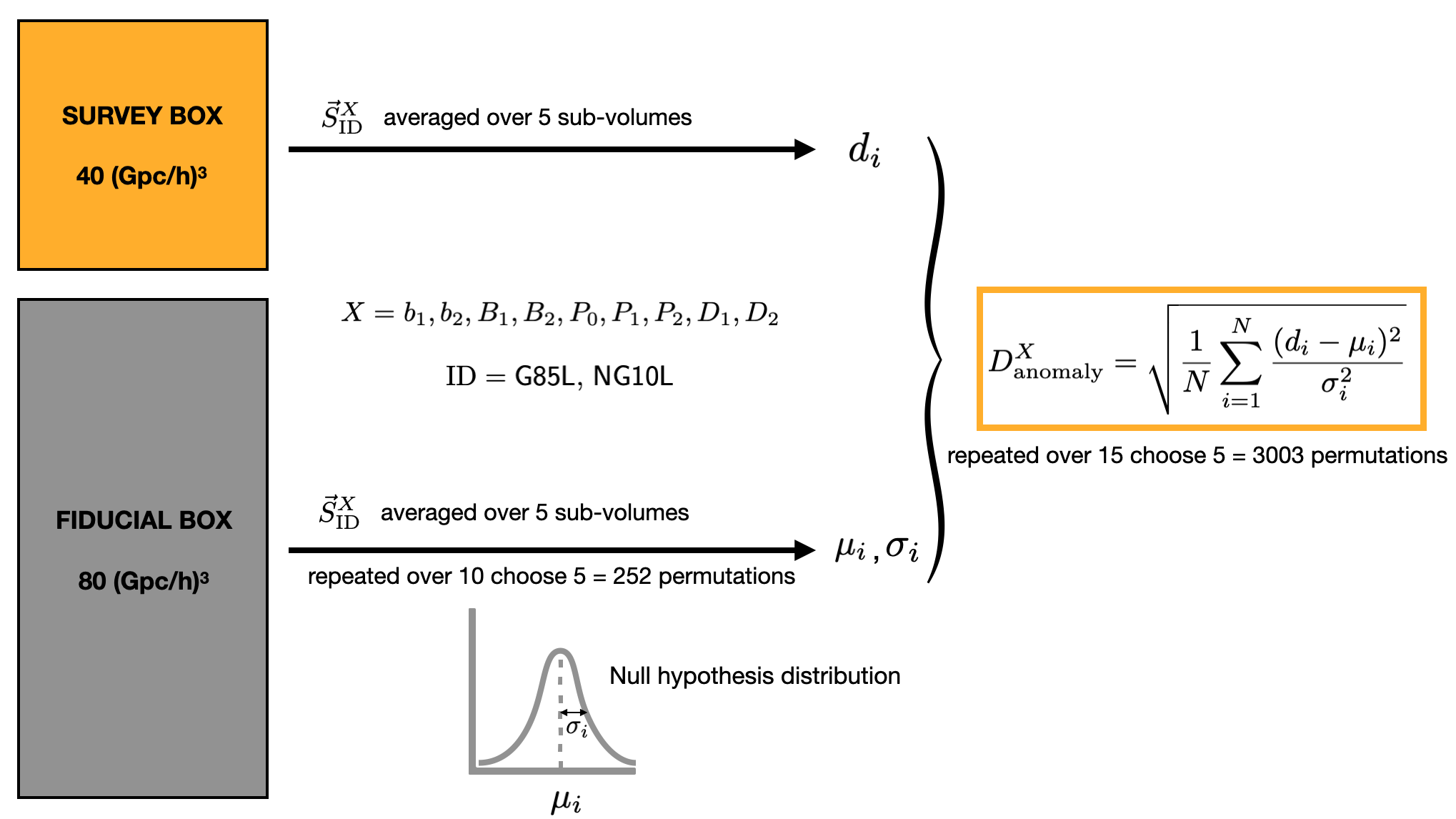}
    \caption{Workflow for determining significance, which we use both for anomaly detection and for template fitting. First, we split the total volume into a \textsf{survey box} and a \textsf{fiducial box}. We then evaluate our statistics in both boxes, by averaging over realizations in each of the two splits. In the \textsf{fiducial box}, we model the distribution of the average statistic (possibly computed against a template) for every combination of $5$ realizations from the total $10$  ones. This model is used to evaluate a $\chi^2$ function on the survey data. To establish robust confidence levels, we run the workflow twice, once using the fiducial cosmology for both boxes, and once where we look for primordial non-Gaussianity, \textsf{NG10L}, in the \textsf{survey box}. Using the terminology of \cite{brehmer2018constraining}, this approach in which the likelihood function is modeled via simulations is ``Approximate Frequentist Computation.''}
    \label{fig:workflow}
\end{figure}
Let us go into details on how we split the total volume of \textsc{Eos}.  A \textsf{survey box} is composed of $5$ realizations of $8\, (\SIGpc)^3$ each, totaling $40\, (\SIGpc)^3$ volume. A \textsf{fiducial box} is composed of $10$ realizations, totaling $80\, (\SIGpc)^3$ volume.  We will consider two categories of these data splits. In the first, which we willrefer to as \textsf{G/G}, both the splits follow the fiducial cosmology, which for the case at hand is defined by the set \textsf{G85L} (cfr. Table \ref{tab:eos}). Thus, both the \textsf{survey box} and the \textsf{fiducial box} have the same cosmology in this case. This sort of split is designed to model deviations between different realizations due to cosmic variance. These deviations are therefore used to set a baseline, in the sense that differences in cosmologies must be larger than this baseline to be considered significant. 
In the second split, which we call \textsf{G/NG}, we assign the non-Gaussian  simulations with $\fnl=10$, \textsf{NG10L}, to the \textsf{survey box}, while leaving the fiducial cosmology, \textsf{G85L}, for the \textsf{fiducial box}. In this context, the survey data has a different underlying cosmology than the fiducial one, which we would like to detect via topological methods.

For each split, we apply the following procedure: first, we evaluate each statistic, i.e. each data vector $\vec S^X_{\rm ID_i}$, on each of the $5$ realizations composing the \textsf{survey box}, and then average over them. Then, we evaluate the same data vector on all the $10$ choose $5 = 252$ combinations of $5$ realizations in the \textsf{fiducial box}, also averaging each time over $5$ realizations in each combination. In this way, we construct a null hypothesis distribution computing its mean and variance.

\subsubsection{Significance estimation}
The first step in searching for deviations from the fiducial cosmology is to identify statistically significant deviations of the survey data from the fiducial cosmology, in other words \emph{anomalies}.\footnote{We imagine perfect modeling of observational effects, so that statistically significant deviations are due to differences in cosmology. To some extent, a specific observational uncertainty would be accounted for when we determine the baseline for false detection using the \textsf{G/G} split, provided that the mock catalogs properly include that uncertainty.} To identify anomalies, we start by considering the \textsf{G/G} split. This is used to understand the impact of cosmic variance on our topological statistics. We calculate the weighted distance of a topological statistic from the \textsf{survey box} to the mean of the \textsf{fiducial box}'s topological statistic. More precisely, we compute the $\chi^2$ variable
\begin{equation}\label{eq:danomaly}
		D^X_{\rm anomaly}=\sqrt{\frac 1N\sum_{i=1}^N\frac{(d_i-\mu_i)^2}{\sigma_i^2}}
\end{equation}
where we recall that each topological statistic $X$ is represented by a vector of dimension $N$. Here $\vec{d}$ is the survey data vector for that statistic, $\vec{\mu}$ is the mean of the fiducial data vector, and $\sigma_i^2$ is the variance of the $i$-th component of the statistic in the mock data.\footnote{  Note that to ensure the validity of our model, we must additionally truncate bins in which the distribution of the fiducial data has exceedingly small variance.}

If the statistic of the fiducial data were sampled from a multivariate normal distribution with mean $\vec{\mu}$ and variance $\vec{\sigma}^2$, $D$ would directly encode the fraction of this distribution lying closer to the mean and therefore the statistical confidence that an observation of $D$ was not sampled from this distribution. Fitting models with normal distributions is tempting from the perspective of the Central Limit Theorem but suspect in our case because (i) our fiducial data includes only $252$ data points\footnote{If we allow ourselves to resample the fiducial data (i.e.\ include repeat copies of an individual realization), this number can be enlarged to 2002. We have checked that this only affects our results to within a few percent.} and (ii) the data points are not independent. On the other hand, rather than using the standard ``$D\to$ significance'' rules, we will determine significance by comparing $D$ computed for a \textsf{G/NG} split to $D$ computed for a \textsf{G/G} split, as we illustrate below. This allows us to assign confidence to various data points without assuming Gaussianity.
In addition, we are able to account for the presence of cosmic variance.
\begin{figure}
    \centering
    \includegraphics[width=0.32\textwidth]{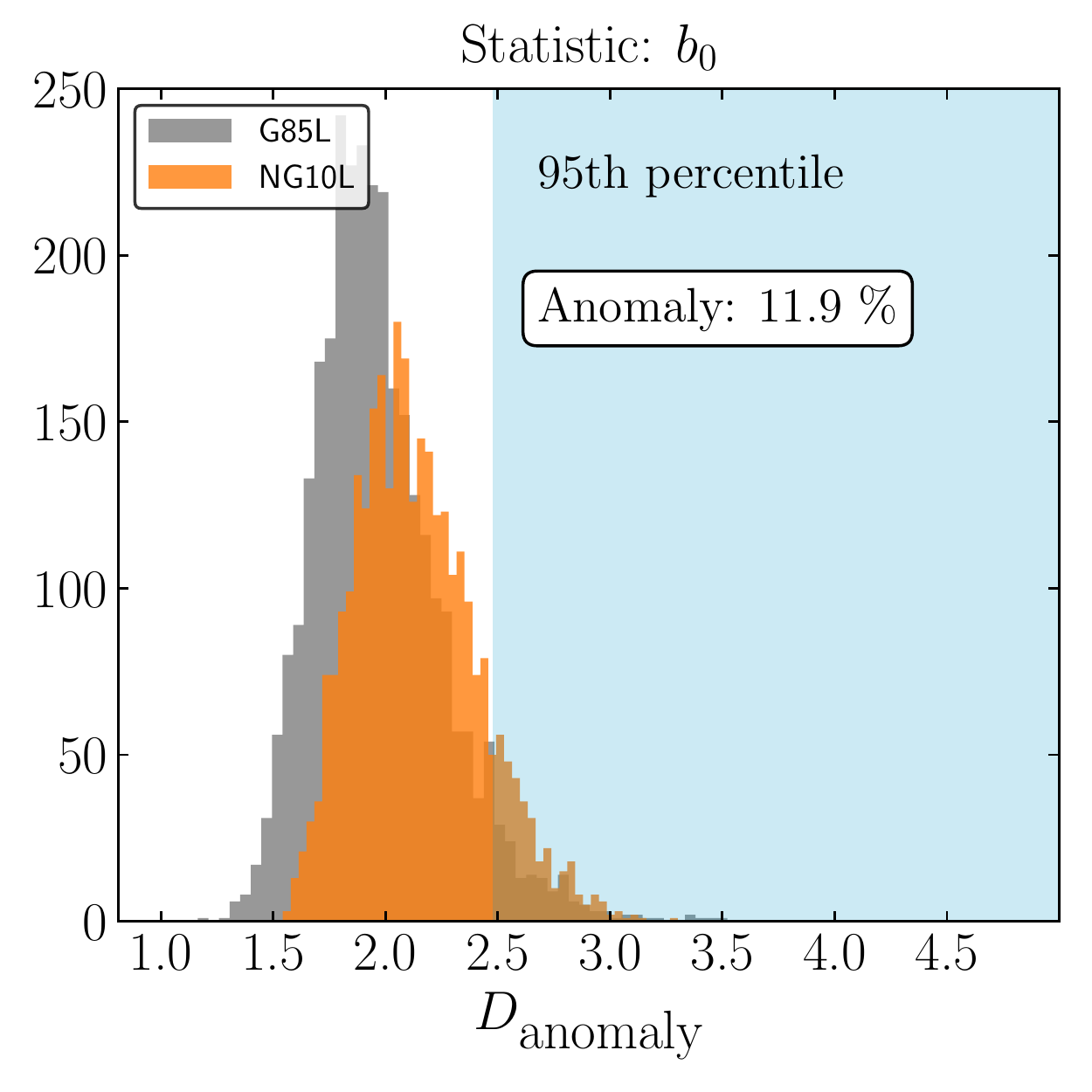}
    \includegraphics[width=0.32\textwidth]{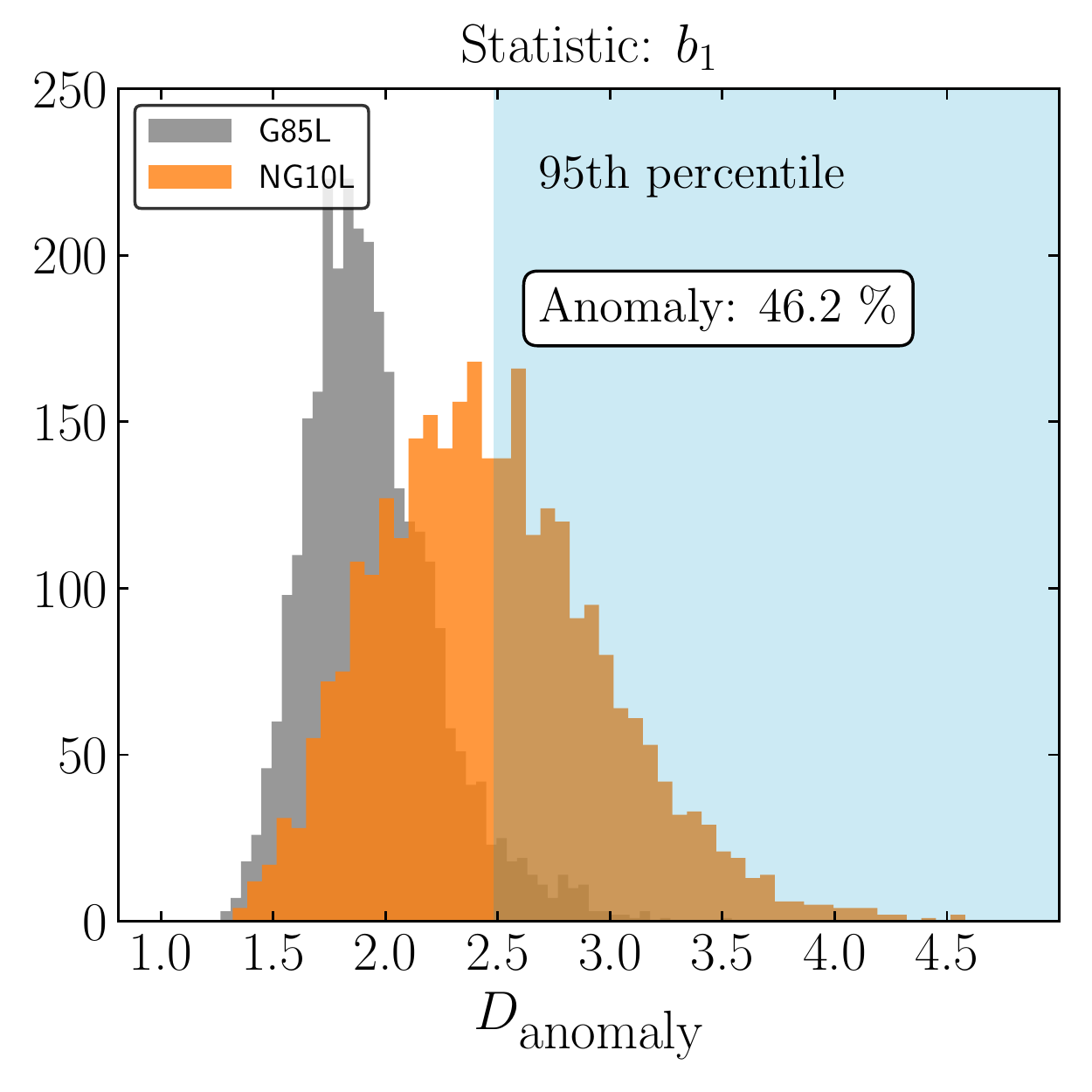}
    \includegraphics[width=0.32\textwidth]{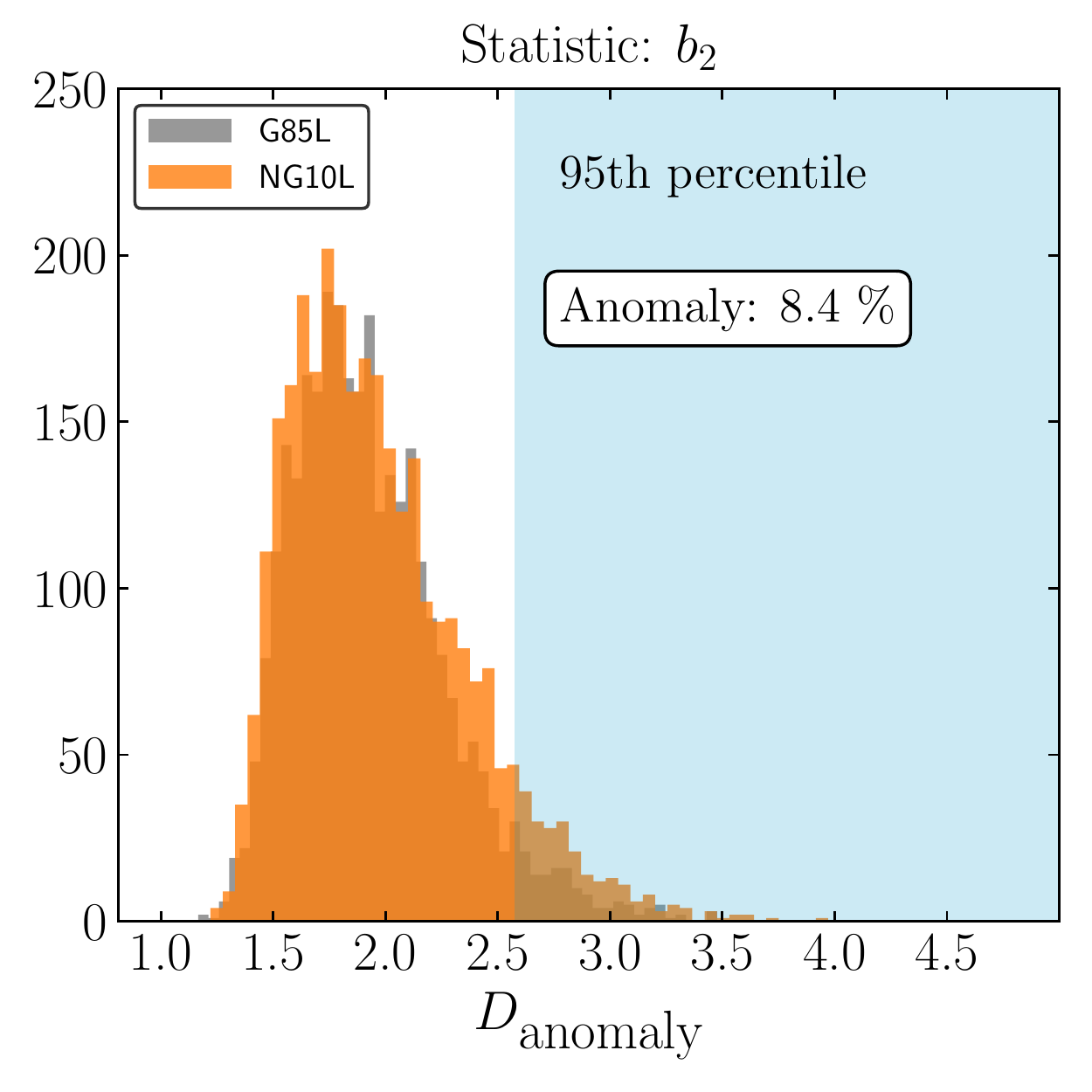}\\
    \includegraphics[width=0.32\textwidth]{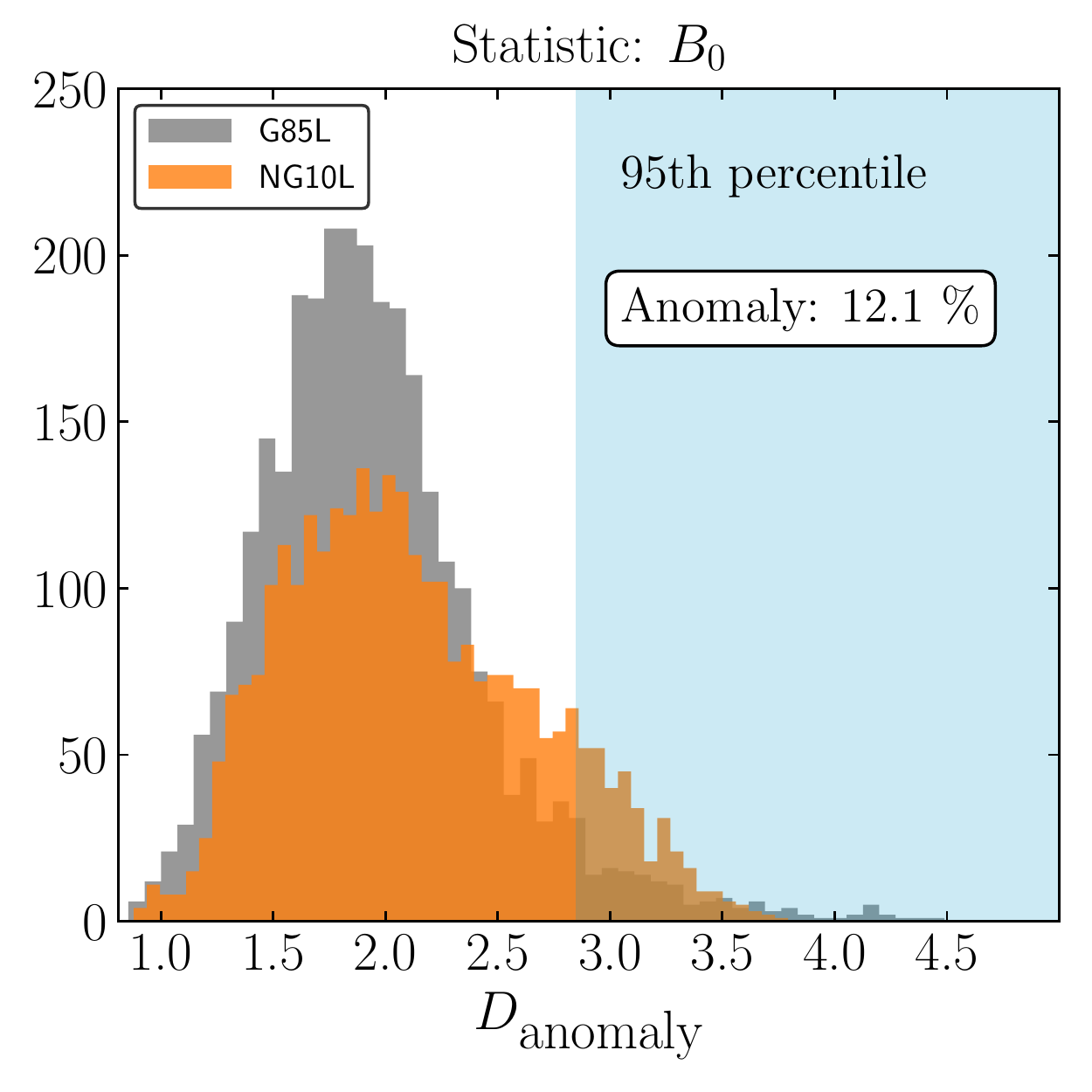}
    \includegraphics[width=0.32\textwidth]{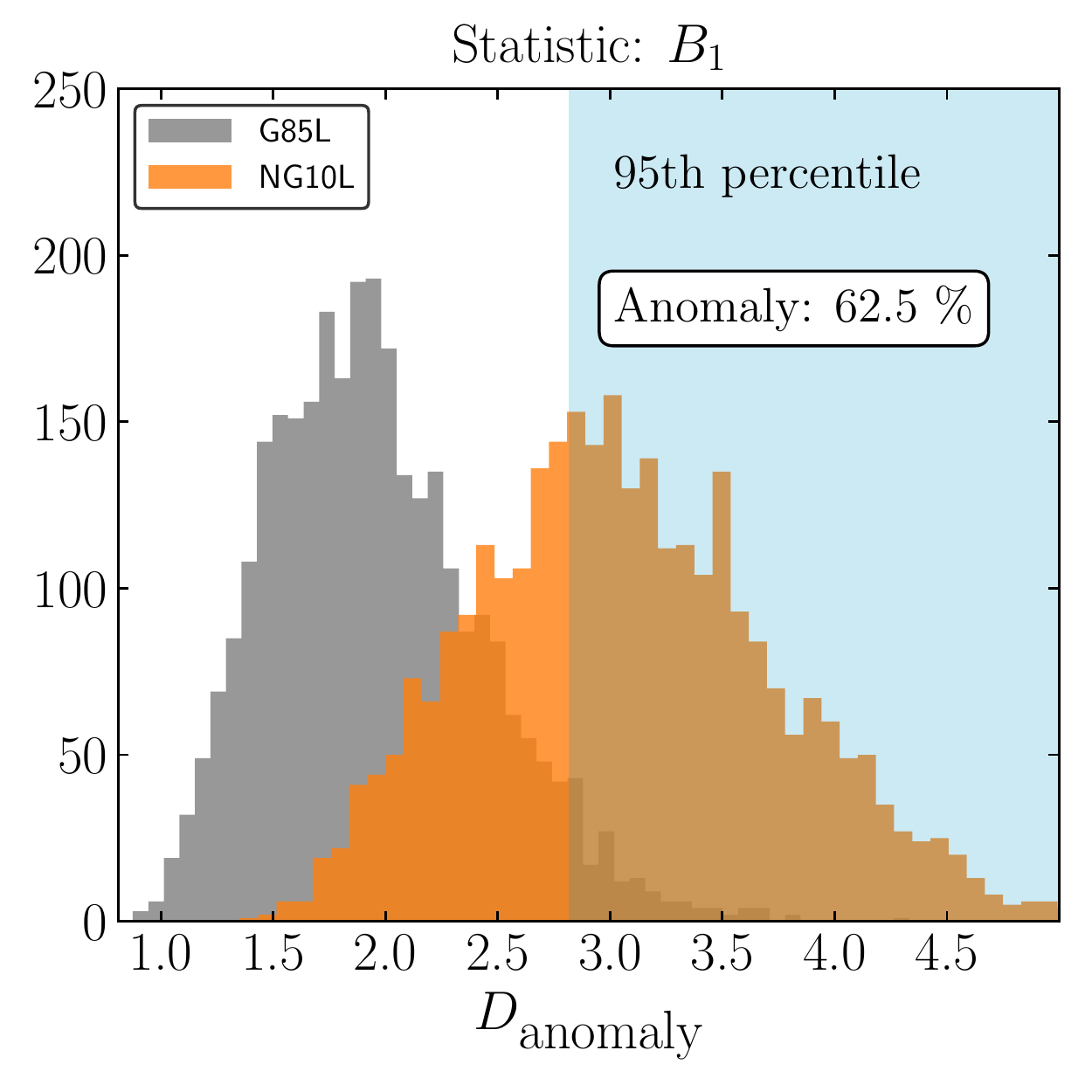}
    \includegraphics[width=0.32\textwidth]{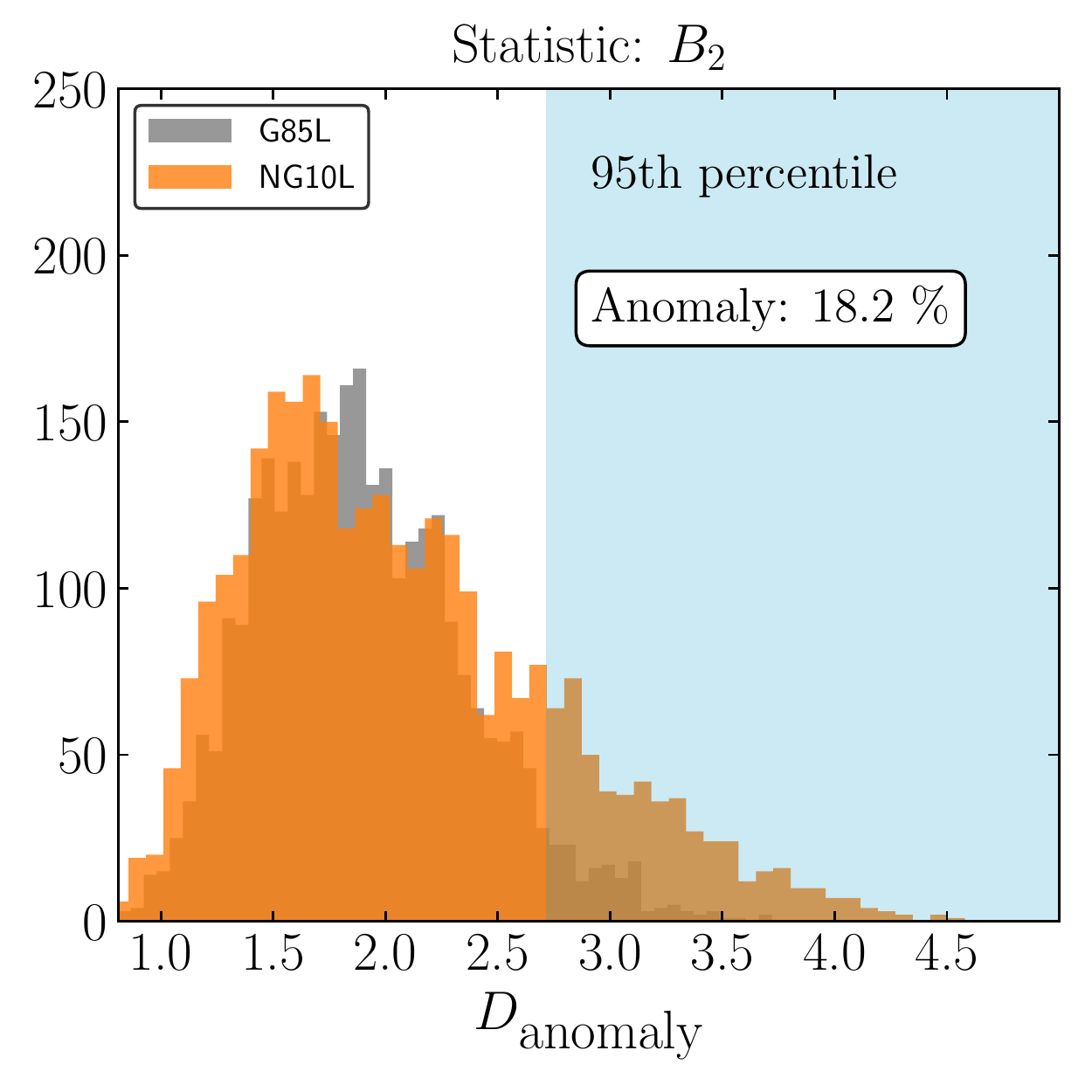}\\
    \includegraphics[width=0.32\textwidth]{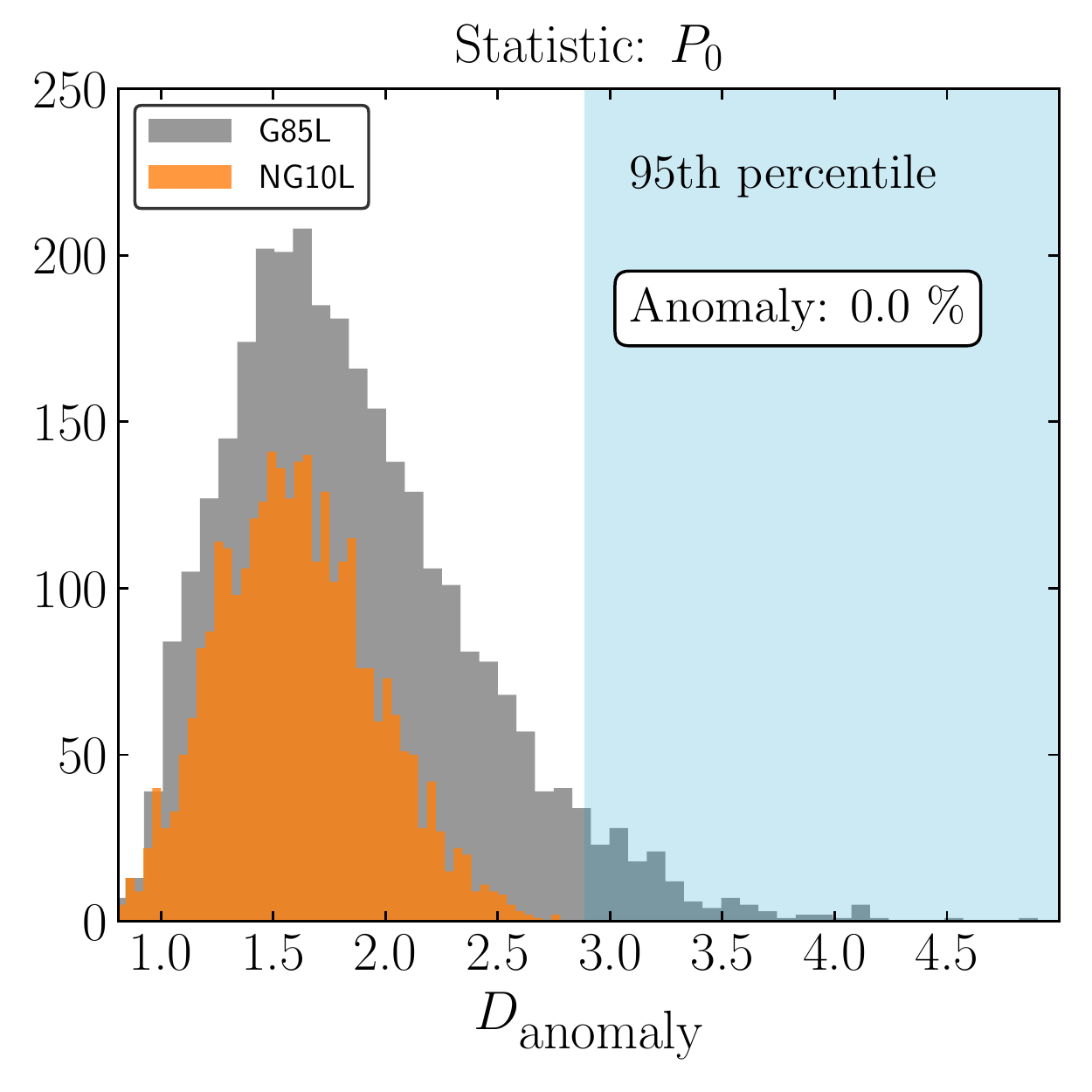}
    \includegraphics[width=0.32\textwidth]{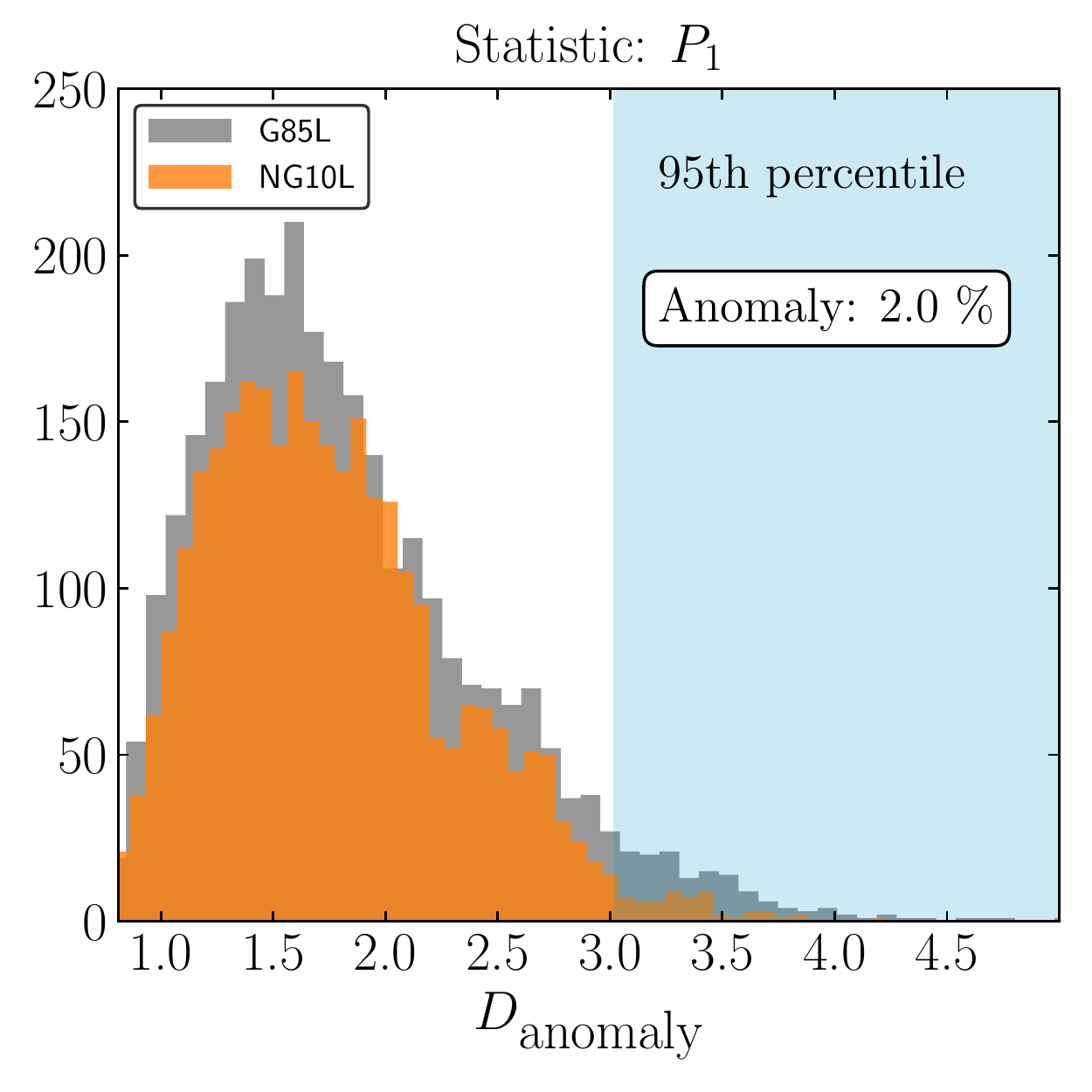}
    \includegraphics[width=0.32\textwidth]{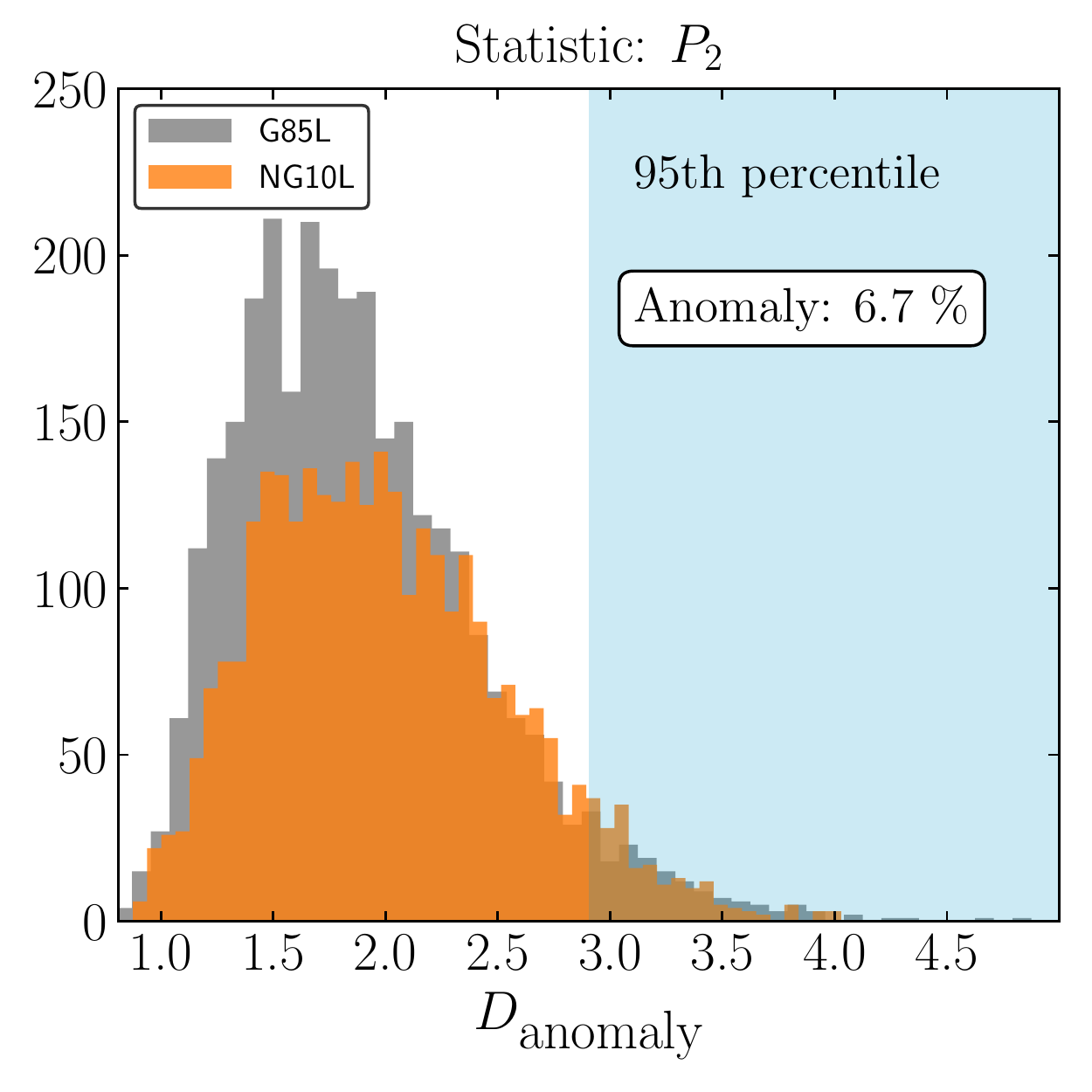}\\
    \includegraphics[width=0.32\textwidth]{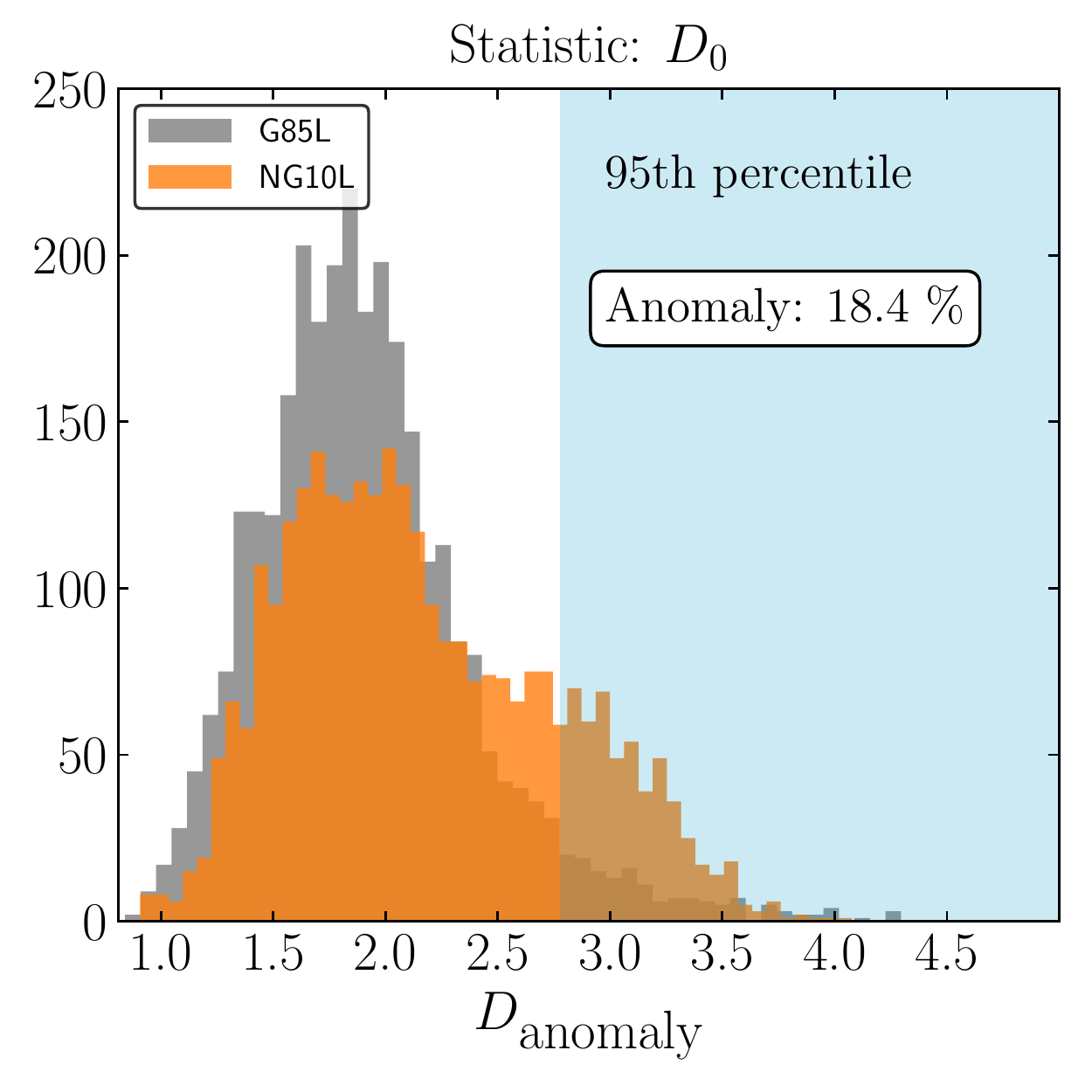}
    \includegraphics[width=0.32\textwidth]{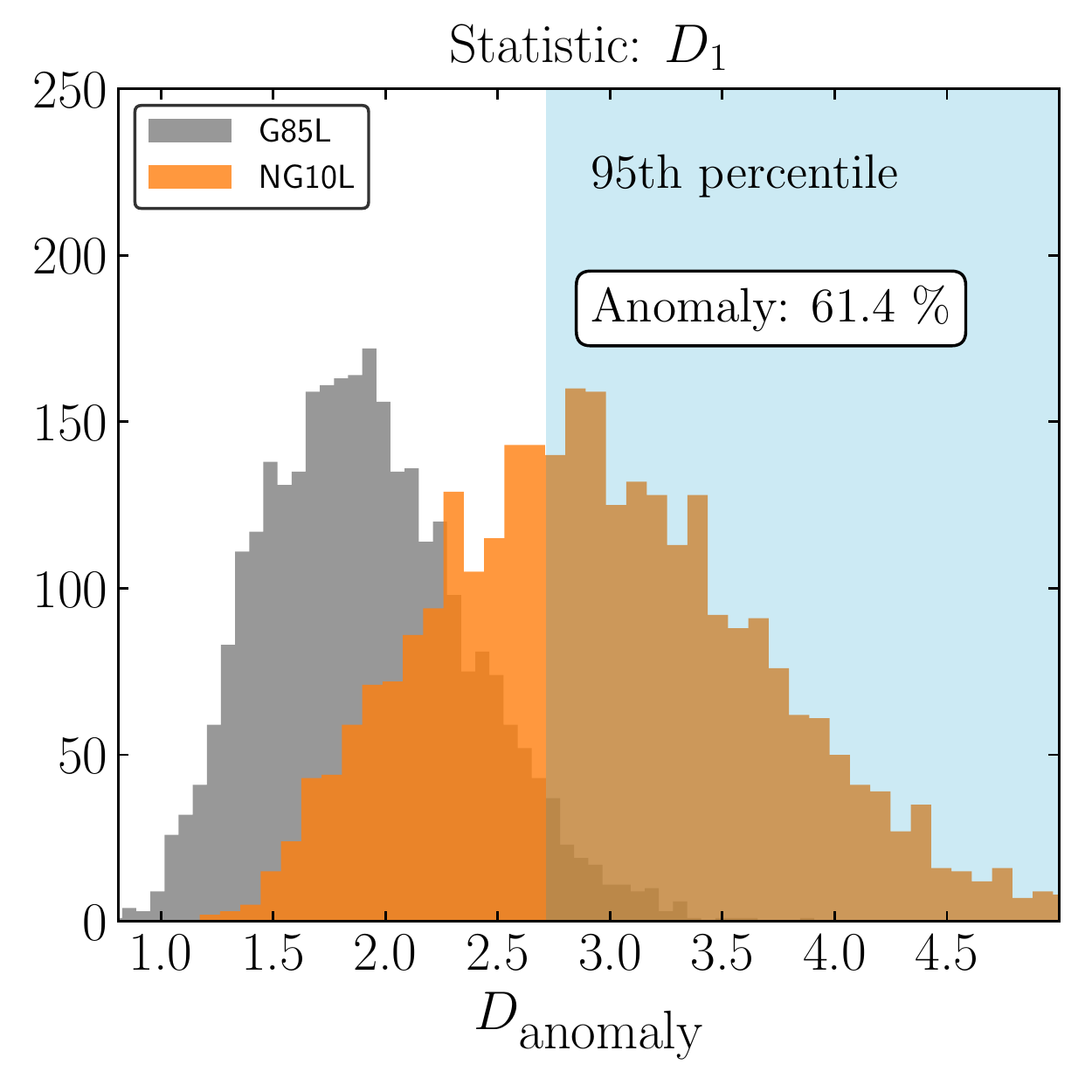}
    \includegraphics[width=0.32\textwidth]{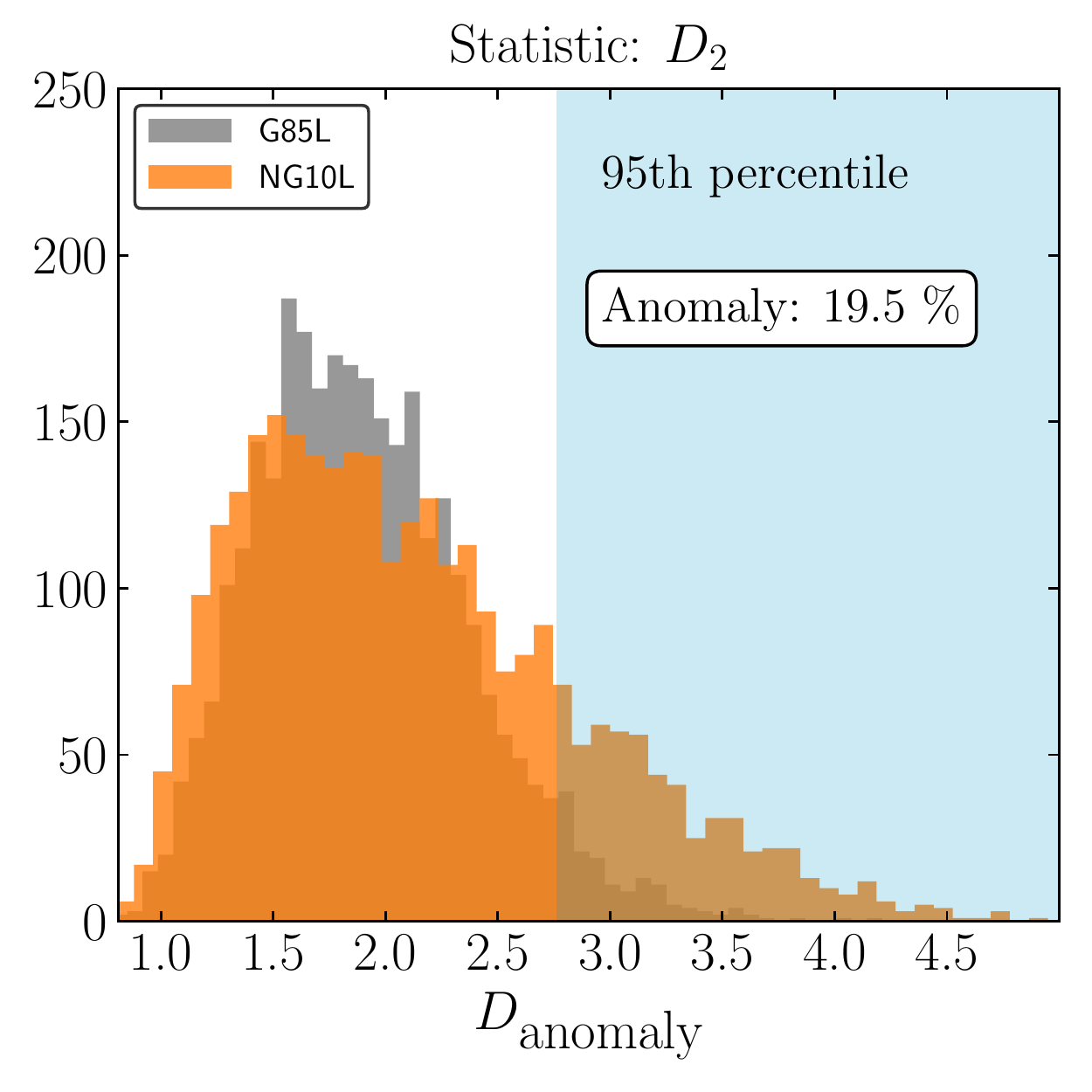}
    \caption{Estimates of log-likelihood for $z=1$, $\alpha$DTM$\ell$-filtration, subsampling the simulations for all the statistics. Each data point in the histograms corresponds to one of the $15$ choose $5=3003$ ways of splitting the data. The fraction of \textsf{NG10L} data points lying beyond the 95\% threshold of the \textsf{G85L} distribution is quoted as anomaly.}
    \label{fig:loglikely}
\end{figure}
\afterpage{\FloatBarrier}
We compute $D$ for every \textsf{G/G} split and every \textsf{G/NG} split. This amounts to computing  $D$ for all the $15$ choose $5=3003$ possible ways to split the realizations\footnote{To have intuition of the effective volume of data considered as compared to a galaxy survey, the Euclid survey will observe a volume of $\sim 50 (\si{Gpc/h})^3$. Our split for the survey is using $5$ realizations, which correspond to $40 (\si{Gpc/h})^3$, which is therefore roughly similar to the Euclid volume.}. We summarize the procedure in Figure \ref{fig:workflow}. This models cosmic variance in the sense that the degree to which non-Gaussianity might exhibit itself in a particular volume of the universe can vary.  The results for all the statistics are shown in Figure \ref{fig:loglikely}. We quantify the  probability of measuring an anomaly by counting the fraction of data points in the \textsf{G/NG} split that lie beyond $95\%$ of the \textsf{G/G} split data points. Note that if positive and negative $\fnl$ affect a statistic oppositely, $D_{\rm anomaly}$ will be the same for $\fnl=\pm 10$. In other words, $D_{\rm anomaly}$ is like a ``distance'' rather than a ``displacement.'' On the other hand, the template method detailed in the next section depends on the sign of $\fnl$. Therefore while we use $95\%$ as a threshold in this section, in the next section we instead take a threshold of $97.5\%$.
With this threshold, several statistics detect an anomaly at the order of tens of percent, with the best statistics giving detection rates larger than $60\%$. Since data points in the histograms represent different combinations of the data splits, the detection rate represents the fraction of simulation boxes in the survey on which $\fnl=10$ is detected with $95\%$ confidence. Notice that, as expected from the study of PIs in the previous Section \ref{sec:png}, the best constraints come from statistics involving the birth and/or the death of features, but not the persistence. 
Another way of testing the anomaly, for the statistics that are defined as empirical distribution functions, i.e. $B_p$, $D_p$ and $P_p$, but not the Betti curves $b_p$, is using the Kolmogorov-Smirnov test \cite{kolmogorov1933sulla,smirnov1939estimation}. Results for this type of test are compatible with the ones using (\ref{eq:danomaly}), therefore we do not show them here.

\subsubsection{Anomaly in the halo power spectrum}
\begin{figure}[t]
    \centering
    \includegraphics[width=0.429\textwidth]{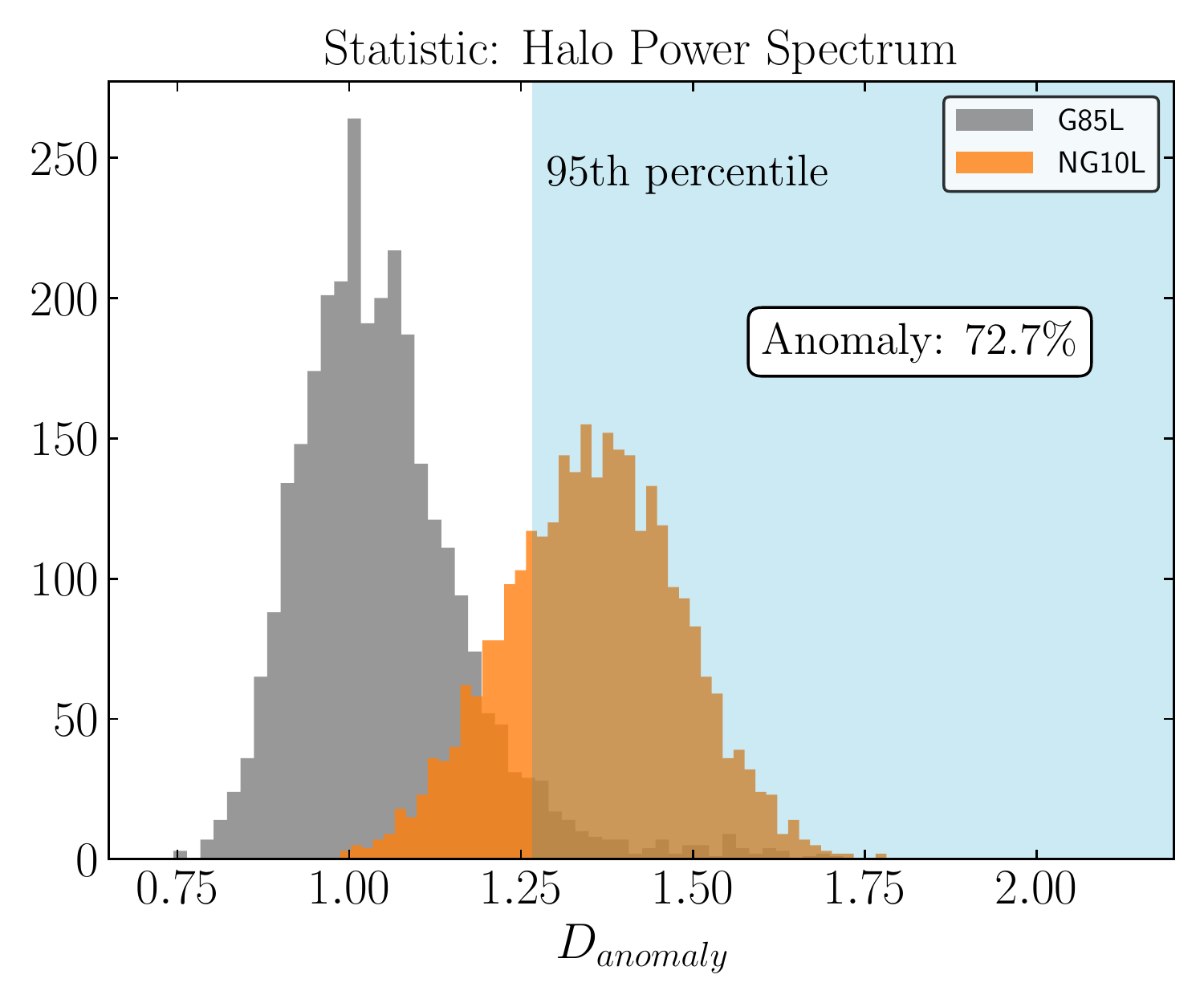}
    \caption{Estimates of log-likelihood following Eq. \eqref{eq:danomaly} for the halo power spectrum at redshift $z=1$ up to $k_{\rm max}=0.07$ h/Mpc. Each data point in the histograms corresponds to one of the $15$ choose $5=3003$ ways of splitting the data. The fraction of \textsf{NG10L} data points lying beyond the 95\% threshold of the \textsf{G85L} distribution is quoted as anomalous.}
    \label{fig:halops}
\end{figure}
For comparison, we can run the computation taking the halo power spectrum as a function of $k$ as our statistic. In this case, we form data vectors taking as each component the value of the power spectrum in a $k$-bin, therefore the indices $i$ in Eq \eqref{eq:danomaly} represent these bins, up to a maximum wavenumber $k_{\rm max}=0.07$ h/Mpc. The reason for looking at large scales only is that we expect the signature of primordial non-Gaussianity to peak in this regime, because of the $1/k^2$ scaling induced by the scale dependent bias   (cfr. Eq.\eqref{eq:pngloc}  and Figure \ref{fig:halopspng}).\footnote{In principle, we could use the halo power spectrum at much higher $k$, well in the nonlinear regime, since in the anomaly method we are not relying on any modeling. However, in that regime, shot-noise becomes dominant. Even subtracting Poisson-shot noise, in the case of local-type primordial non-Gaussianity it is known that the shot-noise receives additional contributions linear in $\fnl$ (see e.g. \cite{Chan:2016ehg}). These contributions are difficult to distinguish from other contributions coming from non-Poissonian deviations without detailed modeling. We additionally note a statistical fluke in which including these scales leads to a larger detected fraction of anomalies. However, such a ``detection'' cannot necessarily be associated with non-Gaussianity, see the right panel of Figure \ref{fig:halopspng}} We show the results of this test in Figure \ref{fig:halops}, determining that the anomaly is identified at a similar rate as those of our topological statistics. It is additionally worth noticing that the topological features live at relatively small scales (cfr.\ Figure \ref{fig:alpha1000sub-boxValid}), where the effect of local primordial non-Gaussianity in the halo power spectrum is negligible. Therefore, our topological features provide a complementary probe with respect to the halo power spectrum.

Note that while neglecting off-diagonal covariance is a good approximation for the likelihood function of the power spectrum (see \cite{Wadekar:2020hax} for a recent study using analytical methods), there is no theoretical reason to neglect this for our topological statistics other than necessity. In fact, for our statistics that can be regarded as empirical distribution functions, one necessarily has nonzero off-diagonal covariance. However, as previously noted, modeling this effect is not possible with the data at our disposal for this work\footnote{It can be done in principle with volumes of the order of the ones typically generated for galaxy surveys as mock catalogs.}.  One might hope to avoid this issue by doing statistics at the sub-box level (i.e. not maintaining the integrity of whole simulations), so that the mock distribution would sample 80 choose 40 $\sim 10^{23}$ configurations. However, in this strategy, extra configurations would break the integrity of individual realization. Since adjacent sub-boxes share primordial information and are in addition related via dynamics, this approach seems incorrect. Therefore, taking a conservative approach, we perform statistics at the level of entire simulations, at the expense of detailed modeling of the likelihood function. In the next section, we include physical priors in our statistical analysis via templates. This amounts to a well-motivated projection of our data, reducing the dimensionality of our test statistics such that simple models appear sufficient.

\subsection{\textsf{Step 3}: Templates}
The results from the previous section show that anomalies are detected with a degree of confidence even in the case where they are expected to be weak, as when looking for $\fnl=10$, and complementary to other more conventional methods as looking at the halo power spectrum. However, we are in the unrealistic case in which we know what anomaly we are looking for, while in a more realistic scenario we certainly
need to work harder to show that the anomaly comes from $\fnl\neq 0$. One way would be to run as many fiducial cosmologies as possible and check anomaly detections among them and the survey data to be able to tell apart anomalies coming from different cosmological parameters. This solution is quite impractical, as it would involve running a very large number of simulations\footnote{It is possible that smaller simulations could sufficiently reduce the computational cost. In this case, one might be able to perform an interpretable classification via a simple machine learning architecture. We expect a more efficient strategy to be using asymptotic templates to narrow down the possible causes of an anomaly before running a reduced set of simulations, perhaps to train some classifier.}.

We therefore proceed with introducing templates to identify the source of anomalies. In a scenario in which modeling of the impact of the parameter we search for is reliable, the template would simply be composed of some theoretical prediction based on the model. In the case of the halo power spectrum, there is a clear prediction for a scale dependence of halo bias at large scales going as $1/k^2$ in the presence of local primordial non-Gaussianity, as explained in the previous Section \ref{sec:png}. In absence of solid modeling, as in our case, the construction of the template again needs to rely on the comparison of datasets, using N-body simulations.

There are several ways to compute a template vector $\vec{T}$. In particular, although we are interested in $\fnl=10$, it is reasonable to expect that the corresponding topological signature will have some overlap with larger non-Gaussianity, say $\fnl=250$. 
We therefore construct a template defined as
\begin{equation}
\vec{T}^X_{\rm NG} \equiv \frac{1}{N_r}\sum_{i=1}^{N_r}\vec{S}_{\textsf{NG250S}_i}^X-\vec{S}_{\textsf{G85S}_i}^X,
\end{equation}
where we average over $N_r=5$ realizations and we use small box simulations \textsf{G85S} and \textsf{NG250S} (cfr. Table \ref{tab:eos}). The subtraction is made between cosmologies with the same initial condition, since here we want to minimize cosmic variance to clearly distinguish the signature of $\fnl$\footnote{While reviewing the manuscript for journal submission, we realized that in the code we use to generate initial conditions for the simulations, \textsf{2LPTic}, the random numbers used to generate the initial displacements are treated differently for the Gaussian and non-Gaussian versions of the code. Namely, for Gaussian initial conditions, the random numbers are used to generate the amplitudes of the density field, whereas for non-Gaussian runs these are used to generated the potential field. As a consequence, even when using the same initial random seed, the runs have different noise properties by design. This makes the templates noisier than expected, so that our final constraints are worse than expected, but qualitatively the same.}.
Using a stronger $\fnl$ signal is useful because less simulation volume is needed to generate a robust template. On the other hand, one must worry that the template $\vec{T}$ changes in shape as $\fnl$ varies, i.e. $\vec{T}=\vec{T}(\fnl)$. We address this in Appendix \ref{app:templates}, where we also show that our templates are fairly stable under downgrade in resolution, see Fig.\ \ref{fig:templateLvsS}.
We therefore compute the distribution $\vec{S}_{\rm mock}\cdot \vec{T}$. The expectation is then that if the survey data has  $\fnl> 0$, it will have larger $\vec{S}_{\rm survey}\cdot \vec{T}$ than the bulk of the distribution $\vec{S}_{\rm mock}\cdot \vec{T}$, the mock data being run with Gaussian initial conditions.  

\subsection{\textsf{Step 4}: Constraining parameters: primordial non-Gaussianity}\label{sec:templatePNG}
\begin{figure}
    \centering
    \includegraphics[width=0.49\textwidth]{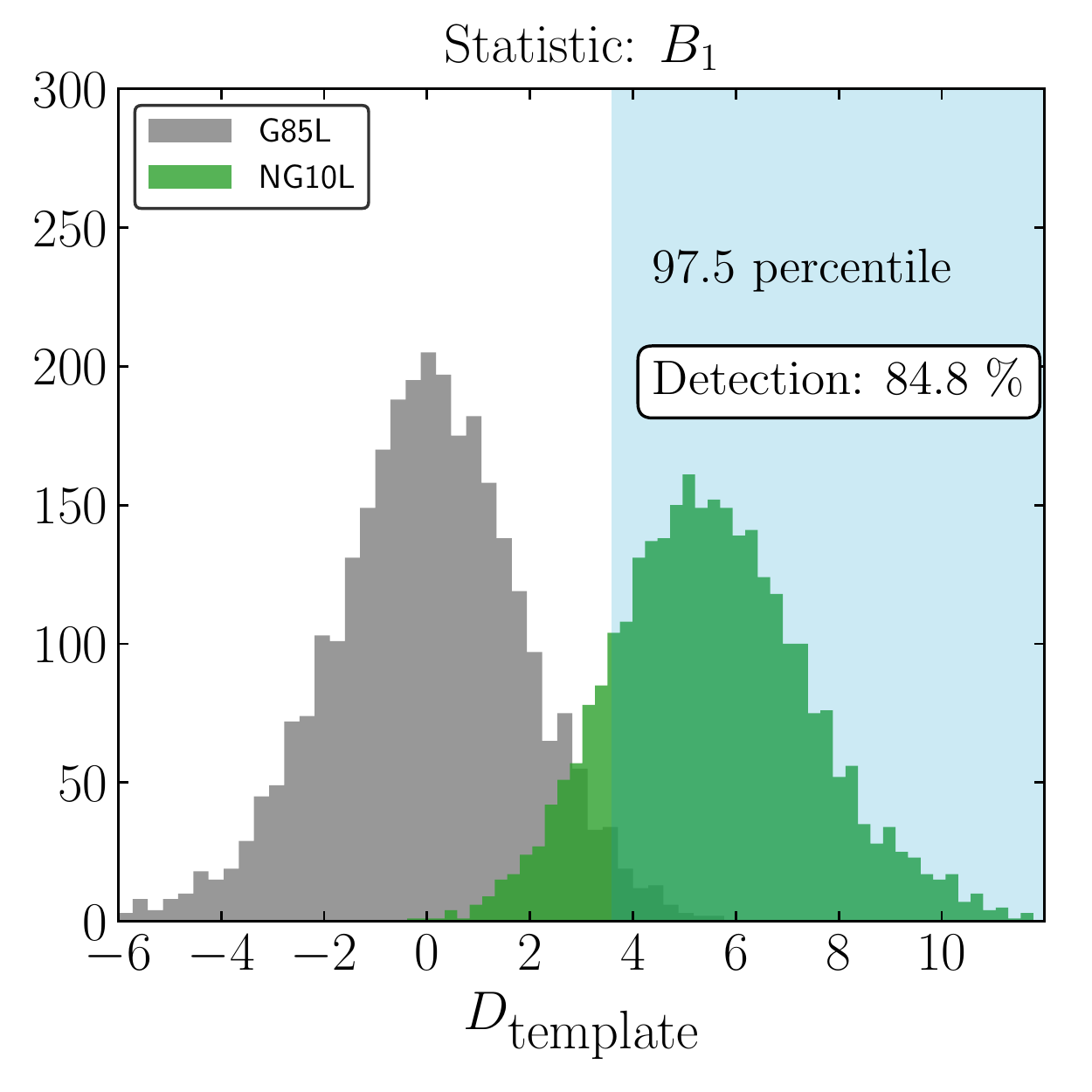}
    \includegraphics[width=0.49\textwidth]{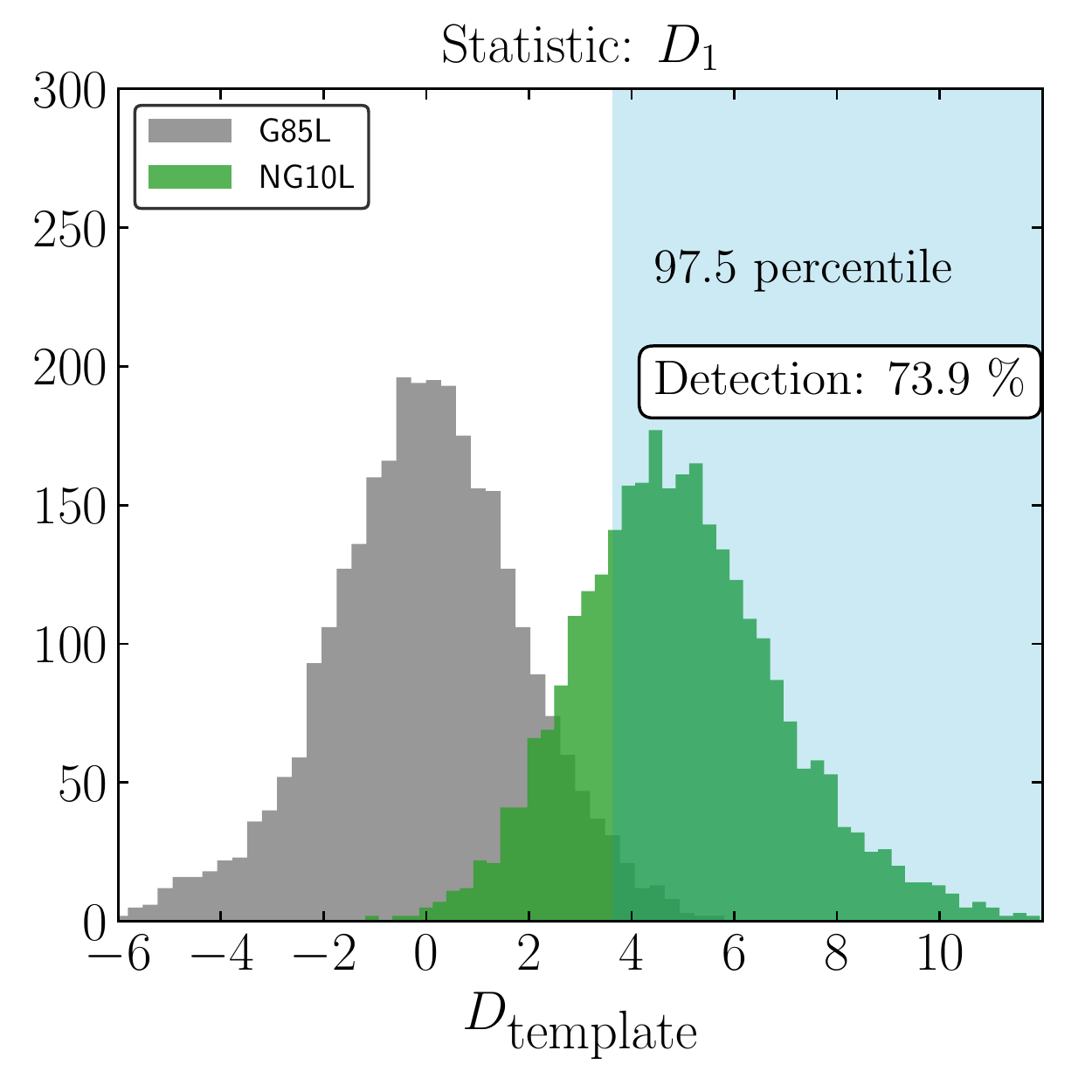}
    \caption{Deviations from the null distribution mean, measured in units of the null distribution's variance. Each data point in the histograms corresponds to one of the $15$ choose $5=3003$ ways of splitting the data.  The fraction of \textsf{NG10L} data points lying beyond the 97.5\% threshold of the \textsf{G85L} distribution is quoted as detection rate of $\fnl=10$.  Results shown for $B_1$ and $D_1$ for $\alpha$DTM$\ell$-filtrations of $z=1$ simulations, computed with subsampling. Templates are computed using the small-box simulations \textsf{NG250S} with $\fnl =250$. }
    \label{fig:TemplateHist}
\end{figure}

With templates in hand, we can test for a particular cosmological reason for a detected anomaly.  Here we study the case of primordial non-Gaussianity of amplitude $\fnl=10$. To this end, we want to compute $\vec{S}\cdot\vec{T}$, where $\vec{S}$ is our survey data vector, $\vec{T}$ is the template and $\cdot$ represents the scalar product of $N$ dimensional vectors, $N$ being the number of bins as explained in the previous section. Using a similar likelihood as for the anomaly detection, we compute 
\begin{equation}\label{eq:dtemp}
D_{\rm template}=\frac{\vec{S}_{\rm survey}\cdot \vec{T}-\vec{S}_{\rm fid,avg}\cdot \vec{T}}{\sigma}
\end{equation}
where now $\sigma^2$ is the variance in the distribution $\vec{S}_{\rm mock}\cdot \vec{T}$.  As before, we perform this computation for both the \textsf{G/G} and \textsf{G/NG} data splits. The \textsf{G/G} results are used to set a threshold for results using the \textsf{G/NG} pipeline. Guided by the results obtained for the anomaly detection (cfr. Figure \ref{fig:loglikely}), we select the most promising statistics, i.e. the ones involving the distribution of births and deaths of $1$-cycles such as $B_1$ and $D_1$ to make the constraints, shown in Figure \ref{fig:TemplateHist}. We now quantify detection by counting the fraction of \textsf{G/NG} data points that lie beyond the $97.5\%$ of the \textsf{G/G} distribution. Note that this threshold is larger than in the case of anomaly detection, since $D_{\rm template}$ depends on the sign of $\fnl$, while (if the effect of $\fnl$ on our statistics is symmetric) $D_{\rm anomaly}$ does not. 
The detection fraction is quite significant at around $85\%$ for the best case. Similarly to the case of the anomaly detection, this represents the fraction of simulation volume in which we are able to detect $\fnl=10$ at $97.5\%$ confidence. It is interesting to notice that the use of a template raises significance with respect to the anomaly detection. As discussed in Section  \ref{sec:anomaly}, this appears to be because our statistical task has been made simpler: when we project onto the physically-motivated template $\vec{T}$, we are simply left modeling a 1-dimensional probability distribution, which in the relevant cases turns out to be well-modeled by a normal distribution. Figure \ref{fig:TemplateHist} is the main result of this analysis, showing that we can detect $\fnl=10$ with a degree of significance using appropriate templates.

\subsection{\textsf{Step 5}: Constraining multiple parameters: degeneracies and template optimality}\label{sec:degen}
In a true observational context, one seeks to constrain multiple cosmological parameters. In general, corresponding constraints will be correlated. These correlations arise from degeneracies between multiple physical scenarios as viewed by the relevant statistics.
In our present context, understanding degeneracies amounts to studying the similarity between templates arising from different physics. If two given templates are very similar, our method cannot distinguish the underlying physics. However, for sufficiently different templates, we should be able to distinguish their signatures in large scale structure.
In particular, we will examine the distinguishability of the topological signatures of varying $\fnl$ and $\sigma_8$. 
We choose these parameters because they vary within the \textsc{Eos Dataset}. We reserve a more detailed study of degeneracies for a companion paper \cite{bcsinprogress}.

In the context of our work, understanding potential degeneracies amounts to a detailed study of template overlap and optimality. The similarity between two templates can be quantified via their normalized inner product, or cosine
\begin{equation}
	\cos(T_1,T_2)=\frac{T_1\cdot T_2}{\sqrt{T_1\cdot T_1}\sqrt{T_2\cdot T_2}}
\end{equation}
If $\cos(T_1,T_2)$ is large, our statistic is not sensitive to the difference between the physics underlying the two templates.

Applying this to survey data, if we find an anomaly, we compute $\cos(S_{\rm survey}-S_{\rm mock},T)$ for all templates. For the optimal template (which must have $\cos(S_{\rm survey}-S_{\rm mock},T)$ beyond some threshold to proceed), we see how $\vec{S}_{\rm survey}\cdot \vec{T}$ compares to the mock distribution, along the lines of the Eq. \eqref{eq:dtemp}. This amounts to removing the normalization factor from the cosine. While the cosine is sensitive only to the ``shape'' details of a particular signal, in the end we are also interested in the signal's magnitude. This magnitude determines the significance of the final result, taking into account as well a thresholding according to the same pipeline run on the data with \textsf{G/G} split.

\begin{figure}
    \centering
	\includegraphics[width=0.5\textwidth]{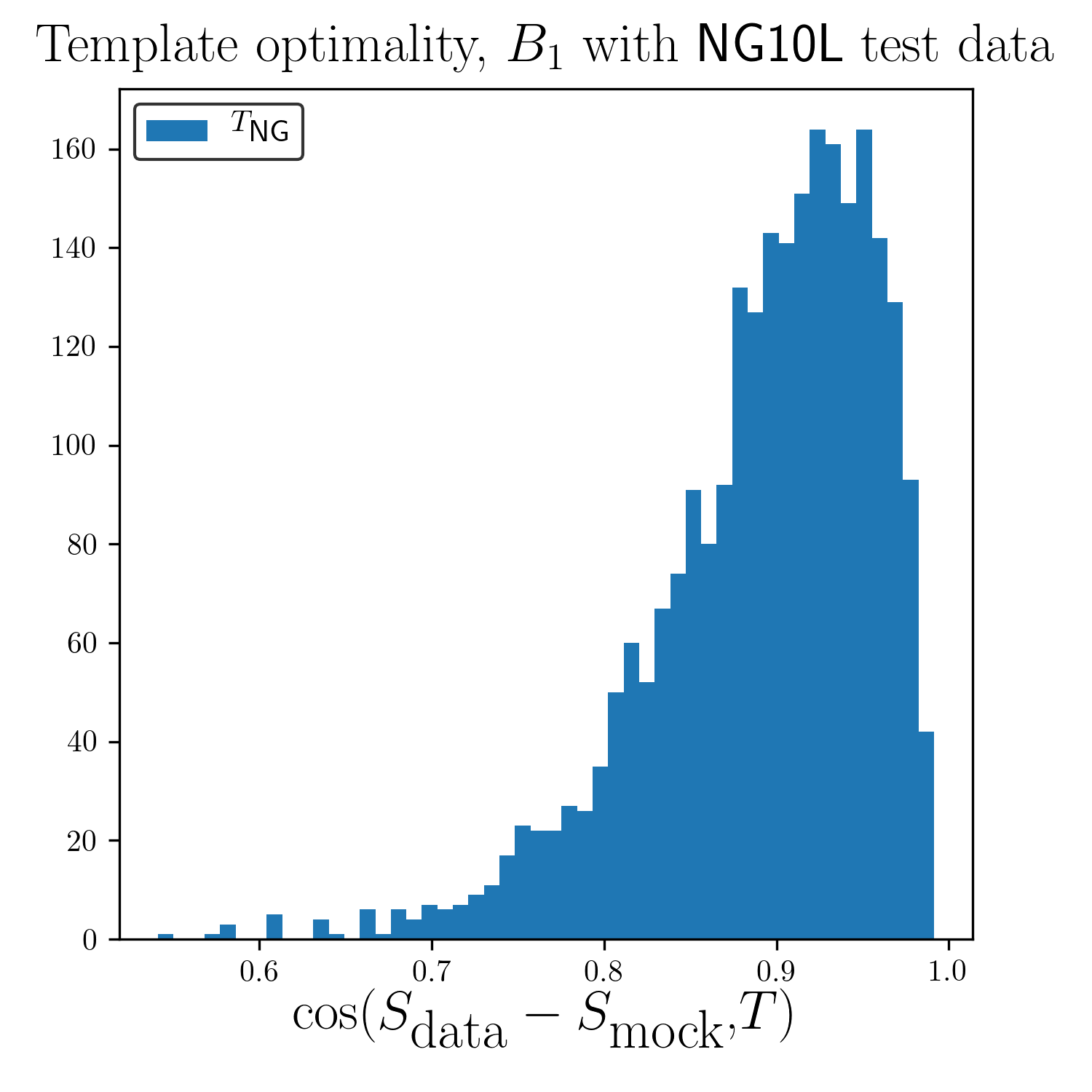}
    \caption{Taking data from \textsf{NG10L} we compute $D_{\rm template}$ with templates corresponding to \textsf{NG250S} and \textsf{G87S}. For those realizations with $D_{\rm template}$ falling beyond 97.5\% confidence, we compute $\cos(S_{\rm data}-S_{\rm mock},T)$ for both templates. Of the 3003 draws from \textsf{NG10L}, 2548 are assigned to the \textsf{NG250S} template and 0 are assigned to the \textsf{G87S} template. }
    \label{fig:optimality}
\end{figure}

We can immediately apply this logic to the \textsc{Eos} simulations, seeing how well we can identify the template derived from the small box simulations with $\fnl=250$, \textsf{NG250S}, as the correct template for the large box simulations with $\fnl=10$ from \textsf{NG10L}. In Figure \ref{fig:optimality} we show the cosine for the statistic $B_1$ computed using templates from \textsf{NG250S} and from the small box simulations  \textsf{G87S} against data from \textsf{NG10L}. Remember that the set \textsf{G87S} has the fiducial cosmology, but for $\sigma_8$ which changes from $0.85$ to $0.87$. Of the $3003$ draws from \textsf{NG10L}, $2548$ lie beyond the $97.5\%$ confidence level for one of the templates. Of the $2548$, all of them are assigned to the \textsf{NG250S} template while none are assigned to the \textsf{G87S} template, indicating no degeneracy.

\begin{figure}
\centering
\includegraphics[width=0.33\textwidth]{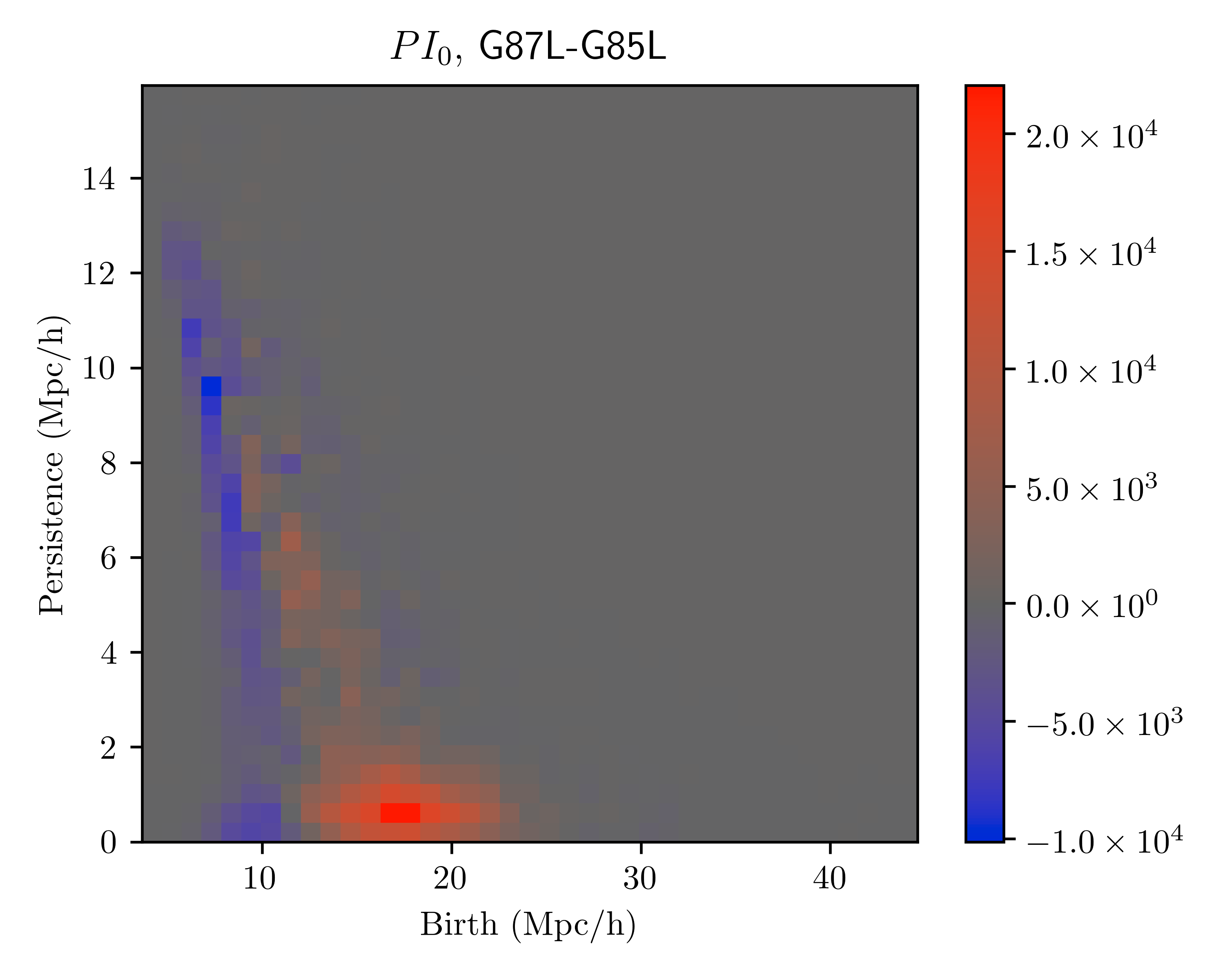}\includegraphics[width=0.33\textwidth]{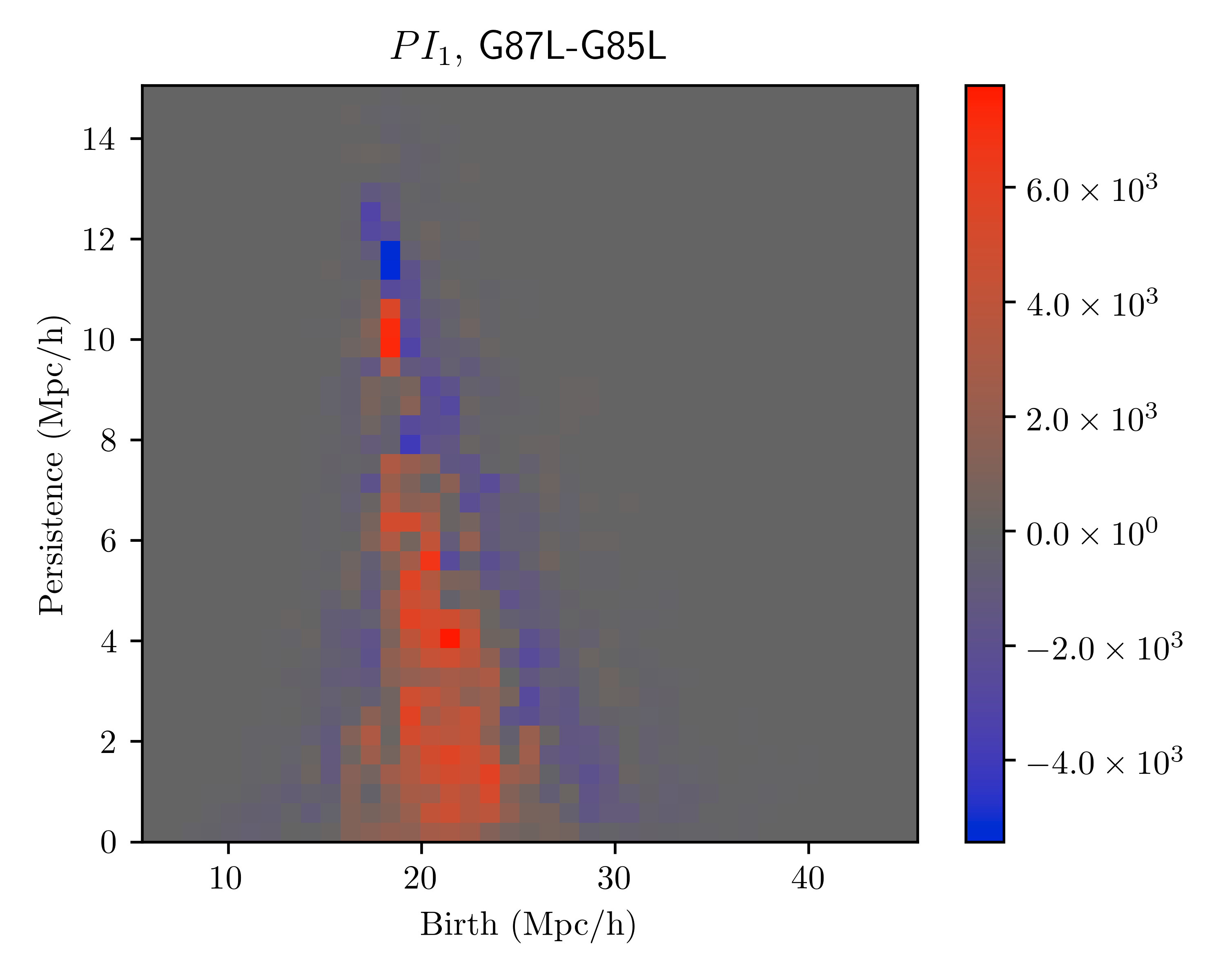}
\includegraphics[width=0.33\textwidth]{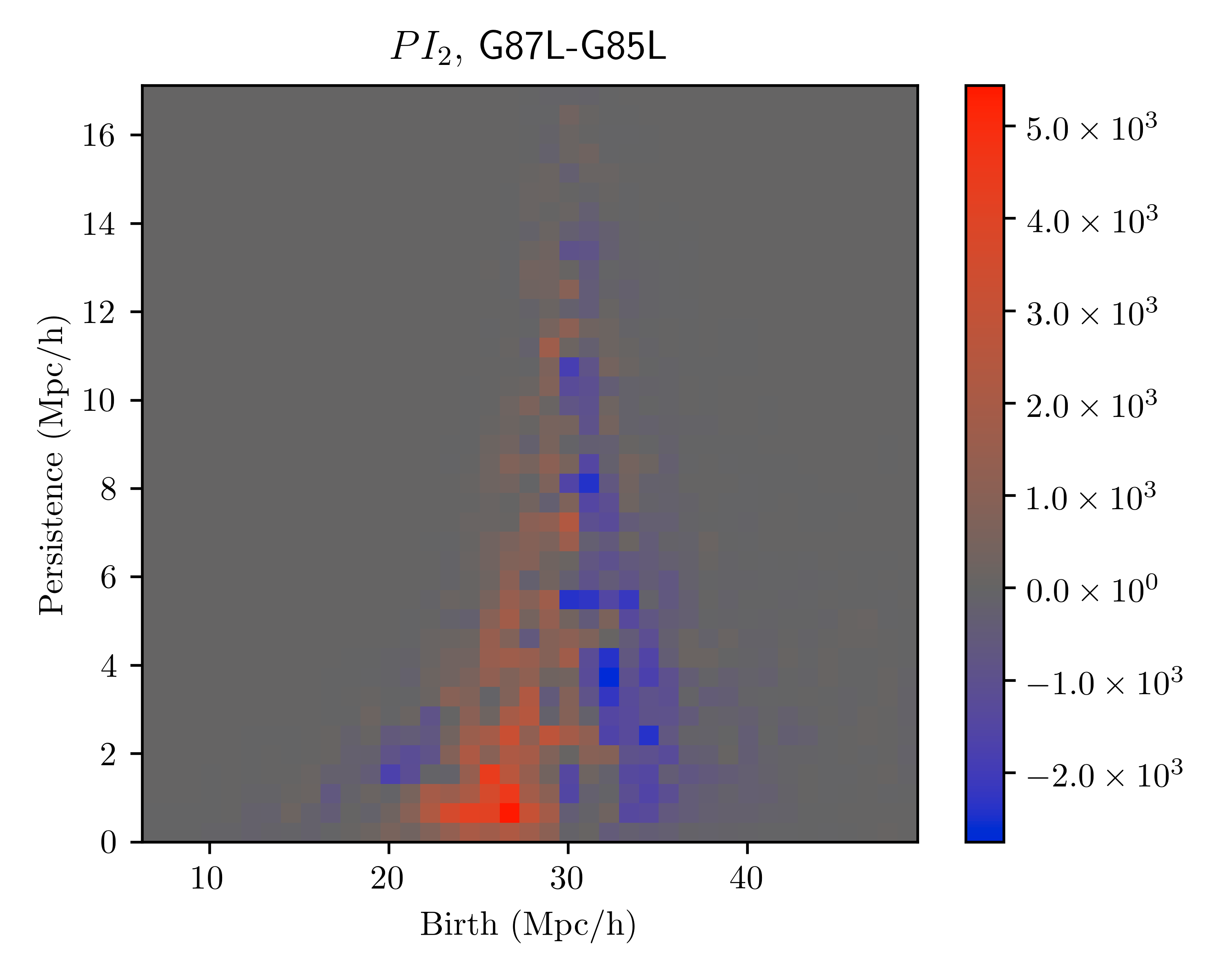}
\caption{Changes in PIs induced by increasing $\sigma_8$ to $0.87$. The difference between these templates and those corresponding to $\fnl\neq0$ (see Figure \ref{fig:PIintuition}) allows us to distinguish between the two cases in practice.}\label{fig:PItemplate87}
\end{figure}

It is encouraging that we can resolve a difference between the two physical scenarios. We can understand this behavior by comparing the PIs generated by $\fnl=250$ (Figure \ref{fig:PIintuition}) and $\sigma_8=0.87$ (Figure \ref{fig:PItemplate87}). The PIs generated by $\fnl=250$ and $\sigma_8=0.87$ are sufficiently different that we can distinguish them by eye. Then we may expect topological curves based on well-chosen slicings of the persistence diagrams will differentiate the two. Recall that although we use the topological curves in practice, in principle one might achieve best results by optimizing parameters in the PIs.  

\subsubsection{Subsampling}
\begin{figure}
\centering
\includegraphics[width=0.33\textwidth]{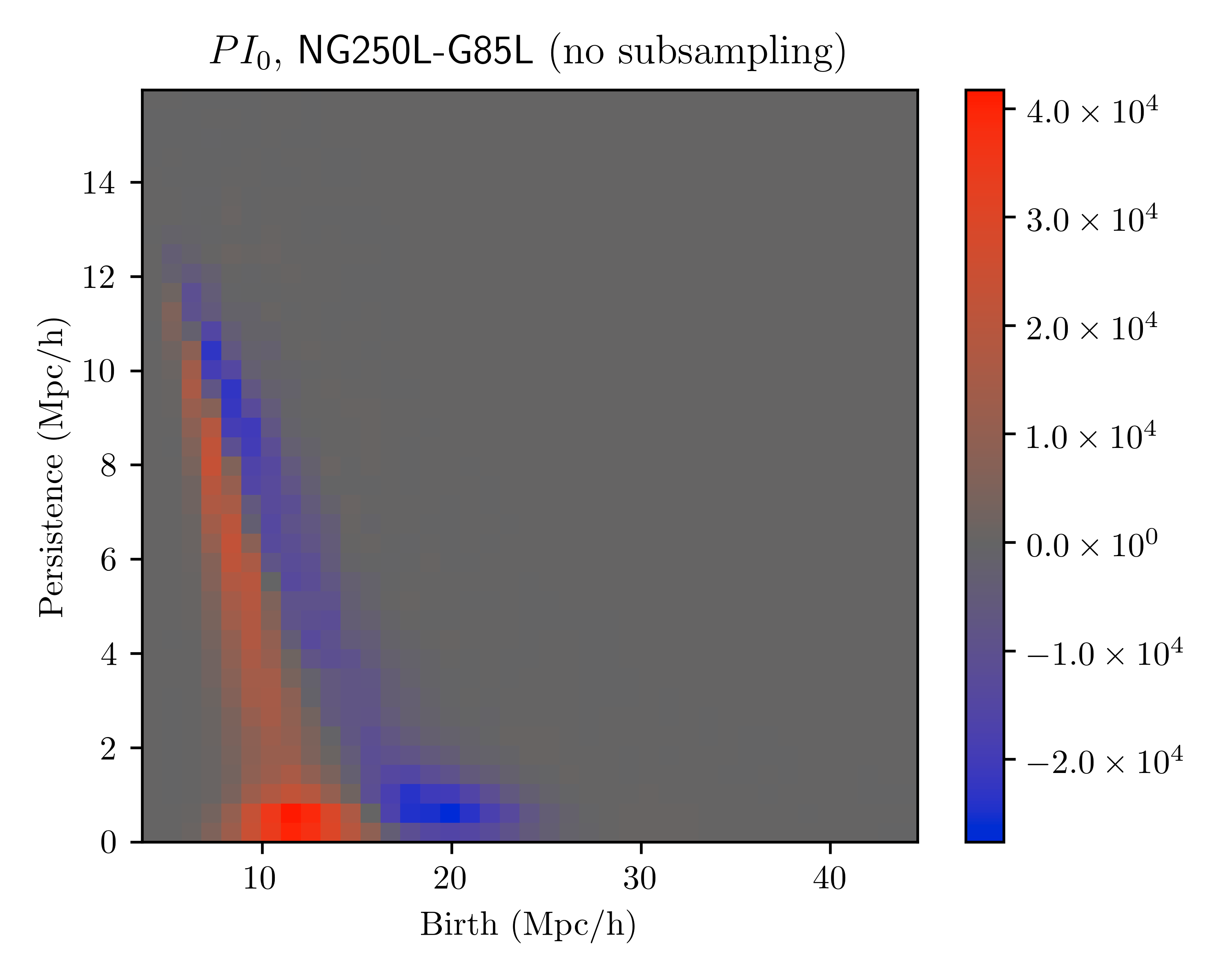}\includegraphics[width=0.33\textwidth]{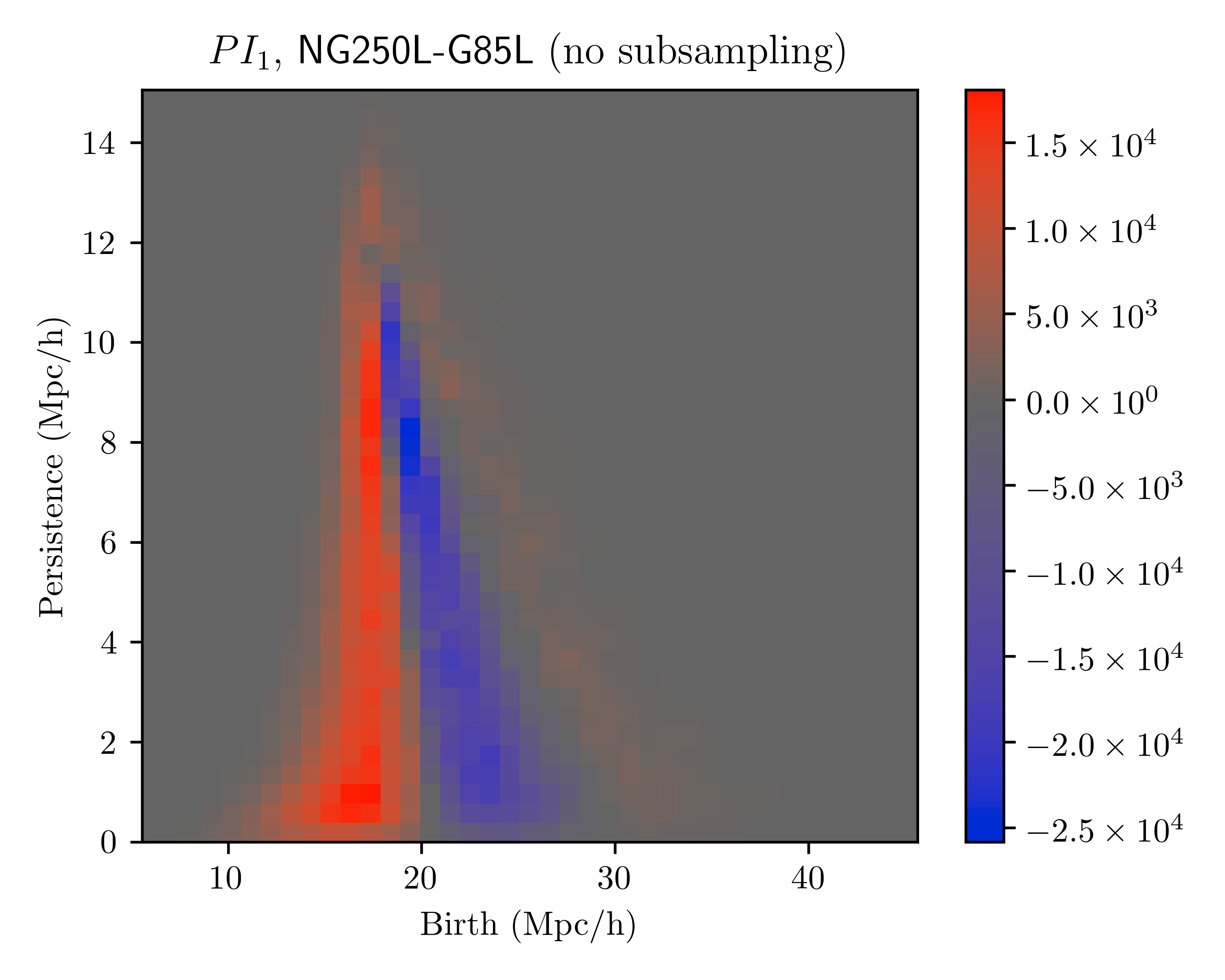}
\includegraphics[width=0.33\textwidth]{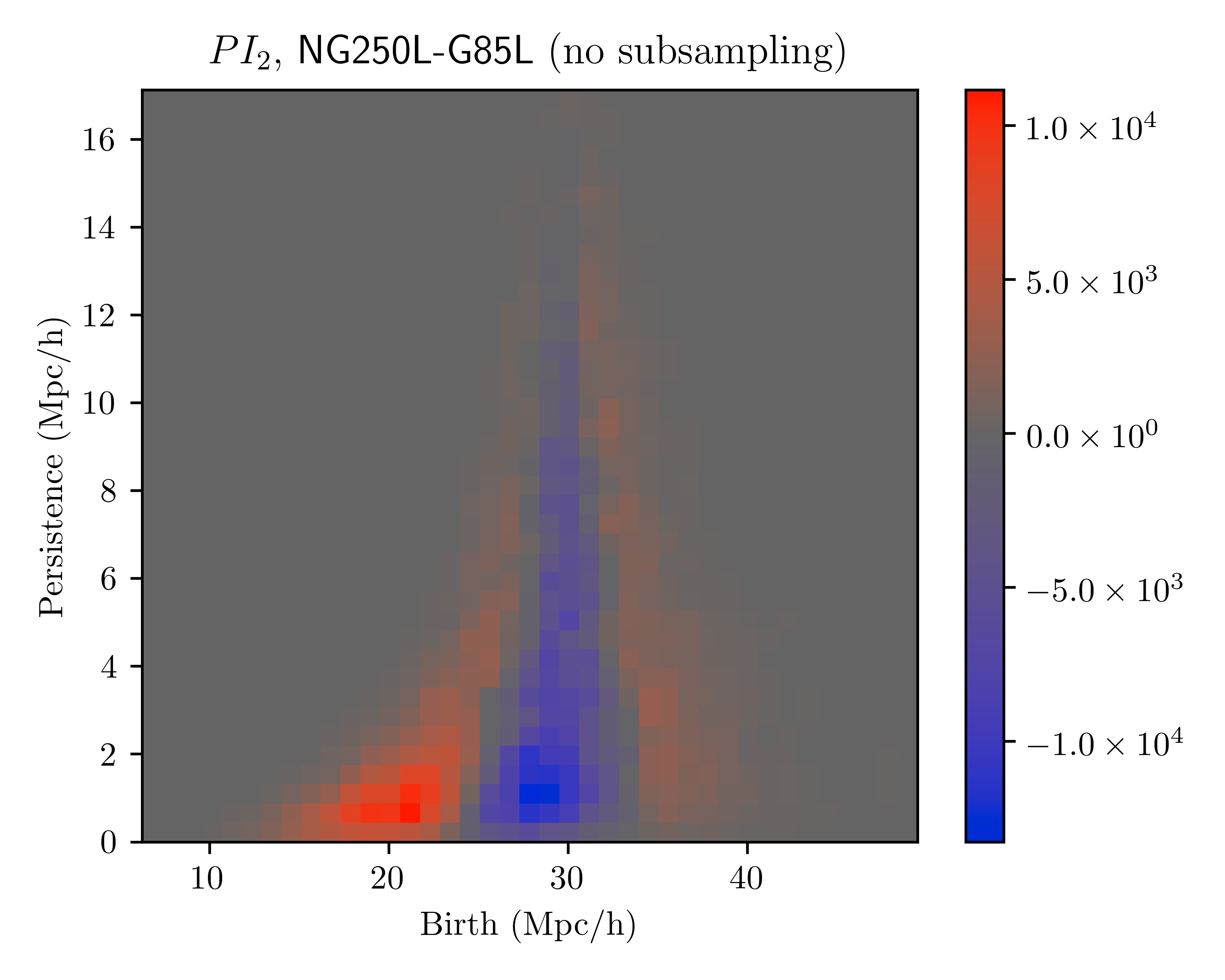}
\includegraphics[width=0.33\textwidth]{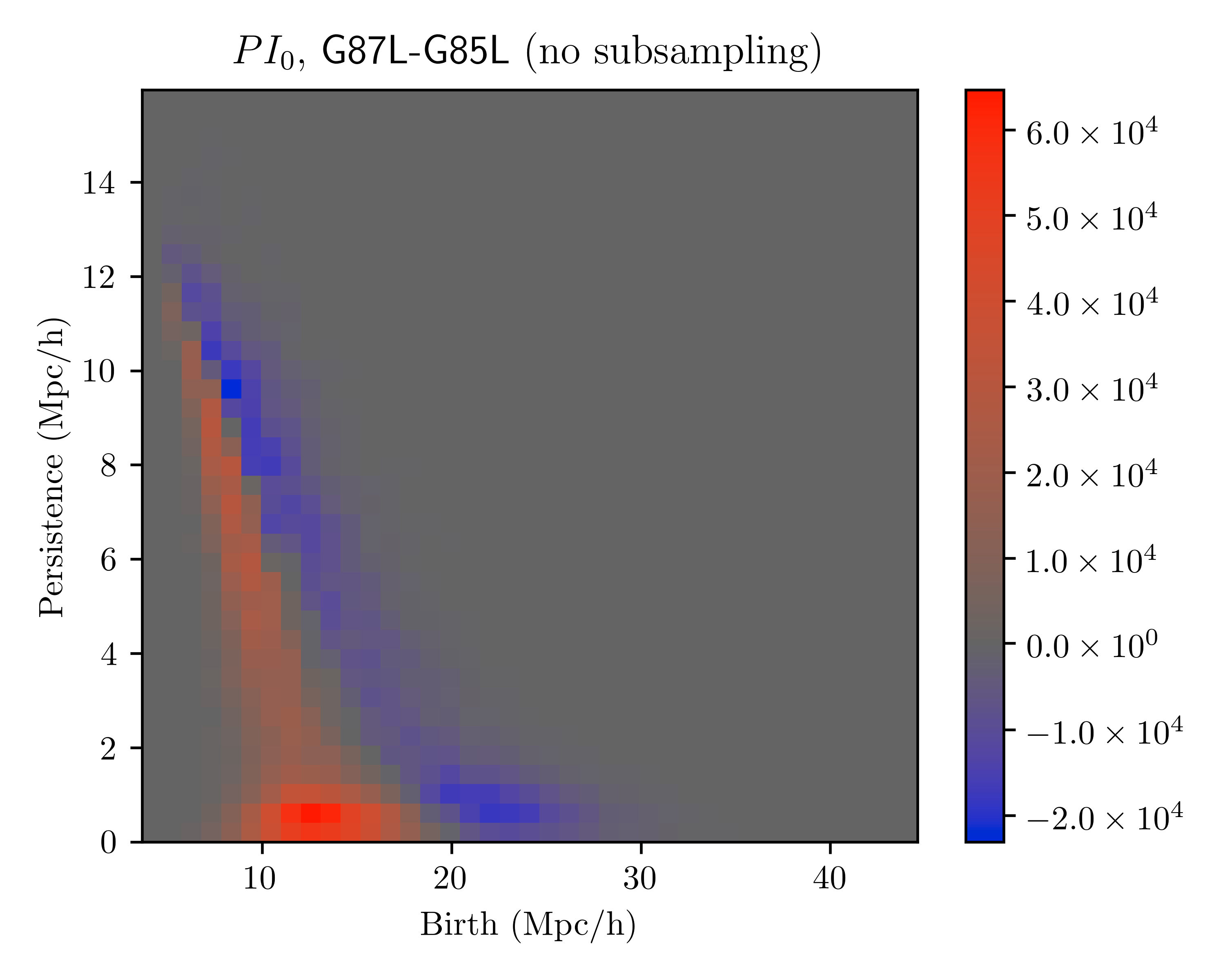}\includegraphics[width=0.33\textwidth]{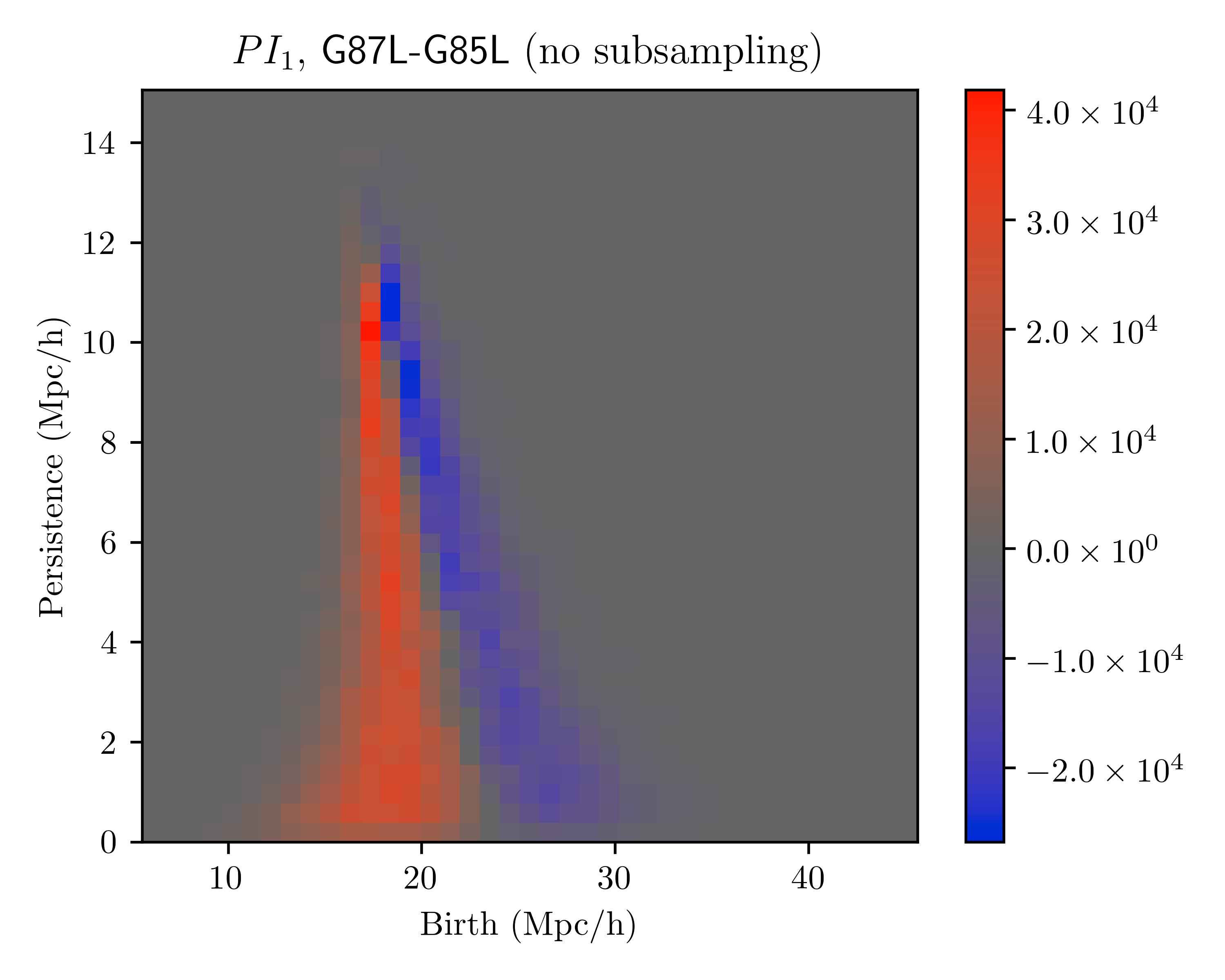}\includegraphics[width=0.33\textwidth]{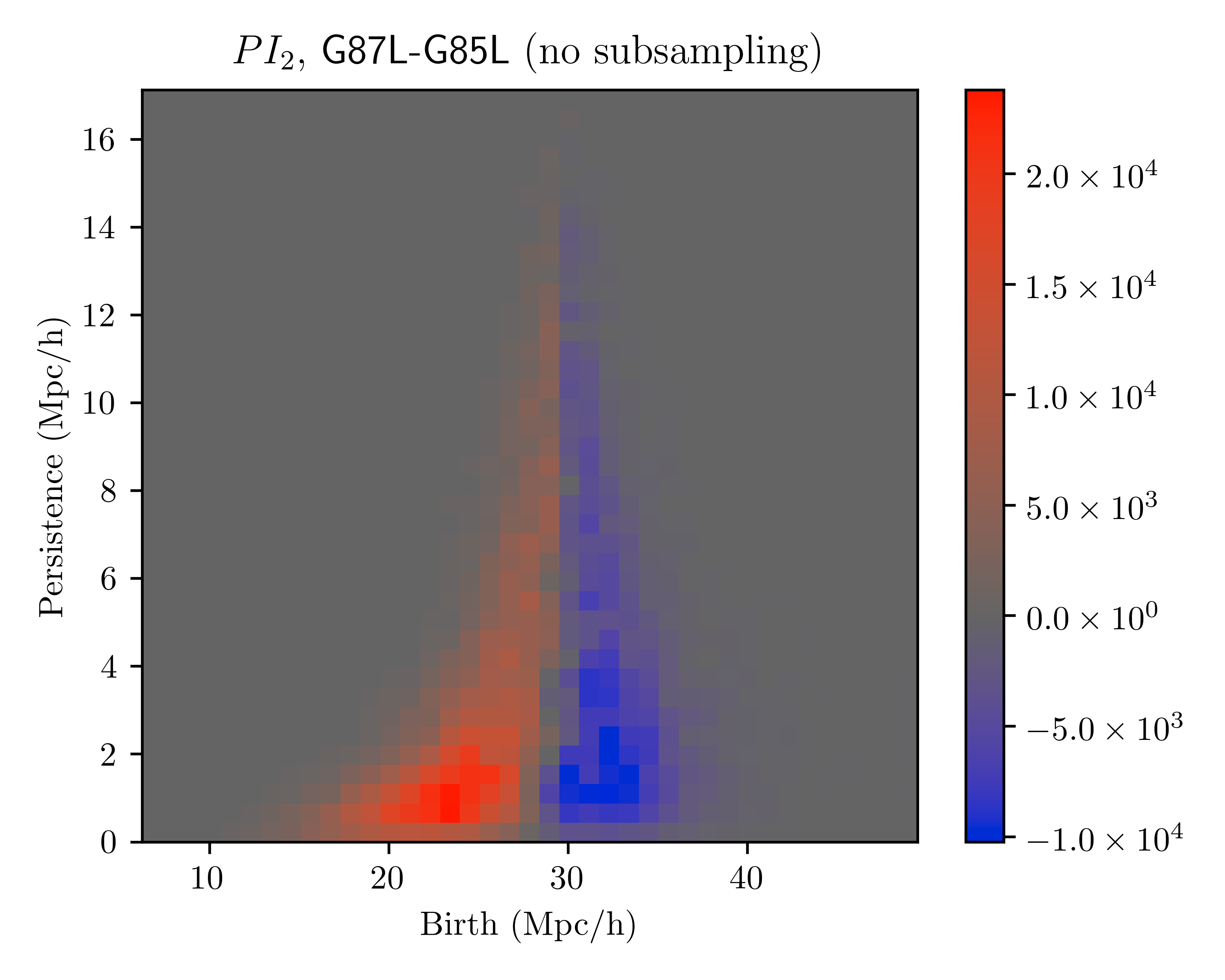}
\caption{Without subsampling, the templates corresponding to \textsf{NG250L} and \textsf{G87L} are very similar. In some sense, the main information that the topological statistics are using is that the total number of halos has changed. This leads to not being able to distinguish between the two scenarios when testing on smaller deviations in cosmology. 
}
\label{fig:templateNoSub}
\end{figure}

The above considerations are the primary motivation for our subsampling procedure. Without subsampling, it is much more difficult to distinguish between survey cosmologies that increase, or decrease, the total number of halos from the fiducial cosmology. When the fractional difference in number density deviates beyond some threshold, this is the dominant feature that our topological statistics see. This would not necessarily be a problem when directly looking for a small deviation from the fiducial cosmology, as for $\fnl=10$. However,  the cosine between templates corresponding to \textsf{NG250S} and \textsf{G87S}  is too large without subsampling for these anomalies to be sufficiently distinguished. In other words, without subsampling the main difference that persistence sees between simulations is the difference in halo number. This is in part due to our reliance on templates derived from strong deviations in cosmology. However, as previously noted using templates derived from small deviations appears to be too costly from a simulation perspective, at least for this current work. In Figure \ref{fig:templateNoSub} we show the templates for PIs computed without subsampling. Without subsampling, the templates for $\fnl=250$ and $\sigma_8$ primarily see information about total halo number, and are therefore more similar than in the subsampled case (cfr. Figures \ref{fig:PIintuition} and \ref{fig:PItemplate87}).

\section{Conclusions}\label{sec:conclusions}

It is important to understand the full sensitivity of higher-point correlations in cosmological observables to primordial physics. 
As such, there is much to gain from developing interpretable summary statistics beyond the galaxy power spectrum and bispectrum. In this paper, we have presented a general method for quantitatively characterizing the multiscale topology of large scale structure using persistent homology. This approach unifies the contributions of clusters, filaments, and voids to the constraint of cosmological parameters. While the pipeline presented here was applied to primordial non-Gaussianity, its applicability is completely general. Primordial non-Gaussianity presents a good application due to not only its physical importance as a probe of initial conditions, but also because topological features including clusters and voids are known to display characteristic signatures as a result of such initial conditions. This is the first time the persistent homology of large scale structure with non-Gaussian initial conditions has been computed. 

The first step in computing persistent homology is to represent the data under consideration via a monotonically increasing sequence of discrete complexes called a filtration. In this work, we studied three different filtrations, the $\alpha$-, $\alpha$DTM-, and $\alpha$DTM$\ell$-filtrations. These all rely on the Delaunay triangulation of a data set. The latter two filtrations employ the so-called Distance-to-Measure function to evaluate the extent to which a particular point in the data set is an outlier. From these filtrations, one computes a persistence diagram, which details the scales at which individual topological features including clusters, filament loops, and voids are created and destroyed. From a persistence diagram, there are various ways to generate tractable statistics. In this work, we primarily relied on the persistence images (PIs) and various one-dimensional curves constructed by slicing the persistence diagram.

In order to rigorously apply this formalism and constrain large scale structure, we first studied how to account for various systematics. In particular, given our desire to rely on asymptotic templates for deviations in cosmology, we subsampled our simulations. This removed the uncontrollable systematic of varying halo number density from our statistics. Additionally, we found that the $\mathcal{O}(10)$ Mpc/h length scales of our topological features allowed us to perform our persistent homology calculations in sub-boxes of the full simulations, giving a significant speedup. Moreover, the interpretability of persistence diagrams allowed us to ignore topological features generated by dynamics at scales where the simulation is not trustable. 

With these systematics addressed, we computed the effect of primordial non-Gaussianity on our statistics. In certain limits, we saw agreement with previously derived results, for example the effect of local non-Gaussianity on the void radius function \cite{Kamionkowski:2008sr}. Additionally, we found new results including a similar pattern in the ``filament loop radius function.'' Given that filament loops are ubiquitous in the cosmic web, following up on this from a first-principles perspective appears worthwhile. Indeed, one of the major results of our analysis is that filament loops are sensitive to primordial non-Gaussianity and present in large enough numbers to provide good constraints even with non-Gaussianity as small as $\fnl=10$.

Turning to statistical power, we then presented the sensitivity of our topological statistics and their competition with cosmic variance. We found that for a simple model for anomaly detection, a volume of $40~(\si{Gpc/h})^3$ with $\fnl=10$ could be identified as anomalous at a $95\%$ confidence level for $\sim 60\%$ of our simulations. To compare, a similar rate is achieved via the scale-dependent bias in the power spectrum, which relies on large-scale correlations. The significant limitation for topological statistics in this case was the simplicity of our model, necessary due to our limited data set. On the other hand, when searching for \emph{particular} physical deviations, our results improved. For this purpose, we introduced templates for the effect of specific physics on our test statistics. We found that using templates for non-Gaussianity, $\fnl=10$ could be detected using a volume of $40~(\si{Gpc/h})^3$ at $97.5\%$ confidence for $\sim 85\%$ of our simulations. Therefore there is a great deal of information to be mined from higher-point correlations of large scale structure at small scales. This reliance on scales of $\mathcal{O}(10)$ Mpc/h provides a bonus that while we still need to sample a large volume, the volume itself does not need to include large-scale information. We confirmed this feature in Appendix \ref{app:templates}, where we show that templates extracted from large and small box simulations are nearly identical (cfr. Figure \ref{fig:templateLvsS}).

We then turned to the question of ruling out other templates and the degeneracy of physical parameters. We gave preliminary results demonstrating that a notion of template optimality can mostly resolve degeneracies between different physics, at least for distinguishing $\fnl \neq 0$ from $\Delta \sigma_8$. In fact, our ability to distinguish these relied crucially on our subsampling scheme. Without subsampling, our asymptotic templates corresponding to $\fnl=250$ and $\sigma_8=0.87$ overlap much more because they both increase the number of halos.
We plan to address degeneracies with more free cosmological parameters in a future work \cite{bcsinprogress}.

While our results are encouraging, there are several relevant effects we did not address. To have complete control over our physics, we studied N-body simulations and dark matter halos. Therefore our results do not directly apply to galaxy surveys. Namely,  there  is  not  a  one-to-one  correspondence  between  halos  and  galaxies  populating  them.   Moreover,  baryon  effects  can complicate galaxy formation, requiring advanced modeling.  Additionally, redshift-space distortions complicate  the  relationship  between  real-space  and  redshift-space  observables. We believe that the results found in this work will be a fundamental baseline for upgrading the study to real data.

\section*{Acknowledgements}
We thank Daniel Chung, Pierluigi Monaco, Cristiano Porciani, Toni Riotto and Emiliano Sefusatti for useful discussions. We also thank Sven Heydenreich and Emiliano Sefusatti for useful comments on a draft. M.\,B. acknowledges support from the Netherlands Organization for Scientific Research~(NWO), which is funded by the Dutch Ministry of Education, Culture and Science~(OCW), under VENI grant 016.Veni.192.210. M. B. also acknowledges support from the NWO under the project ``Cosmic Origins from Simulated Universes'' for the computing time allocated to run a subset of the \textsc{Eos} Simulations on \textsc{Cartesius}, a supercomputer which is part of the Dutch National Computing Facilities. M.B. thanks the Institute for Fundamental Physics of the Universe for kind hospitality during the completion of this work. A.C.\ is supported in part by the Heising-Simons Foundation, the Simons Foundation, and National Science Foundation Grant No.\ NSF PHY-1748958, and gratefully acknowledges support from the UW-Madison Graduate School via a Straka Fellowship. A.C.\ also thanks Mert Besken and John Stout for letting him crash in their apartments while this work was being completed. G.S.\ is supported in part by the DOE grant DE-SC0017647 and the Kellett Award of the University of Wisconsin.
\appendix

\section{Full Results}
\label{app:table}
In this appendix, we include all results for the three filtrations we consider, i.e. $\alpha$-, $\alpha$DTM- and $\alpha$DTM$\ell$-filtrations, for all the statistics we consider (cfr. Figure \ref{fig:curveConstruct}) at all redshifts, i.e. $z=0$, $1$ and $2$, both for anomaly (cfr. Section \ref{sec:anomaly}) and template (cfr. Section \ref{sec:templatePNG}) detection methods. While, performance of constraints vary across statistics, redshifts and filtrations, we typically see that the following features
\begin{itemize}
    \item Among the statistics, births $B_p$ give the best constraints
    \item Among homology classes, $1$-cycles give the best constraints
    \item Among redshifts, $z=1$ give the best constraints
    \item Among filtrations, $\alpha$DTM$\ell$ give the highest detection rate.
\end{itemize}
Our explanations for such features are found in the main text (cfr. Section \ref{sec:phhalo} and Section \ref{sec:pipe}).
\begin{table}
\begin{center}
\begin{small}
\begin{tabular}{|c|c||c|c|c||c|c|c|}\hline
       \mbox{ }&$\boldsymbol{\alpha}$ &anomaly$_{z=0}$&anomaly$_{z=1}$& anomaly$_{z=2}$&template$_{z=0}$&template$_{z=1}$& template$_{z=2}$ \\\hline\hline
  \multirow{9}{*}{\rotatebox[origin=c]{90}{~Statistic}} & $b_1$&2.0&\bf{55.4}&\bf{55.2}&11.3&\bf{86.5}&41.9\\\cline{2-8}
   \mbox{ }&$b_2$ & 2.0&6.8&30.0&0.4&2.1&9.7\\\cline{2-8}
   \mbox{ }& $B_1$&7.1&29.6&48.7&16.9&43.1&\bf{63.7}\\\cline{2-8}
  \mbox{ }& $B_2$ & 8.2&29.6&\bf{52.1}&0.7&15.3&15.7\\\cline{2-8}
   \mbox{ }& $D_1$&5.5&21.8&7.4&8.7&6.2&5.1\\\cline{2-8}
  \mbox{ }& $D_2$ & 4.5&1.7&2.2&4.4&9.7&5.1\\\cline{2-8}
  \mbox{ }& $P_0$ & 0.4&9.8&1.3&0.5&5.3&44.8\\\cline{2-8}
  \mbox{ }&$P_1$&6.7&34.7&\bf{53.2}&13.0&39.1&\bf{65.2}\\\cline{2-8}
  \mbox{ }& $P_2$ & 13.1&22.7&19.1&0.4&18.1&6.6\\\hline\hline
 \mbox{ }&  $\boldsymbol{\alpha}$\bf{DTM}&anomaly$_{z=0}$&anomaly$_{z=1}$& anomaly$_{z=2}$&template$_{z=0}$&template$_{z=1}$& template$_{z=2}$ \\\hline\hline
    \multirow{12}{*}{\rotatebox[origin=c]{90}{~Statistic}} &  $b_0$ & 2.3&2.9&14.8&3.8&12.9&5.9\\\cline{2-8}
   \mbox{ }&  $b_1$&1.7&\bf{67.0}&2.9&3.1&45.0&23.1\\\cline{2-8}
   \mbox{ }&$b_2$ & 4.0&10.6&13.8&0.0&7.8&\bf{63.4}\\\cline{2-8}
   \mbox{ }&$B_0$ & 9.1&0.3&2.7&8.8&9.0&0.8\\\cline{2-8}
  \mbox{ }&   $B_1$&1.5&49.7&\bf{63.0}&2.7&\bf{57.0}&\bf{81.4}\\\cline{2-8}
  \mbox{ }& $B_2$ & 0.2&27.2&9.8&0.0&11.6&33.5\\\cline{2-8}
  \mbox{ }& $D_0$ & 2.3&0.0&5.2&19.0&0.0&26.3\\\cline{2-8}
   \mbox{ }&  $D_1$&2.6&0.0&0.8&0.0&0.0&0.0\\\cline{2-8}
  \mbox{ }& $D_2$ & 1.4&20.2&12.5&0.1&2.4&1.2\\\cline{2-8}
  \mbox{ }& $P_0$ & 14.0&15.1&10.3&14.7&27.1&14.7\\\cline{2-8}
  \mbox{ }& $P_1$&1.2&\bf{51.3}&35.9& 2.1&47.8&\bf{59.1}\\\cline{2-8}
  \mbox{ }&$P_2$ & 0.9&11.1&33.2&0.0&7.6&43.1\\\hline
 \hline
    \mbox{ }& $\boldsymbol{\alpha}$\bf{DTM}$\boldsymbol{\ell}$ &anomaly$_{z=0}$&anomaly$_{z=1}$& anomaly$_{z=2}$&template$_{z=0}$&template$_{z=1}$& template$_{z=2}$ \\\hline\hline
     \multirow{12}{*}{\rotatebox[origin=c]{90}{~Statistic}} & $b_0$&6.6&2.4&46.5&0.0&1.3&\bf{54.8}\\\cline{2-8}
    \mbox{ }& $b_1$&5.8&44.3&\bf{51.8}&6.6&30.1&20.3\\\cline{2-8}
   \mbox{ }&$b_2$ & 6.3&9.9&\bf{79.8}&0.0&6.6&\bf{53.3}\\\cline{2-8}
   \mbox{ }&$B_0$ & 8.7&11.9&\bf{70.8}&18.0&32.1&15.3\\\cline{2-8}
    \mbox{ }& $B_1$&1.6&\bf{63.2}&\bf{68.6}&4.9&\bf{84.8}&\bf{59.2}\\\cline{2-8}
   \mbox{ }&$B_2$ & 11.2&17.9&\bf{72.2}&1.6&16.2&\bf{80.1}\\\cline{2-8}
   \mbox{ }&   $D_0$ & 11.9&0.0&10.1&18.1&2.1&21.7\\\cline{2-8}
   \mbox{ }&  $D_1$&6.4&2.9&\bf{60.9}&1.9&9.5&36.1\\\cline{2-8}
   \mbox{ }&$D_2$ & 10.3&6.5&8.1&0.0&0.5&8.1\\\cline{2-8}
   \mbox{ }&$P_0$ & 7.5&18.2&\bf{74.2}&19.4&31.7&18.9\\\cline{2-8}
   \mbox{ }&$P_1$&2.3&\bf{62.4}&\bf{71.9}&5.2&\bf{73.9}&\bf{57.1}\\\cline{2-8}
   \mbox{ }&$P_2$ & 1.6&19.5&\bf{62.4}&1.5&17.0&\bf{57.4}\\\hline
\end{tabular}
\caption{Summary table of all results both for the anomaly and template methods for constraining $\fnl$. All numbers represent percentages, in the case of anomaly detection they represent the fraction of simulation volume for which we are $95\%$ confident of detecting $\fnl=10$, while for the template method, the confidence corresponds to $97.5\%$ (see Section \ref{sec:anomaly} for details). We highlight all results for which the detection fraction is higher than $50\%$.}
\label{tab:fullResults}
\end{small}
\end{center}
\end{table}

\section{Template Consistency}\label{app:templates}
\begin{figure}
    \centering
    \includegraphics[width=0.33\textwidth]{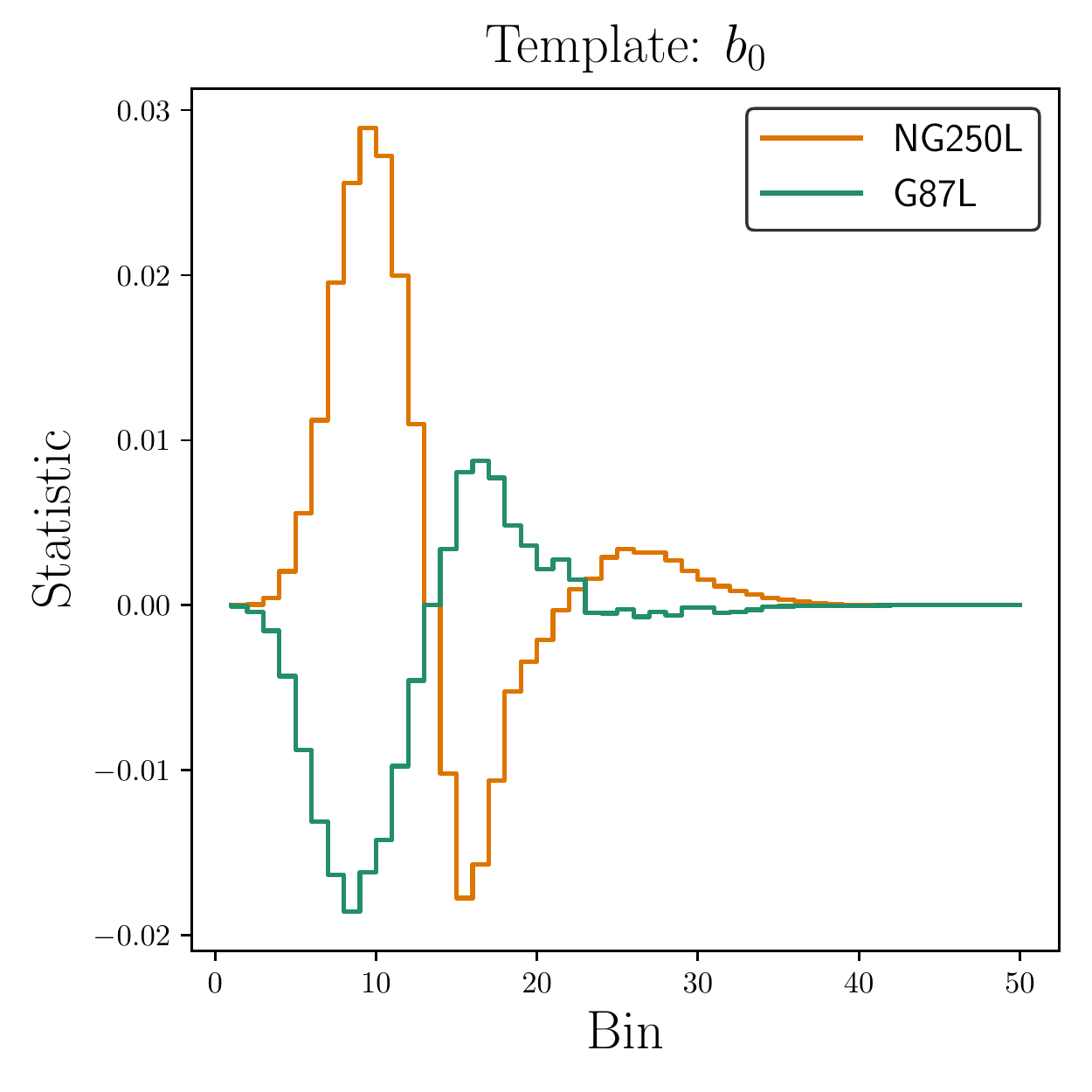}\includegraphics[width=0.33\textwidth]{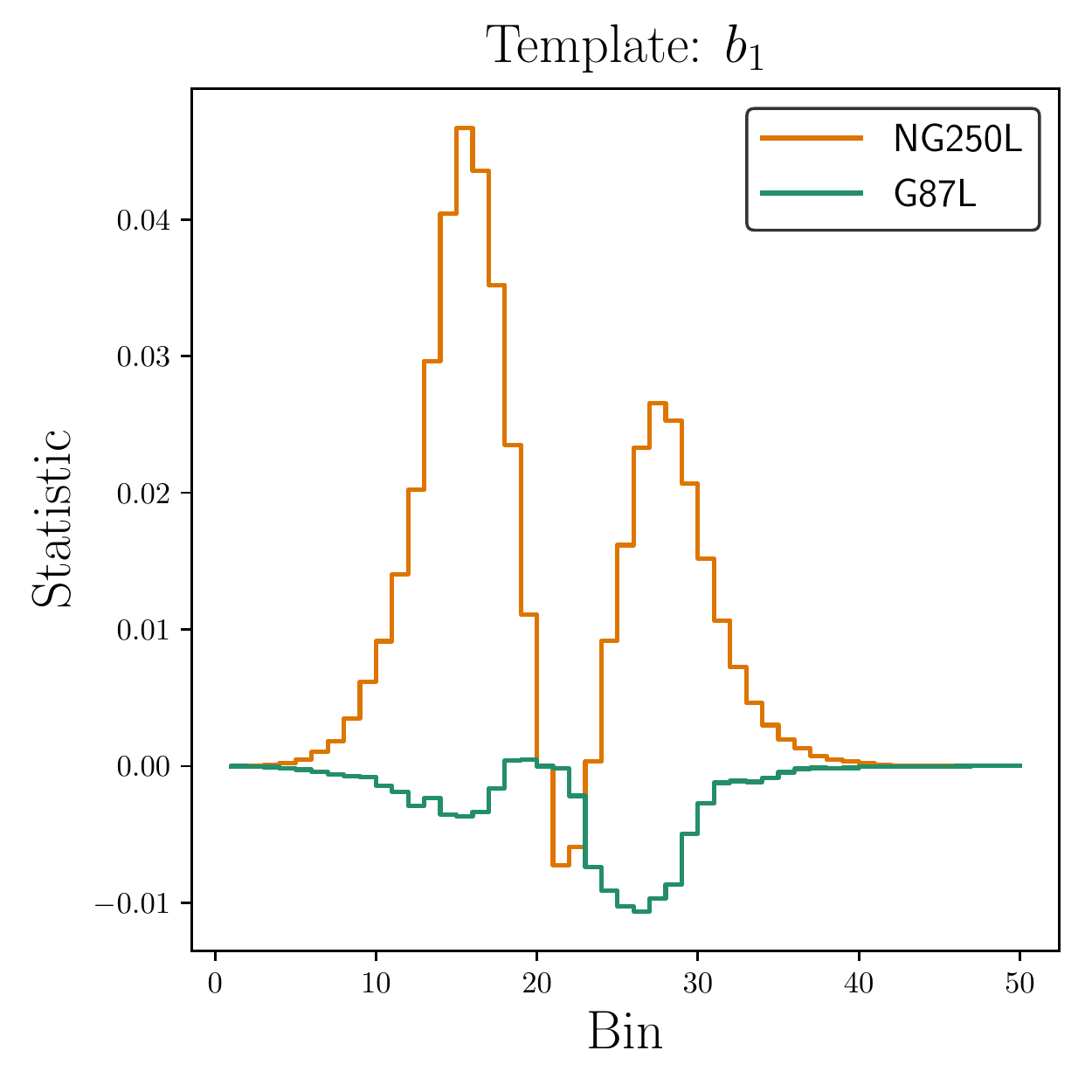}\includegraphics[width=0.33\textwidth]{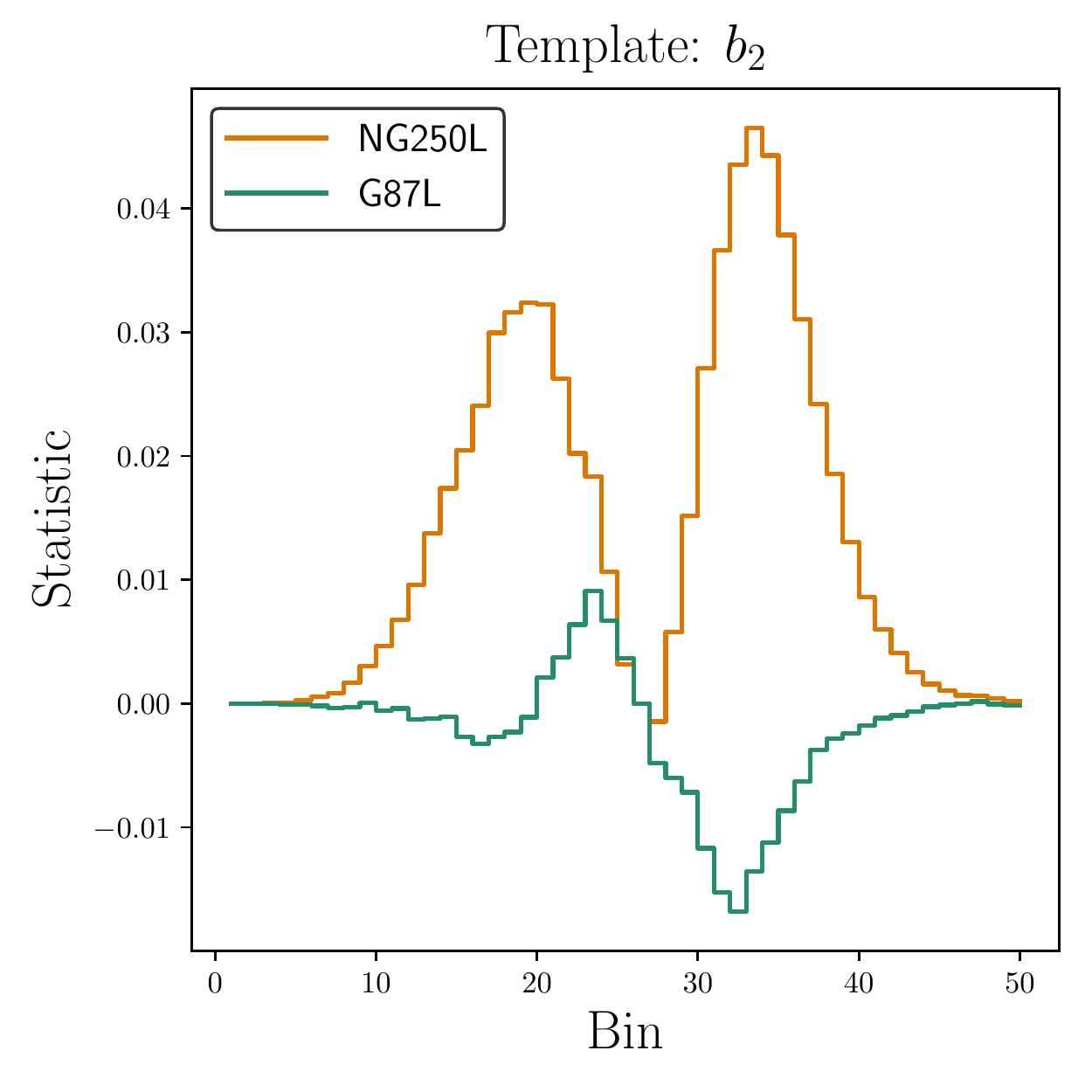}
    \includegraphics[width=0.33\textwidth]{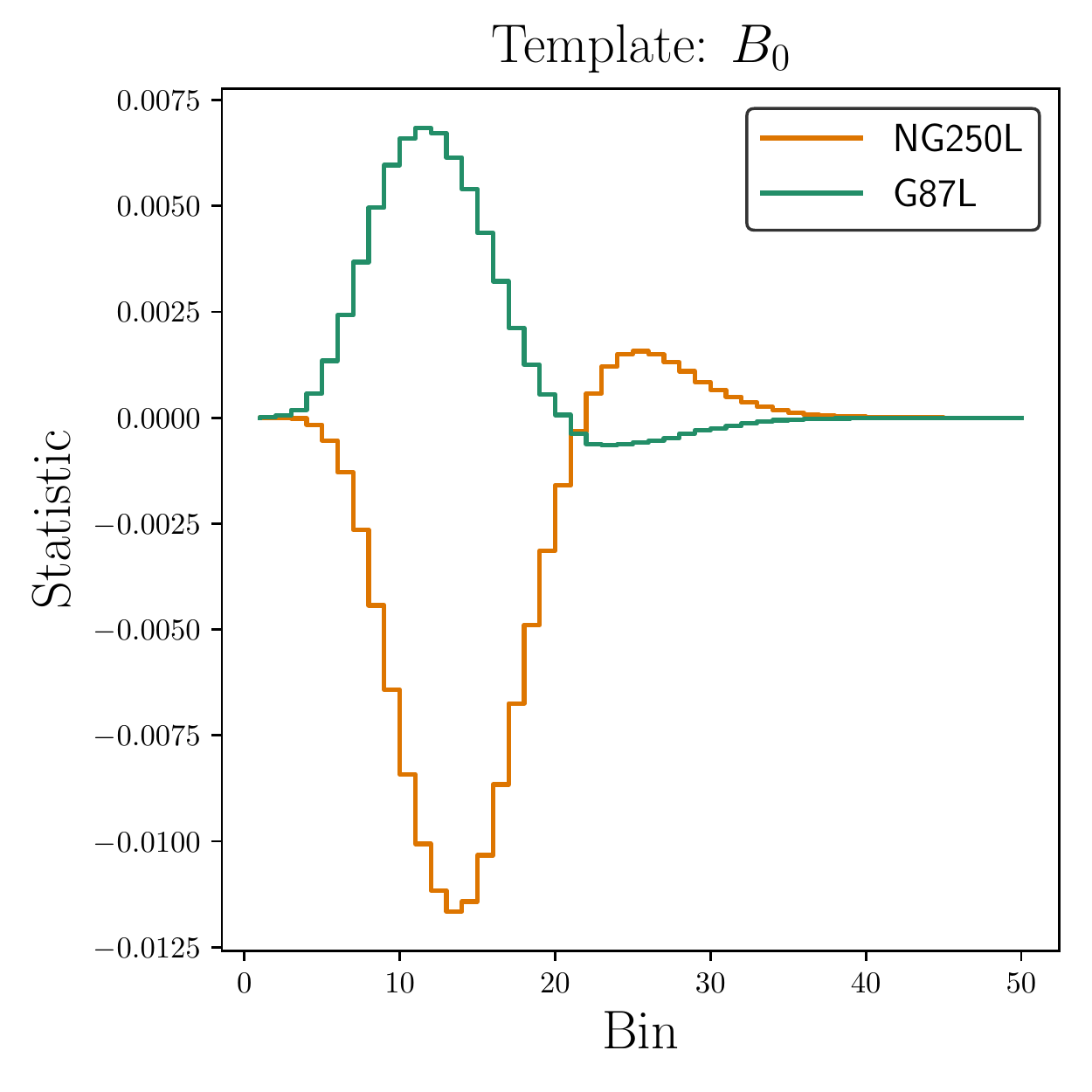}\includegraphics[width=0.33\textwidth]{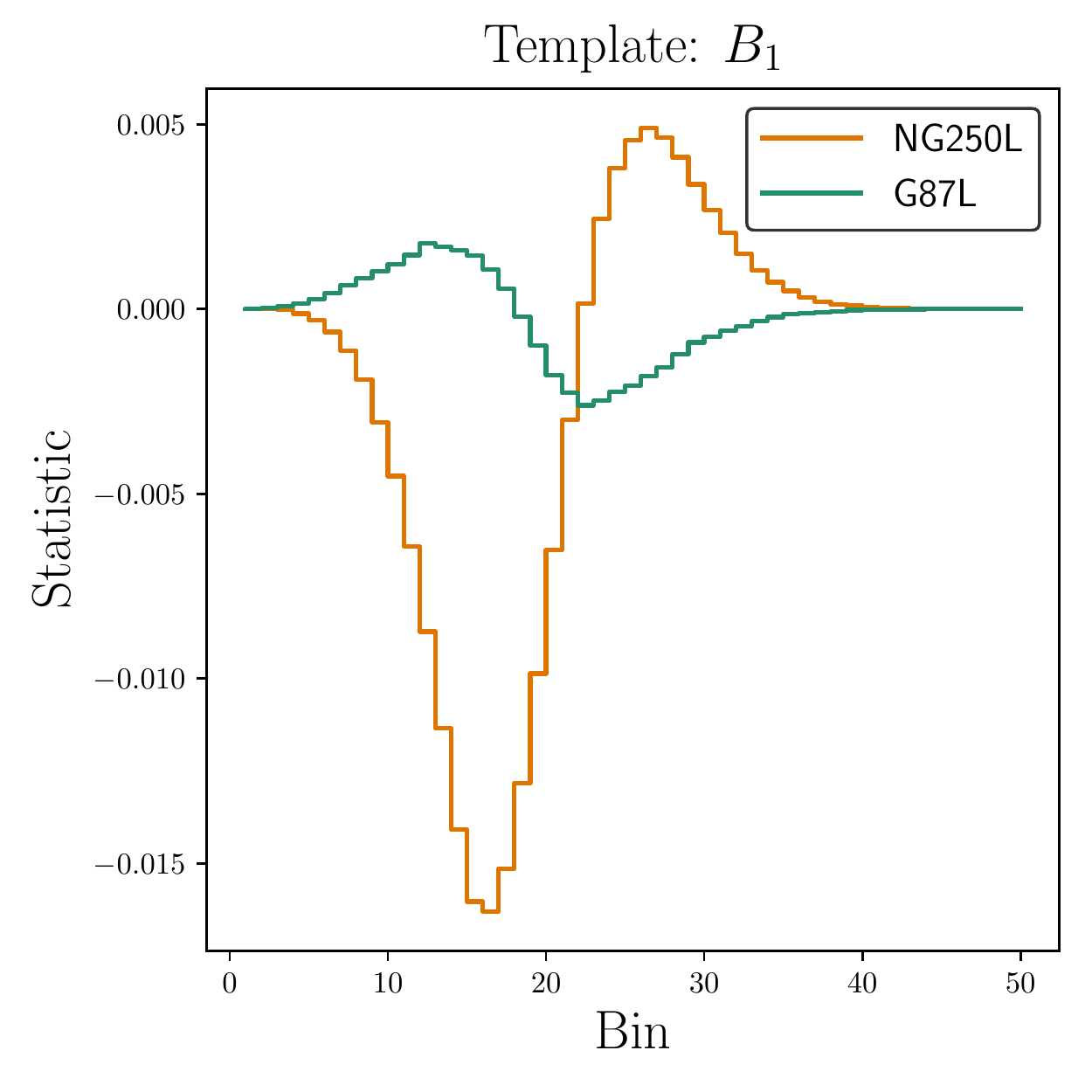}\includegraphics[width=0.33\textwidth]{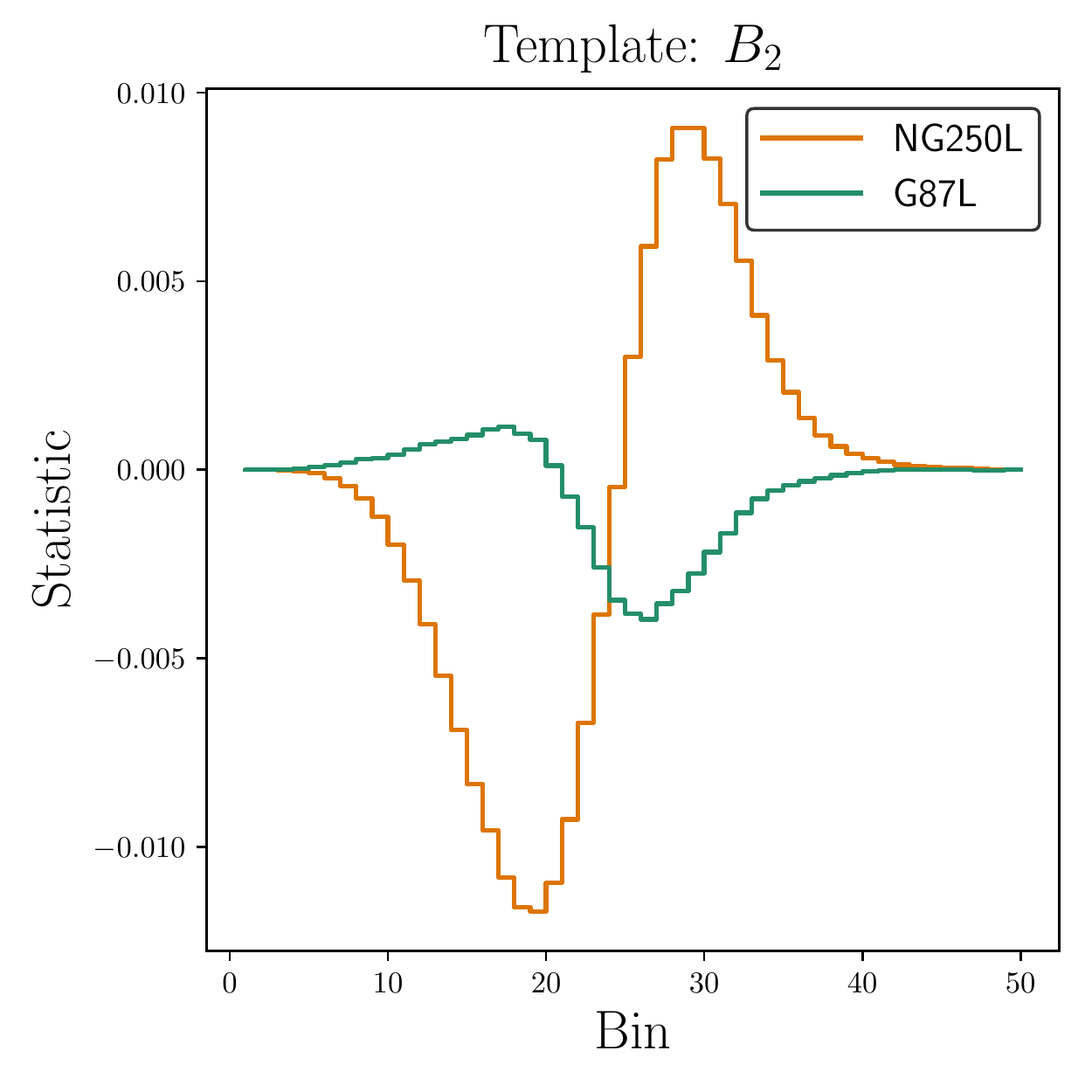}
    \includegraphics[width=0.33\textwidth]{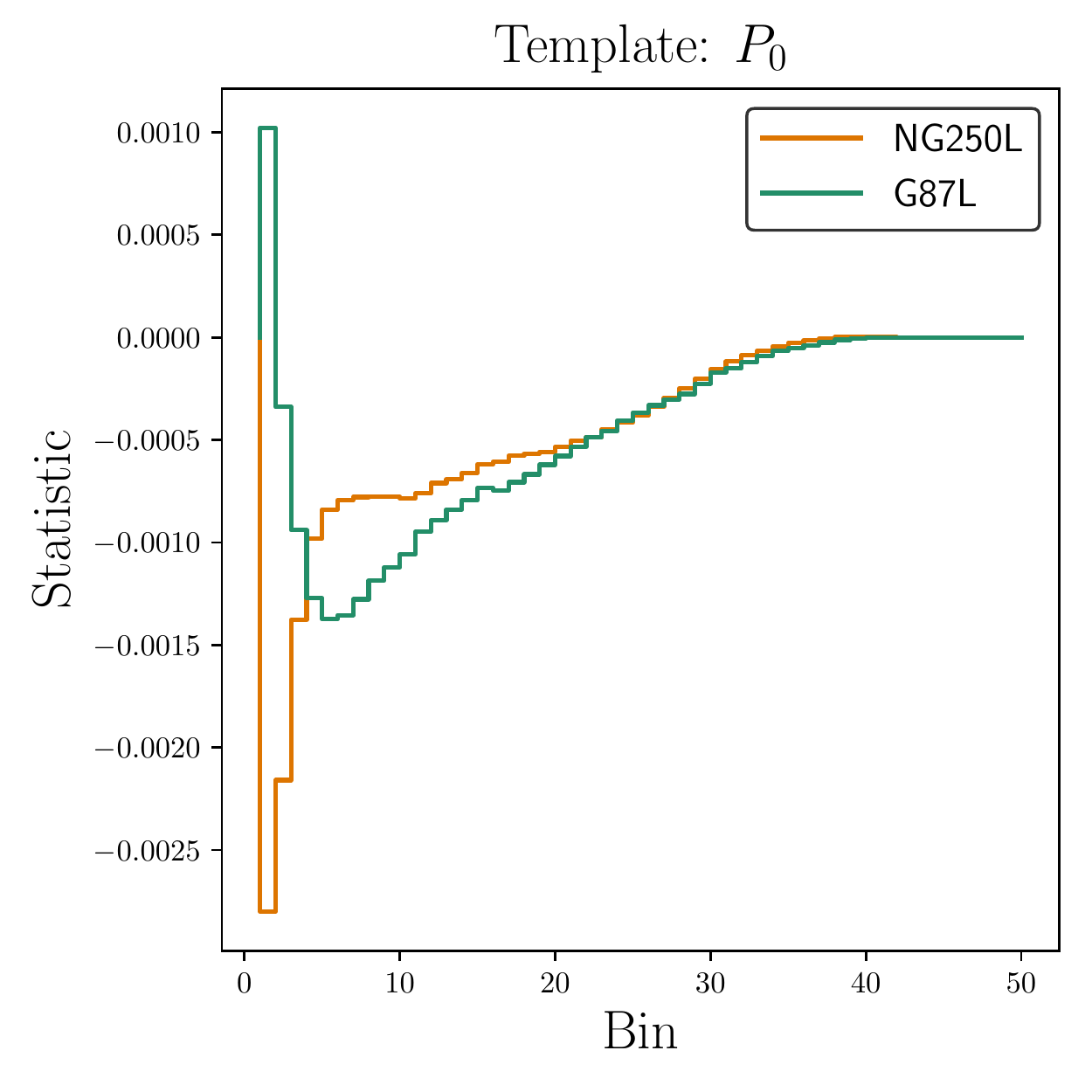}\includegraphics[width=0.33\textwidth]{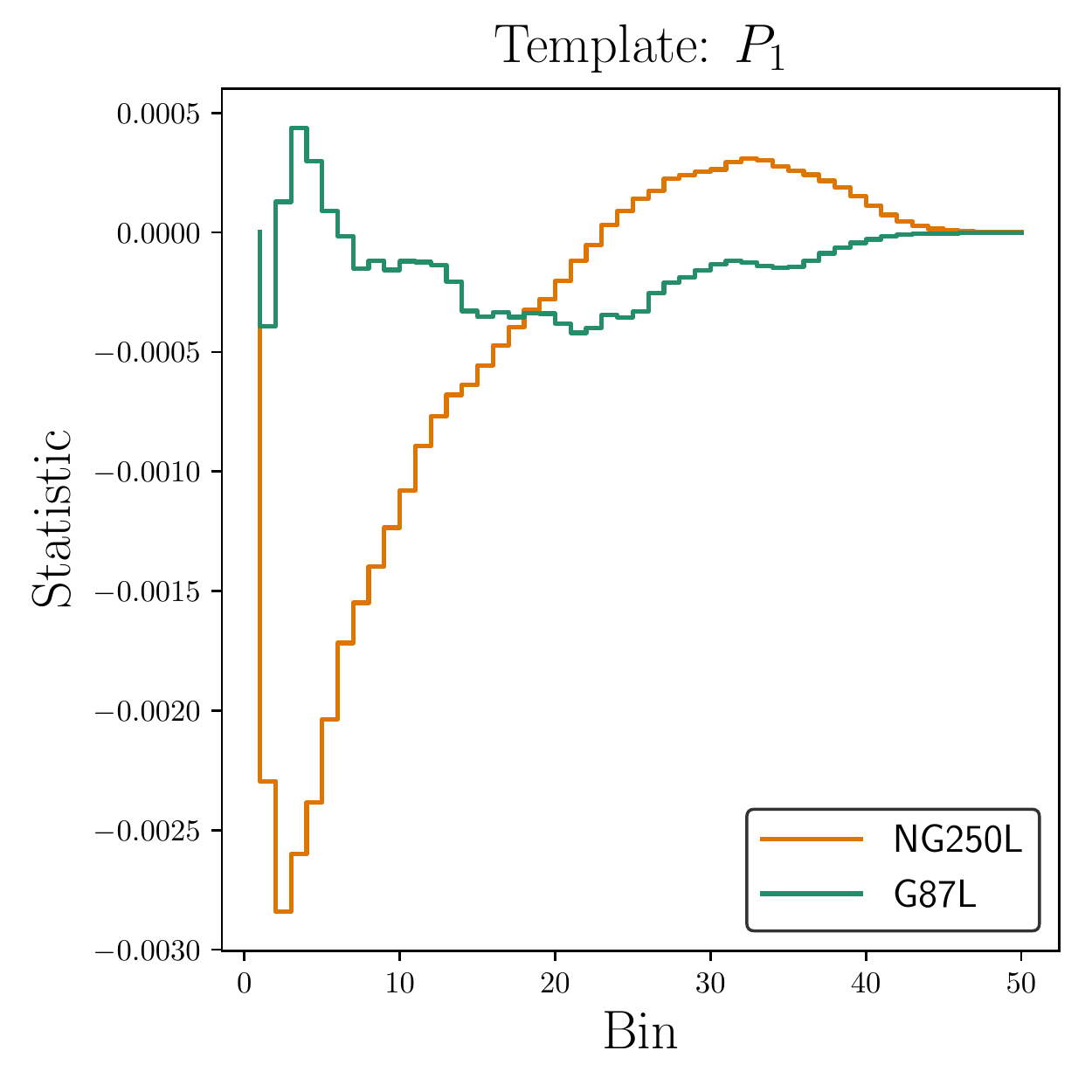}\includegraphics[width=0.33\textwidth]{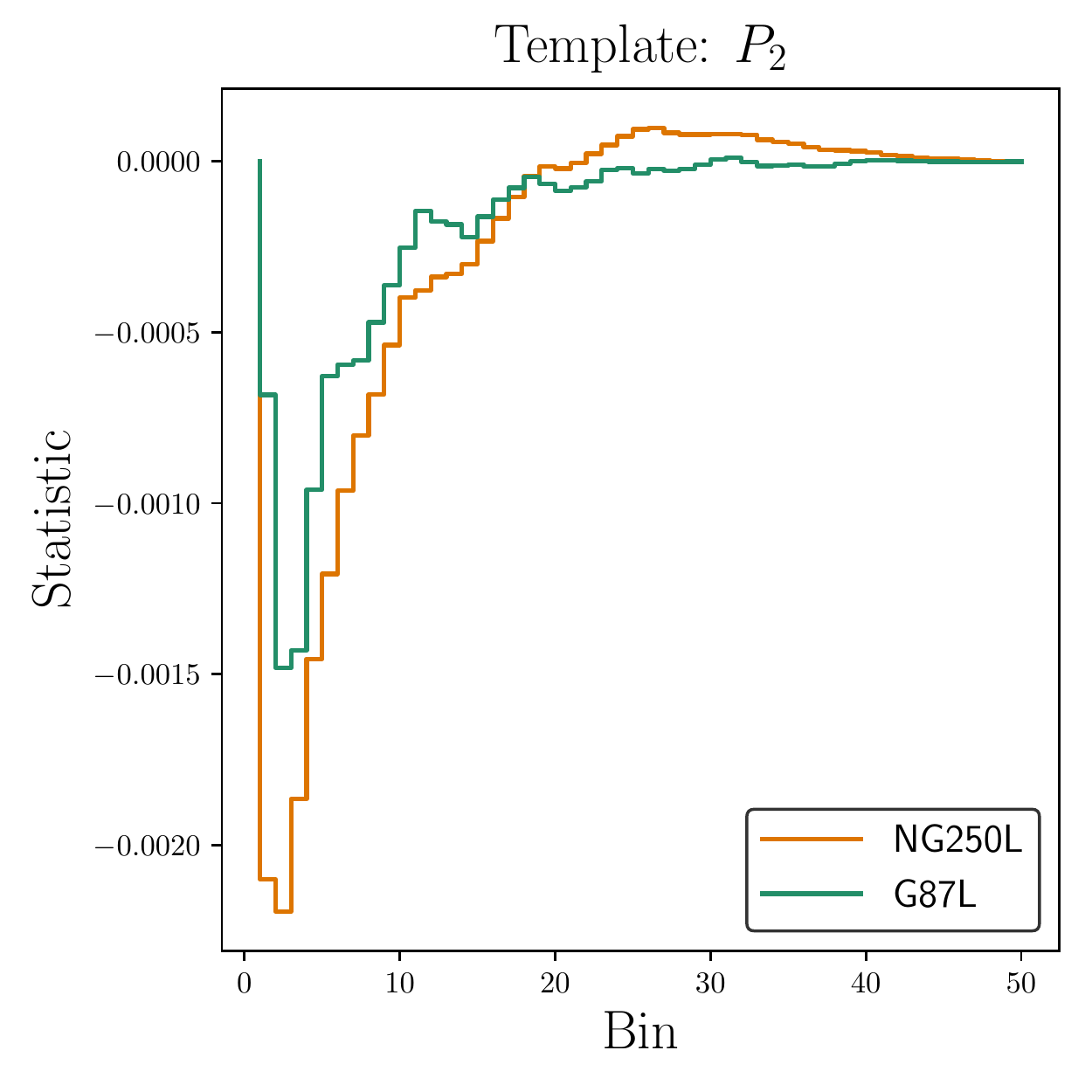}
    \includegraphics[width=0.33\textwidth]{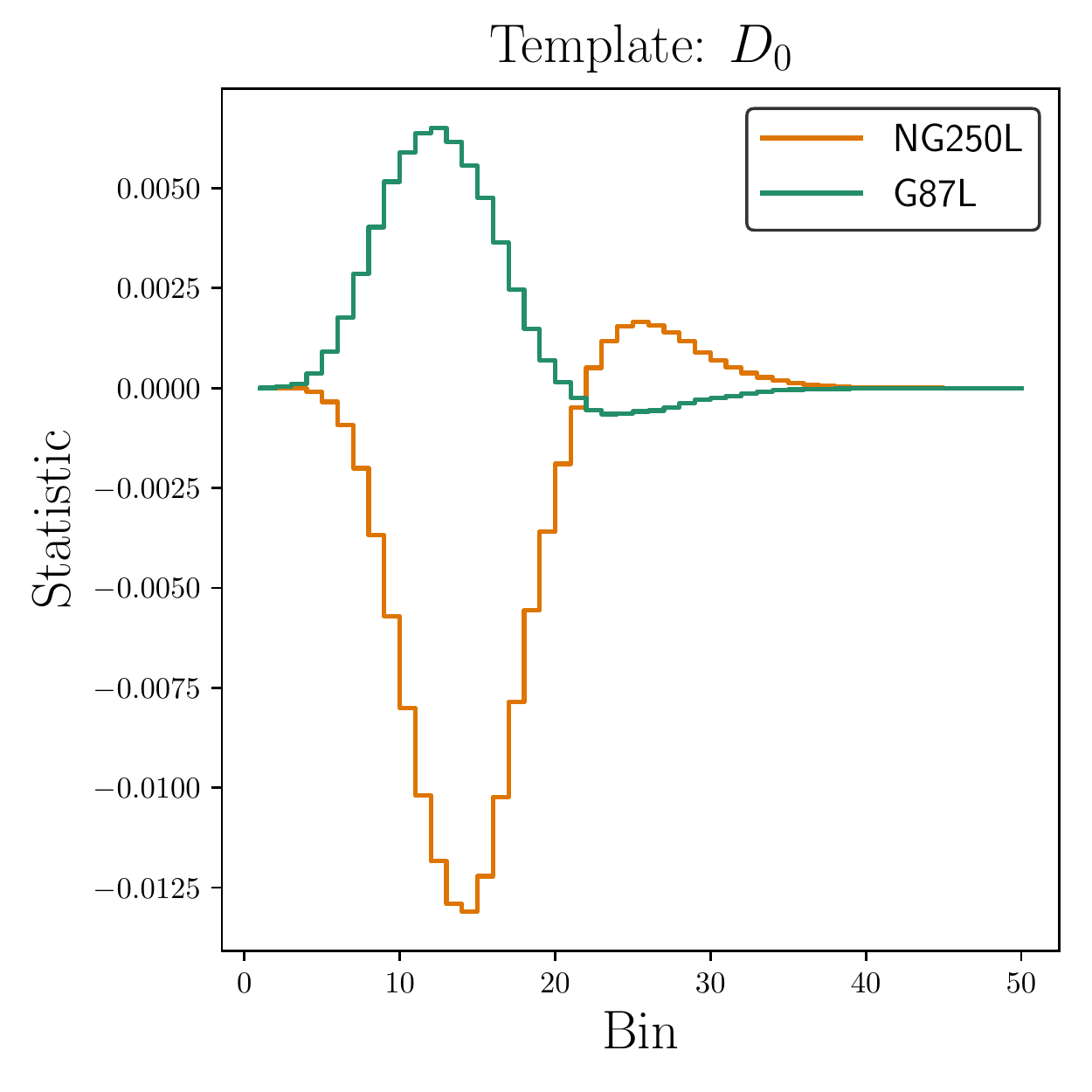}\includegraphics[width=0.33\textwidth]{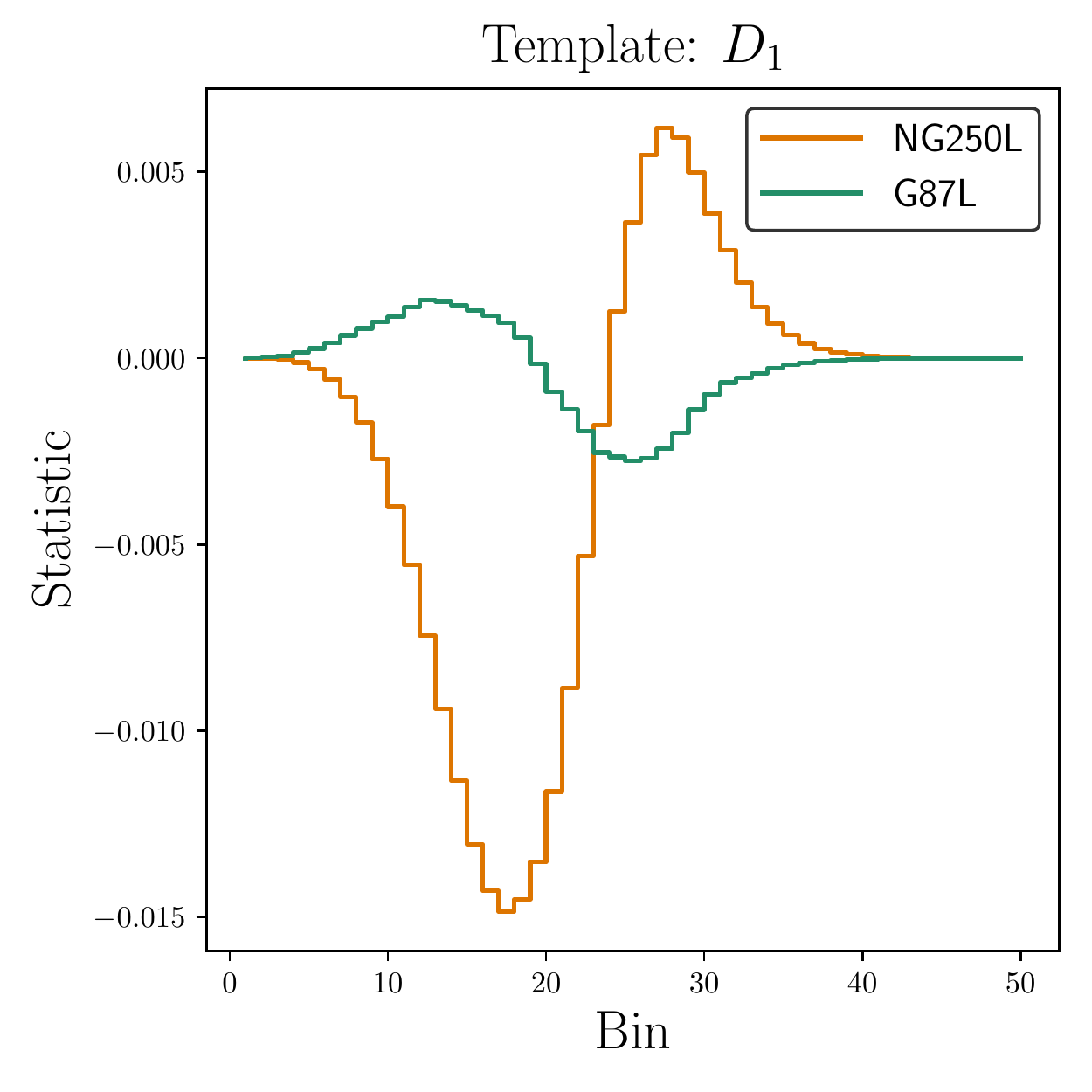}\includegraphics[width=0.33\textwidth]{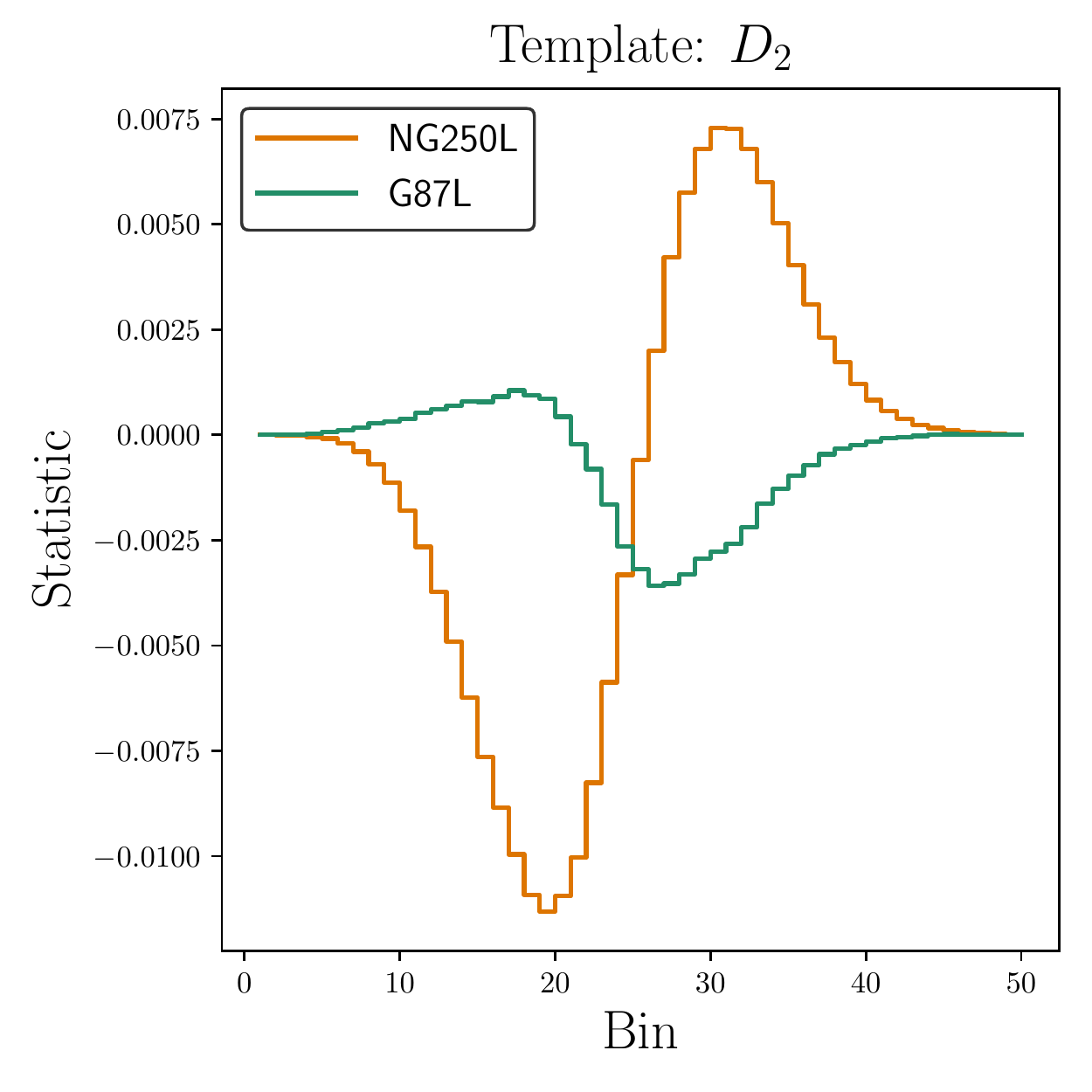}
    \caption{Templates demonstrating the effect of varying $\fnl$ and $\sigma_8$ on our topological statistics. For most of our statistics, the two scenarios are easily distinguished. This was already anticipated by comparing the corresponding PIs, see Section  \ref{sec:degen}.}
    \label{fig:templateAsymp}
\end{figure}
\afterpage{\FloatBarrier}
An important aspect of our statistical pipeline is the use of templates for the change in a topological statistic as some aspect of the underlying cosmology is varied. In this section we demonstrate several aspects of our templates that enhance the efficiency of our pipeline. We show the running of our non-Gaussian templates with the magnitude of $\fnl$ and compare templates generated from simulations with different resolutions.
\begin{figure}
    \centering
        \includegraphics[width=0.33\textwidth]{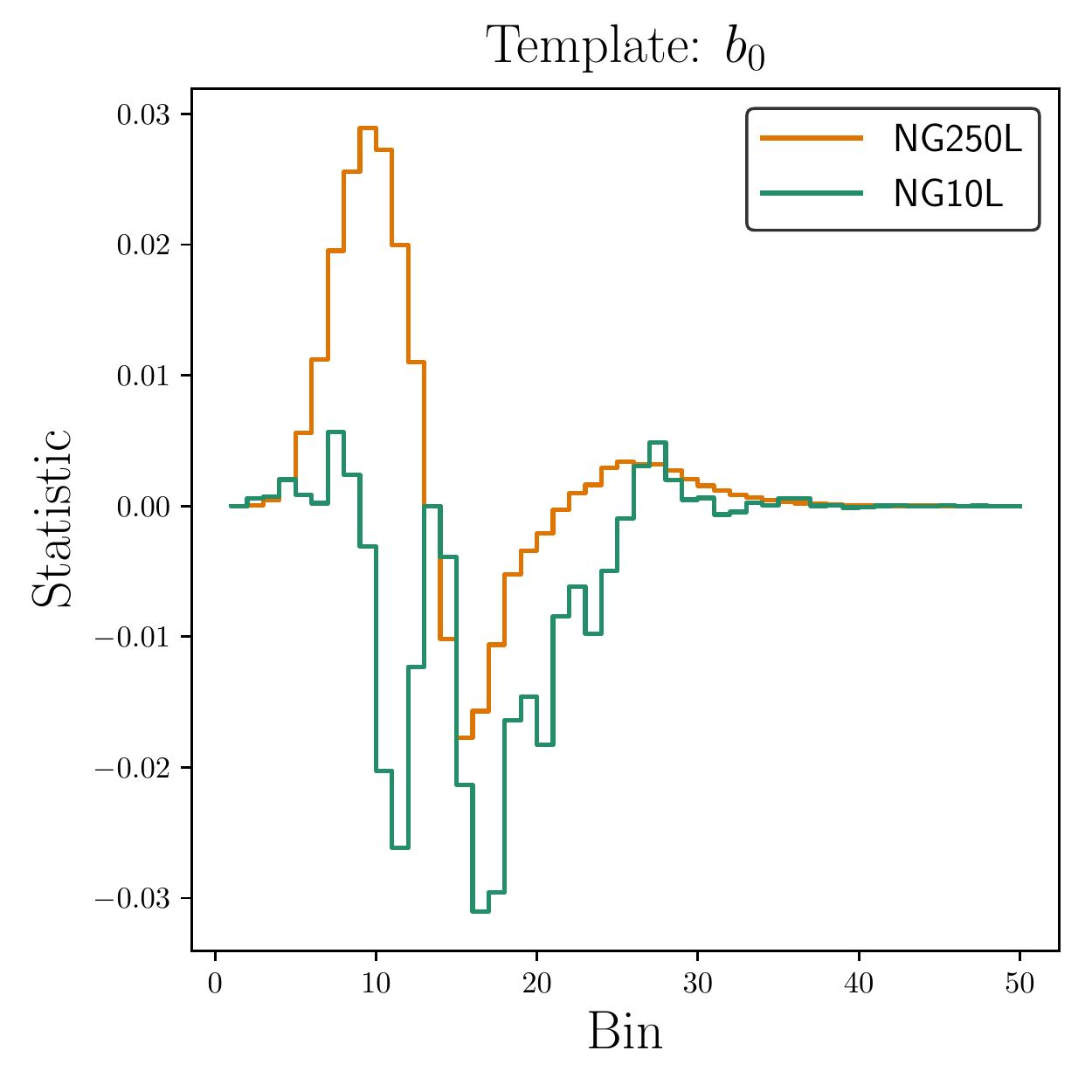}\includegraphics[width=0.33\textwidth]{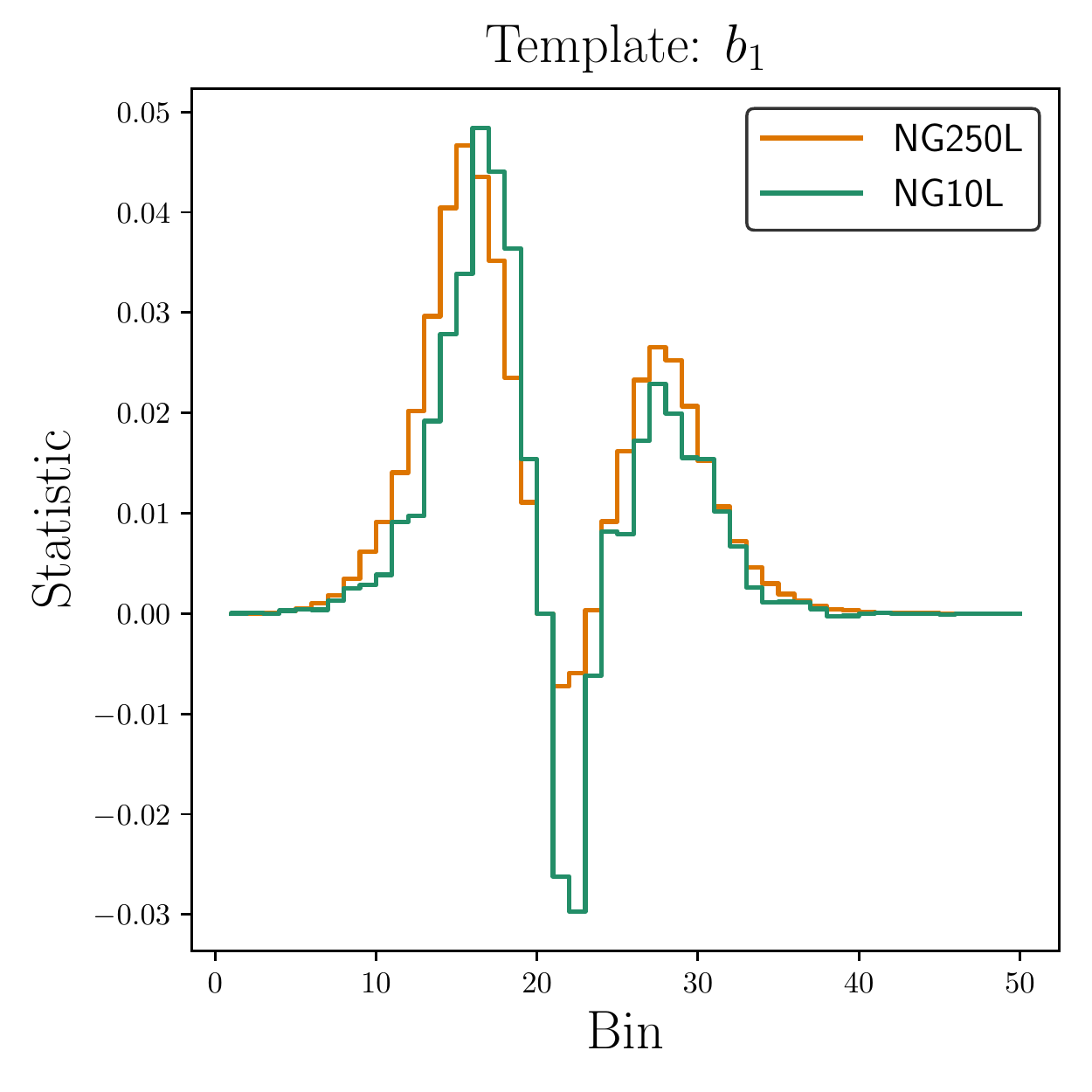}\includegraphics[width=0.33\textwidth]{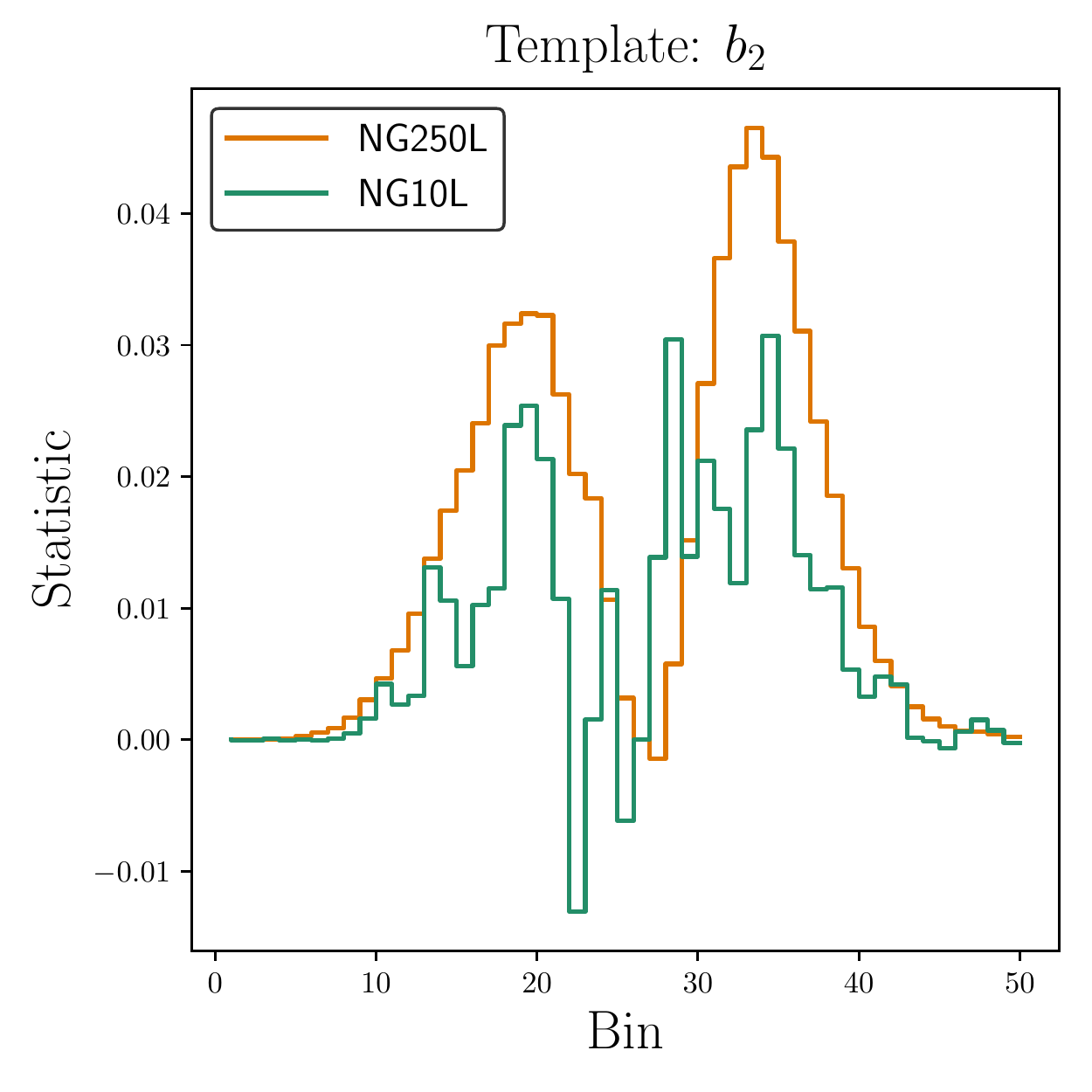}
        \includegraphics[width=0.33\textwidth]{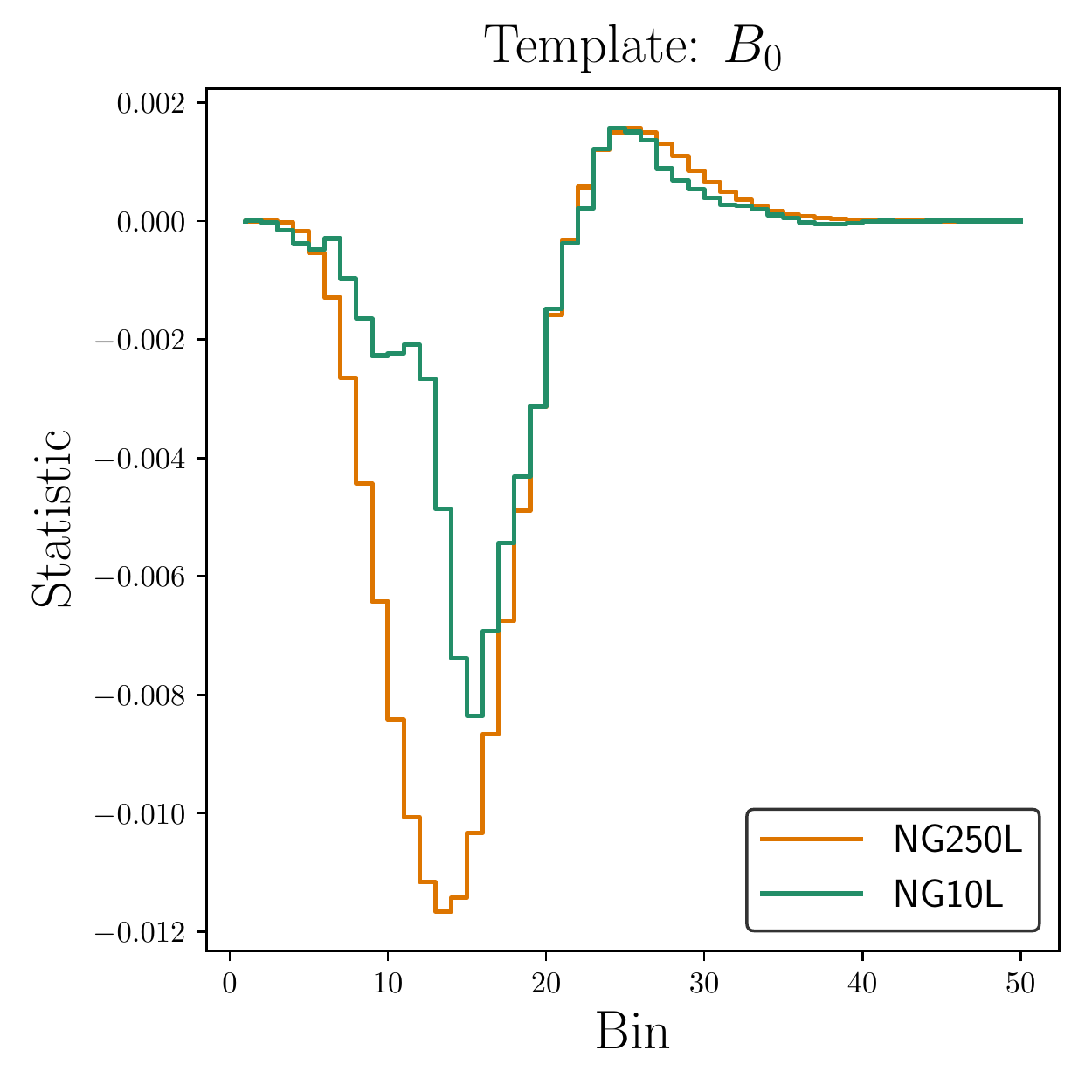}\includegraphics[width=0.33\textwidth]{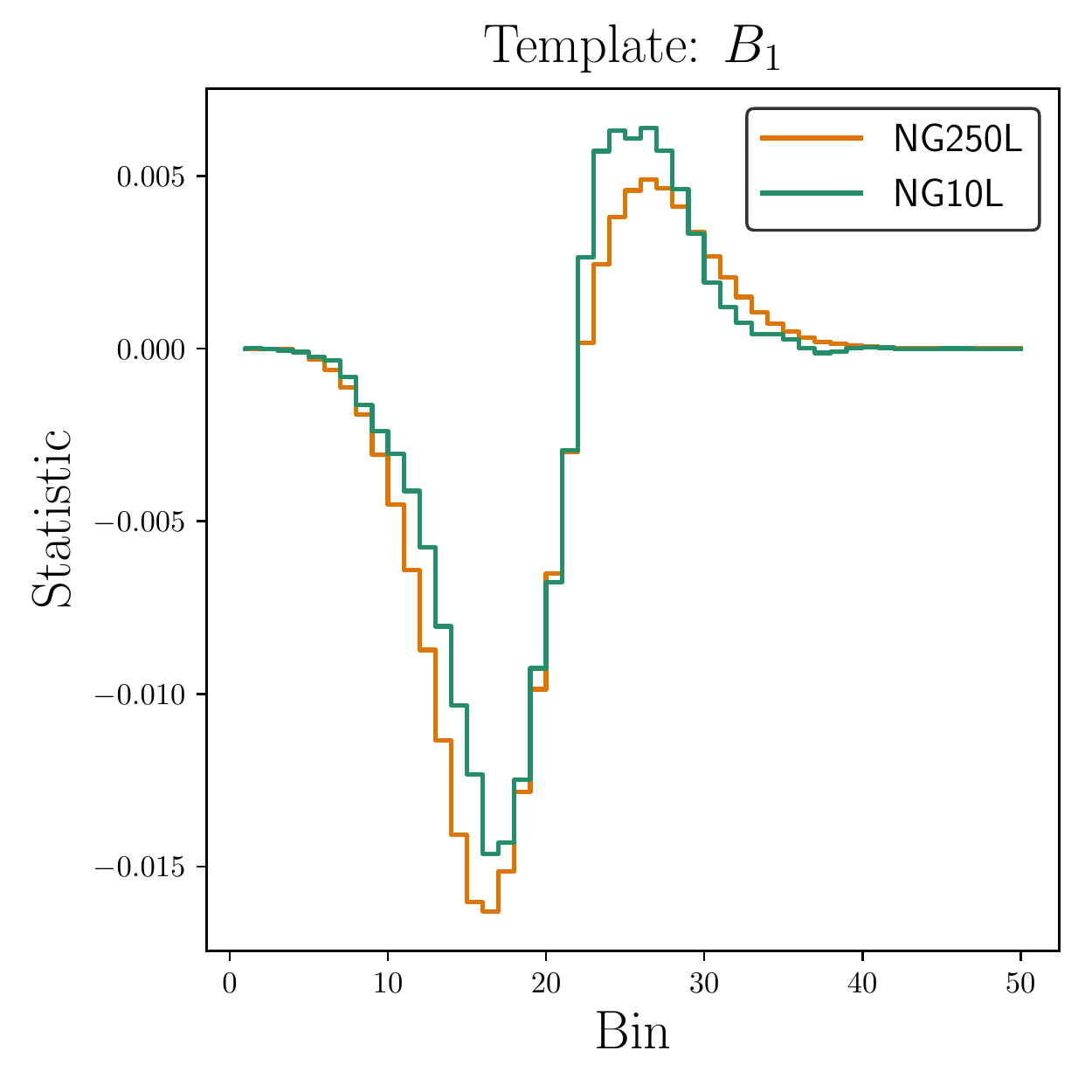}\includegraphics[width=0.33\textwidth]{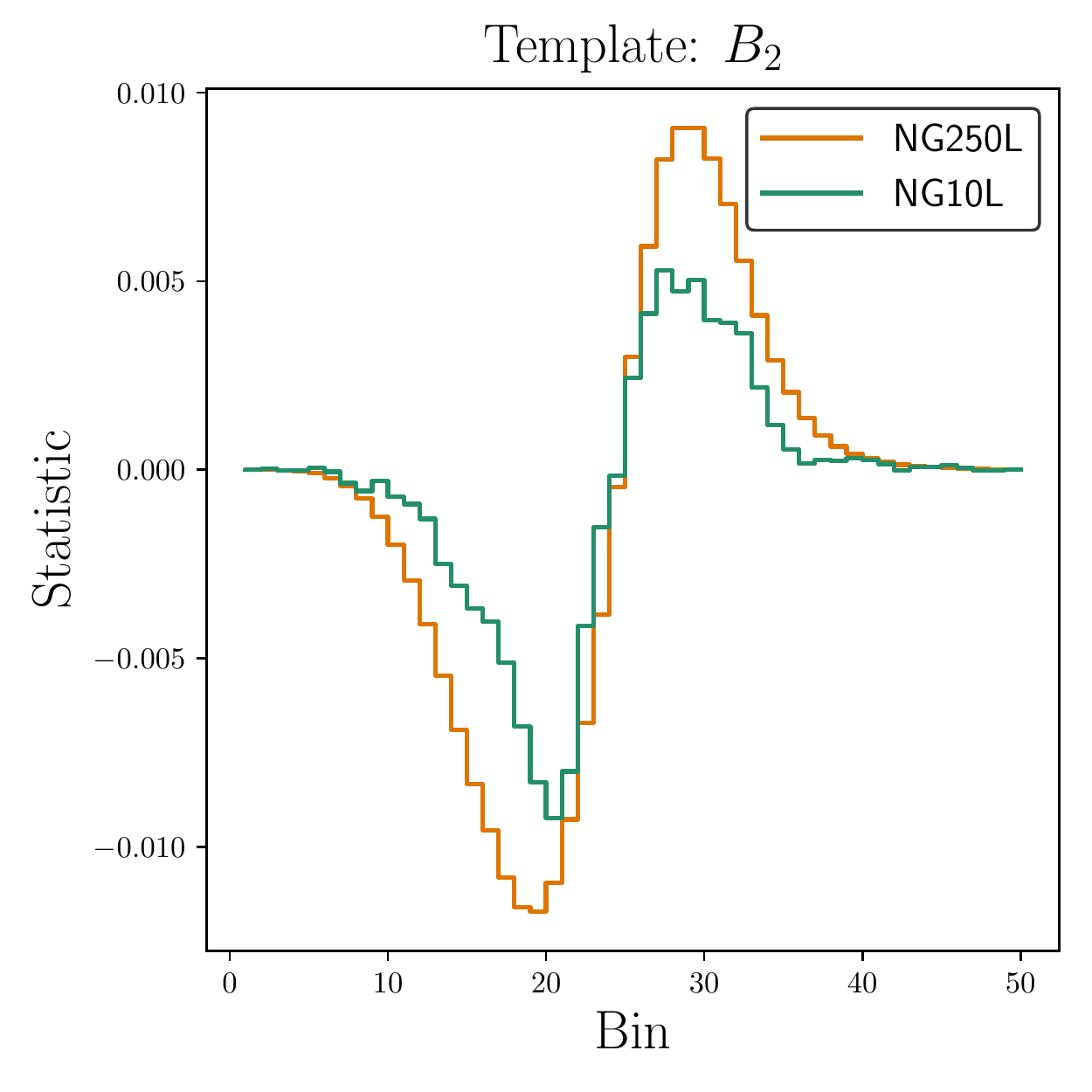}
        \includegraphics[width=0.33\textwidth]{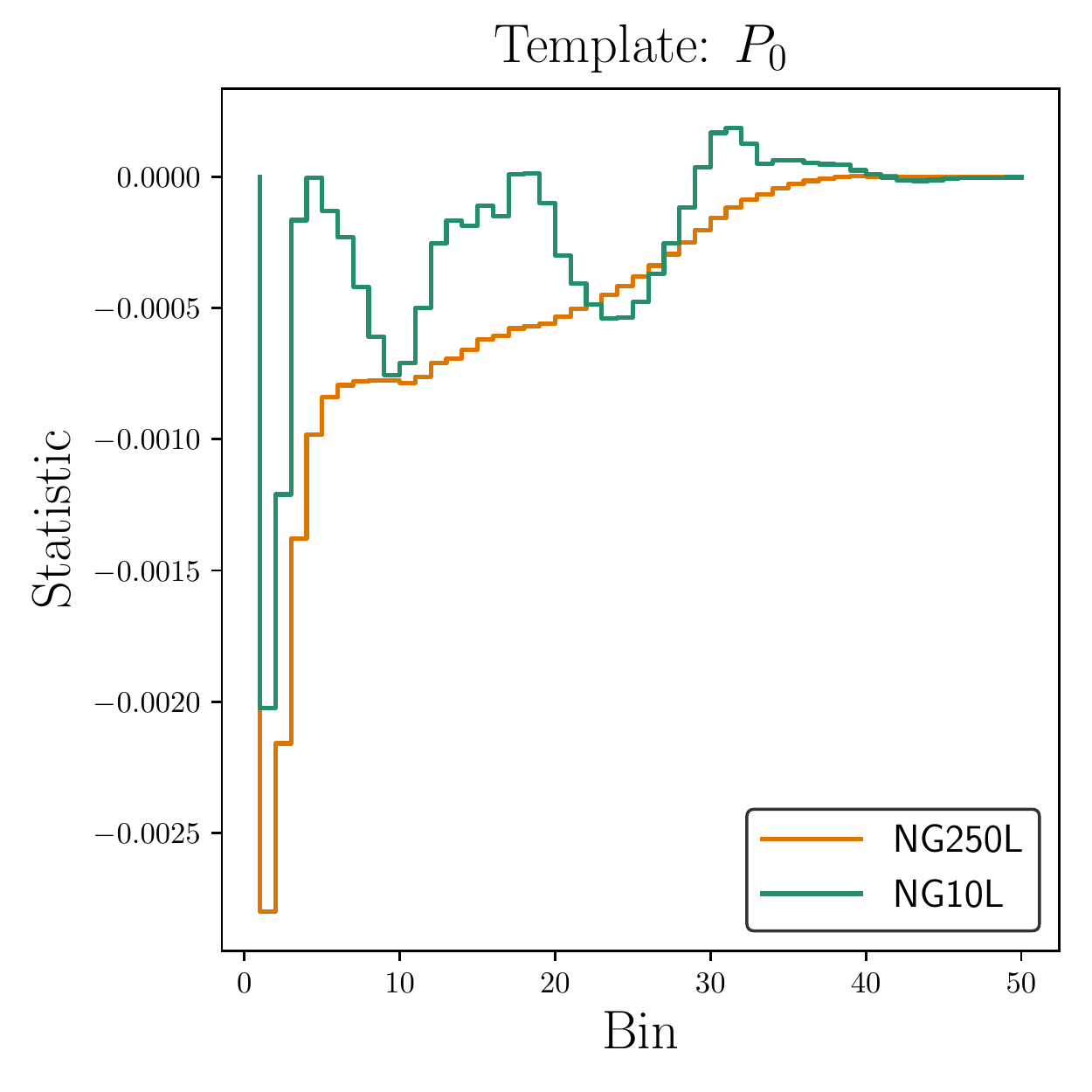}\includegraphics[width=0.33\textwidth]{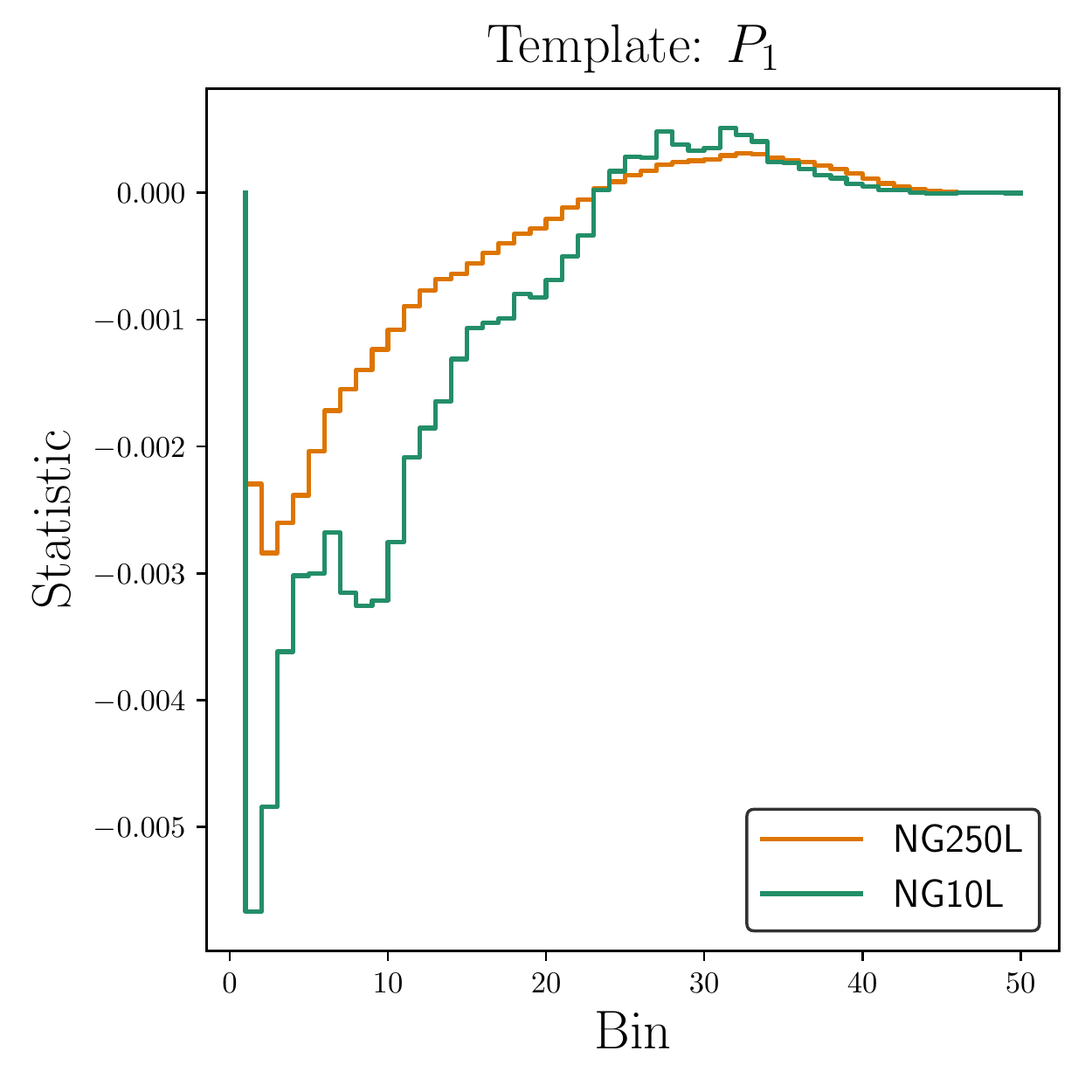}\includegraphics[width=0.33\textwidth]{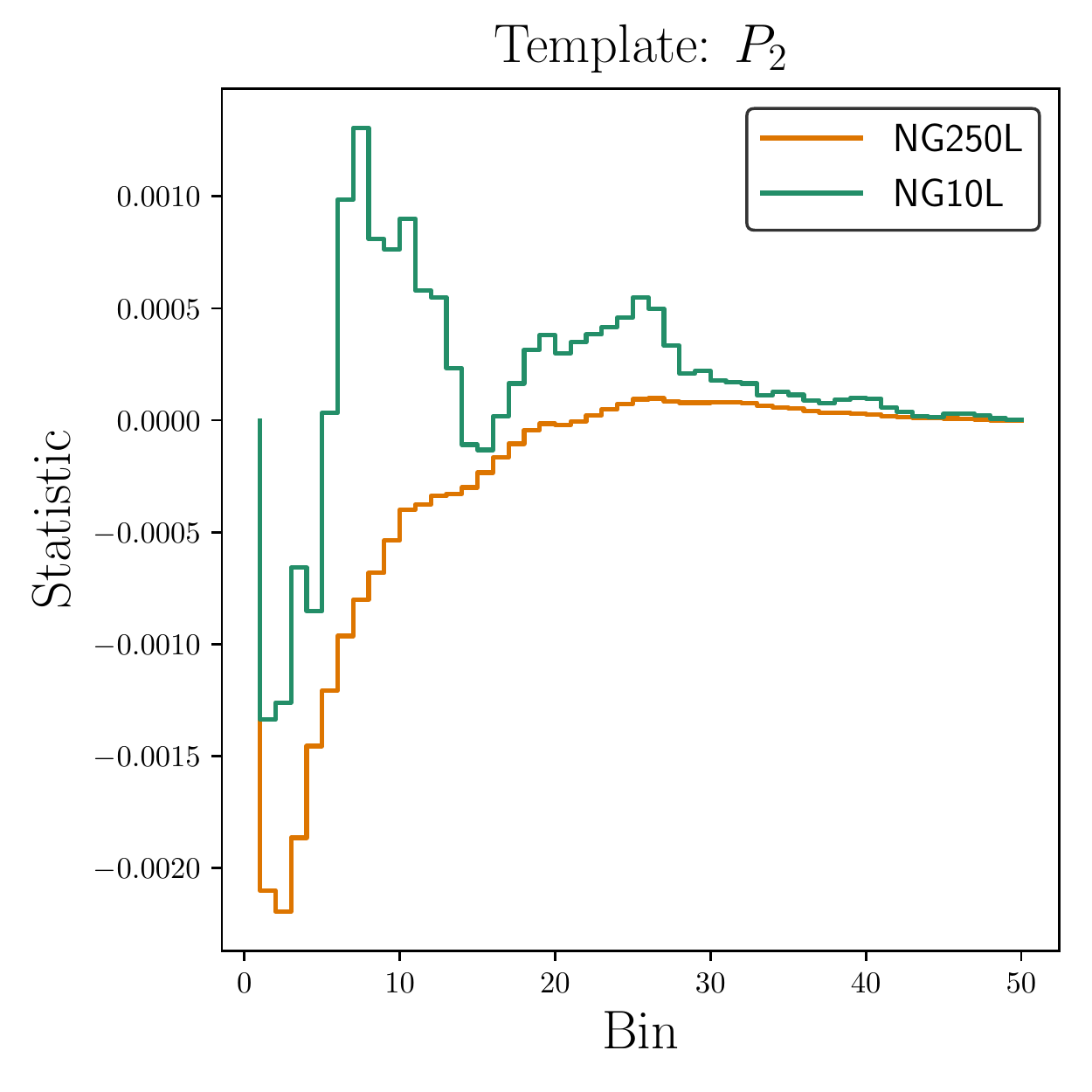}
        \includegraphics[width=0.33\textwidth]{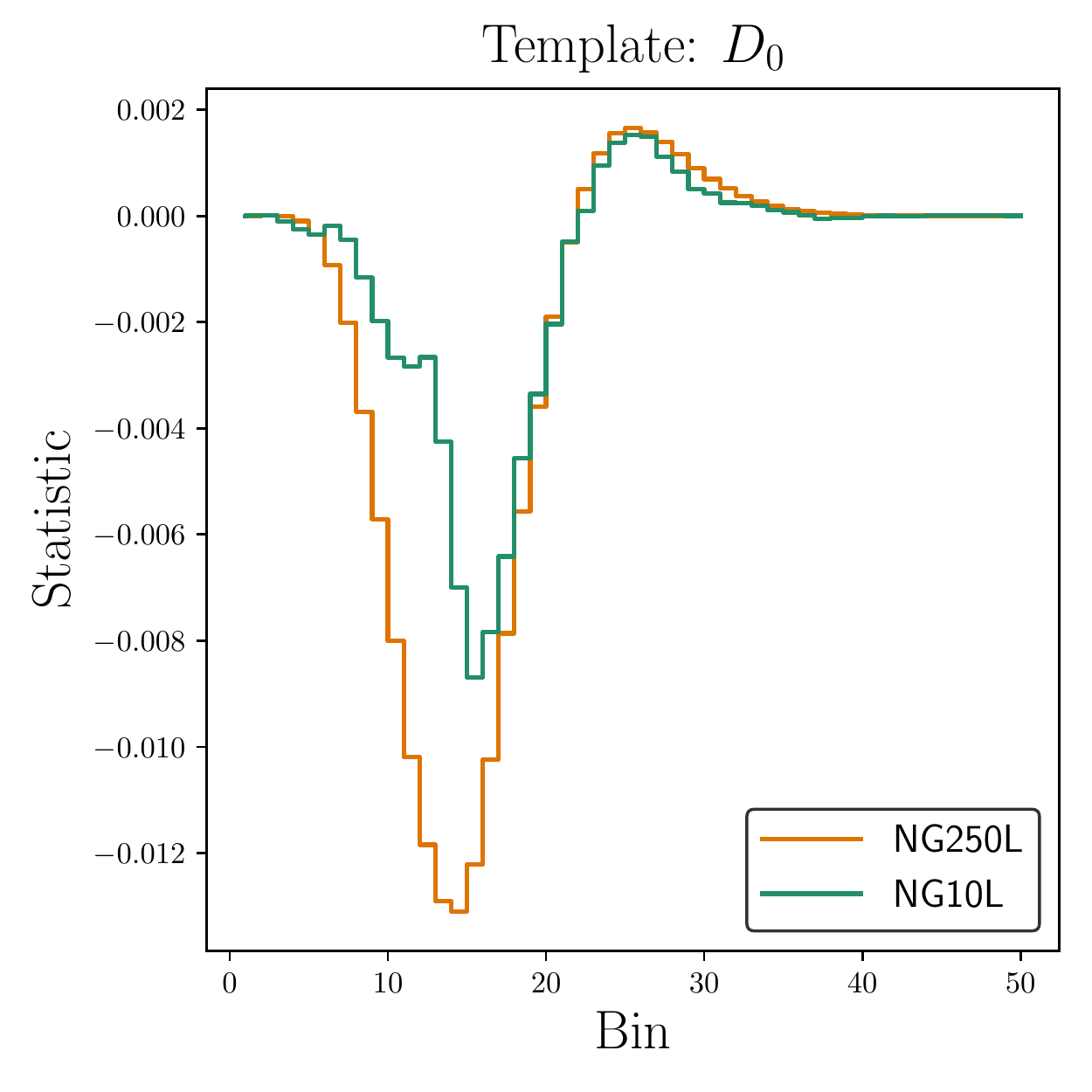}\includegraphics[width=0.33\textwidth]{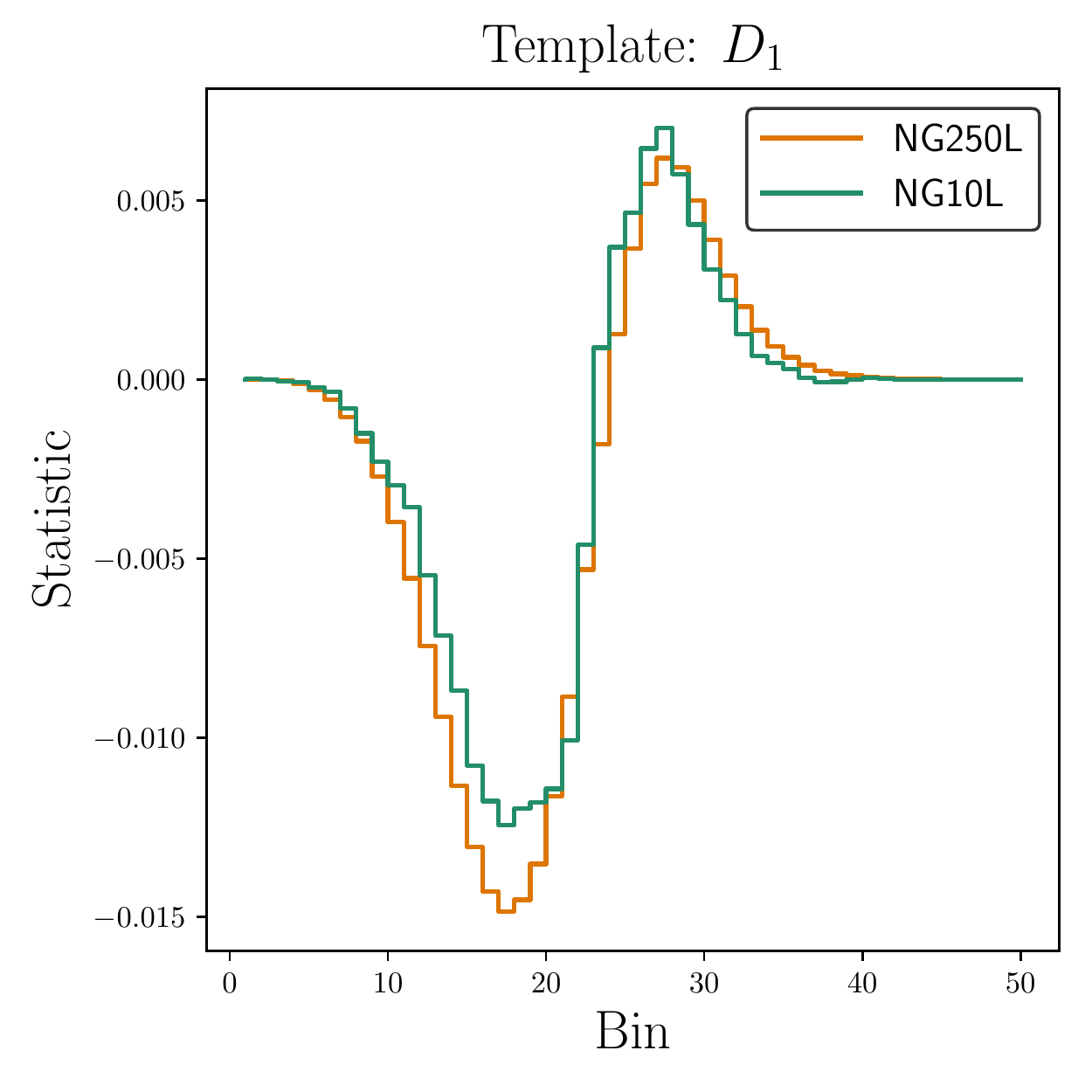}\includegraphics[width=0.33\textwidth]{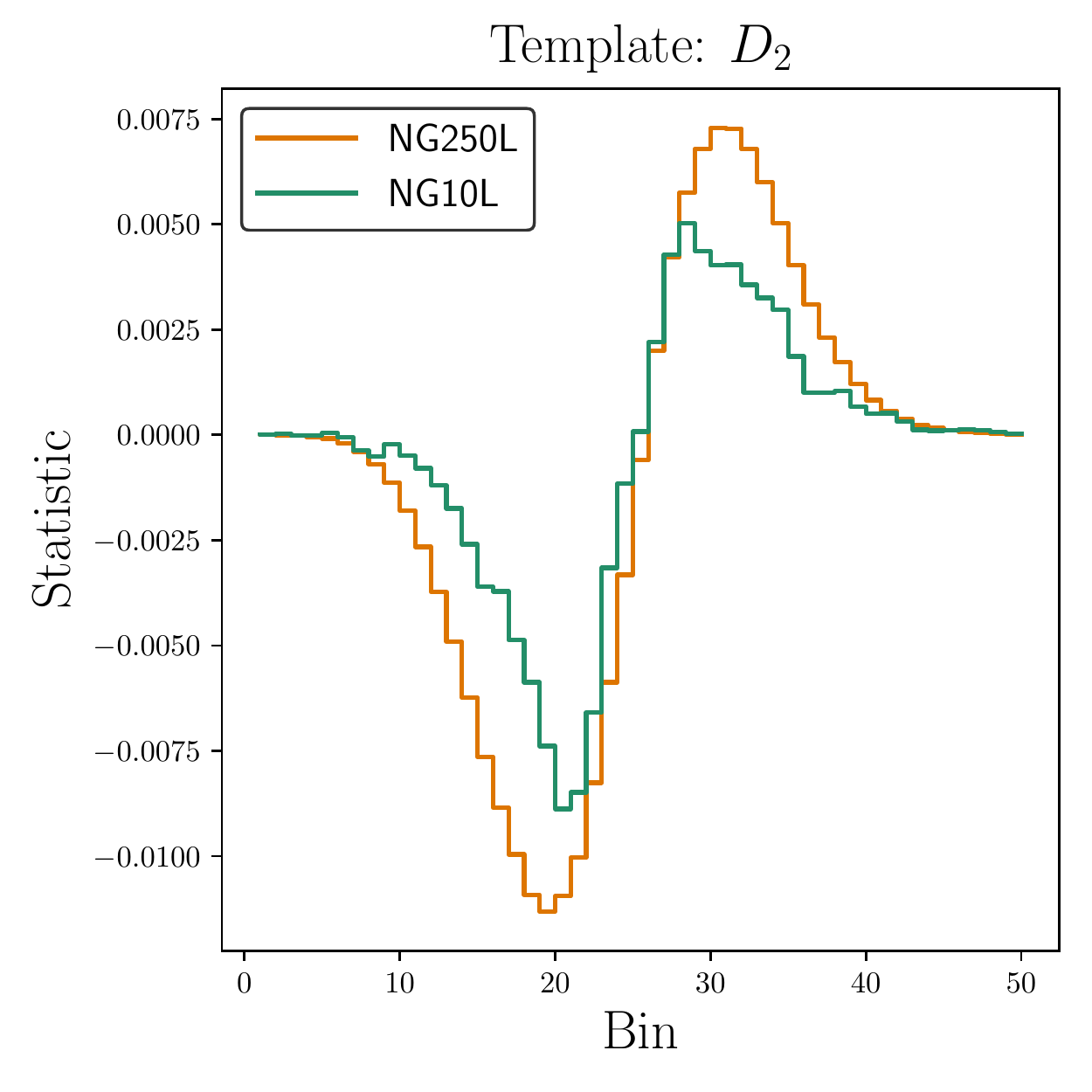}
    \caption{Dependence of non-Gaussian template on magnitude of $\fnl$. Here we show the templates computed by comparing \textsf{NG250L} and \textsf{NG10L} to \textsf{G85L}. The $\fnl=10$ template is scaled up by a factor of 10 for ease of comparison. It is interesting that this scaling is not linear with $\fnl$. The strong overlap between templates with different levels of non-Gaussianity allows us to use templates computed for large non-Gaussianity to detect more realistic $\fnl$ in simulations. }
    \label{fig:templateRunning}
\end{figure}
\afterpage{\FloatBarrier}
In Figure \ref{fig:templateAsymp} we show the templates for $\fnl=250$ and $\sigma_8=0.87$, computed using \textsf{NG250L}, \textsf{G87L}, and \textsf{G85L}. We observe that for most of the templates, the signature of $\fnl$ is easily distinguished from that due to varying $\sigma_8$. This can be traced back to differences in the PIs, as noted in Section \ \ref{sec:degen}. In Figure \ref{fig:templateRunning} we show the templates computed by comparing \textsf{NG250L} and \textsf{NG10L} to \textsf{G85L}.  The overlap between these allows us to use a template computed using large non-Gaussianity to search for more realistic levels of non-Gaussianity. This is useful because it reduces the amount of simulation volume needed to generate a trustworthy template. 
In Figure \ref{fig:templateLvsS} we check the robustness of our templates under change in  resolution in simulations which have otherwise the same cosmological parameters (cfr.\ Table \ref{tab:eos}). We show the templates computed by comparing \textsf{NG250L} to \textsf{G85L} and \textsf{NG250S} to \textsf{G85S}. The latter, being computed in smaller simulation volumes, are less computationally expensive. 
\begin{figure}
    \centering
           \includegraphics[width=0.33\textwidth]{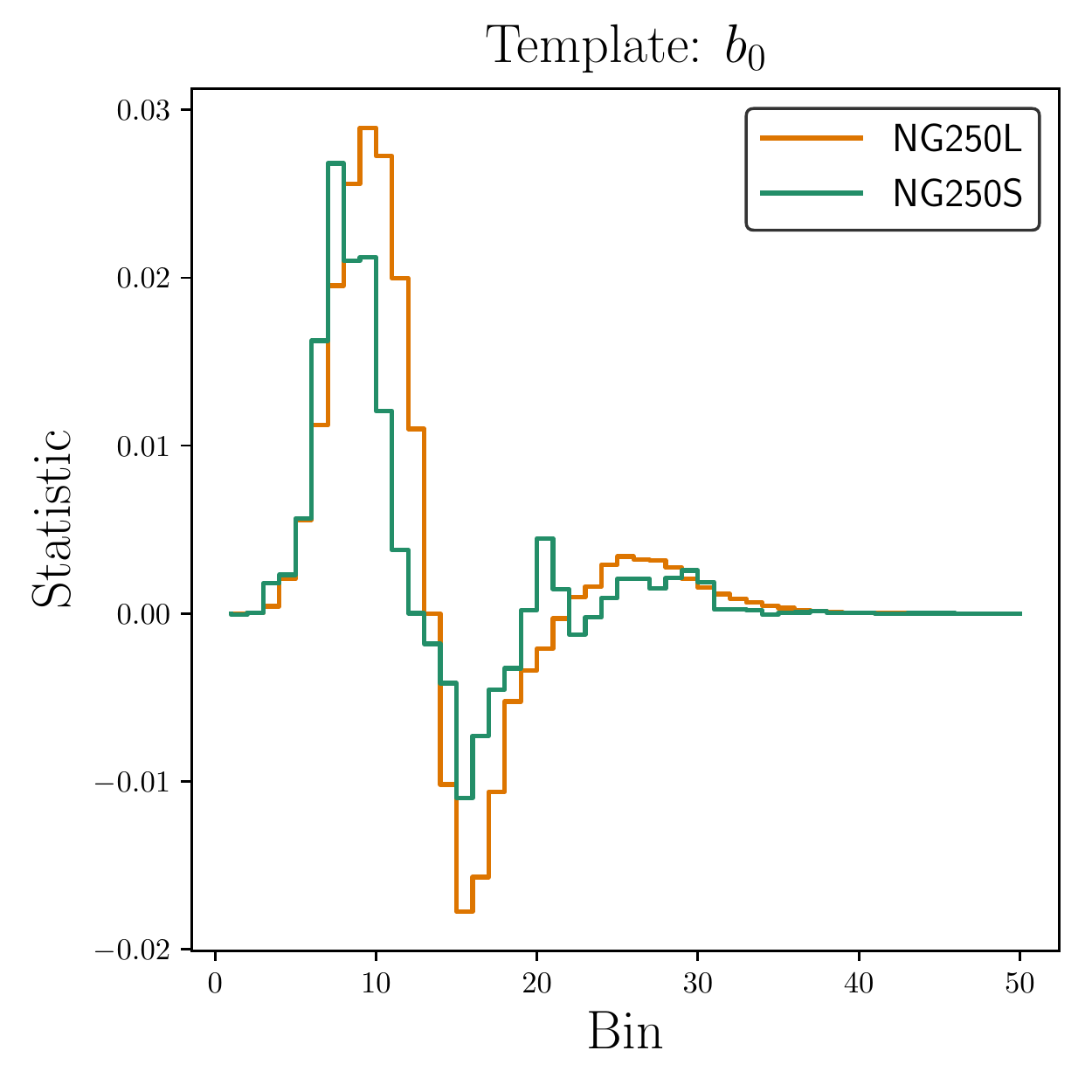}\includegraphics[width=0.33\textwidth]{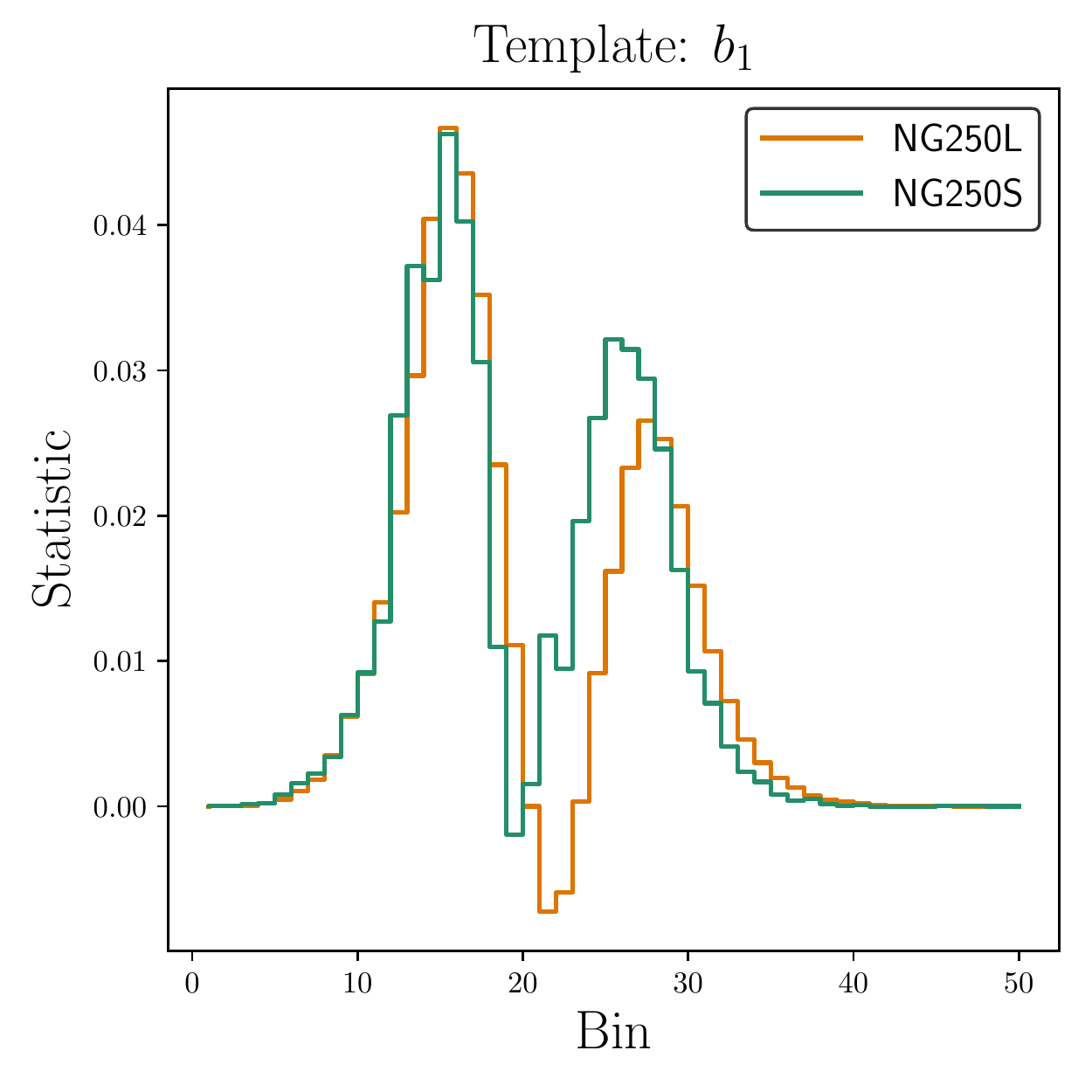}\includegraphics[width=0.33\textwidth]{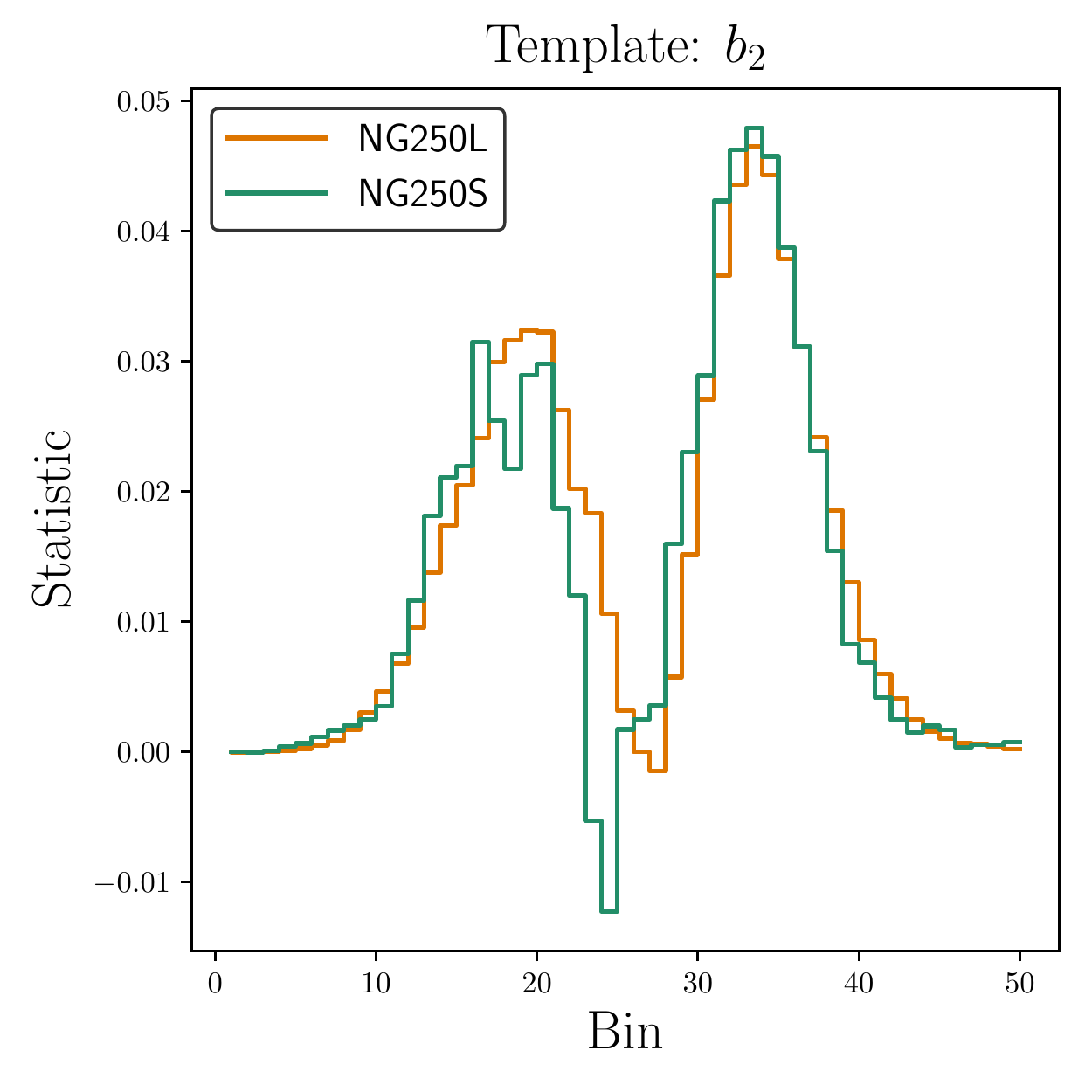}
            \includegraphics[width=0.33\textwidth]{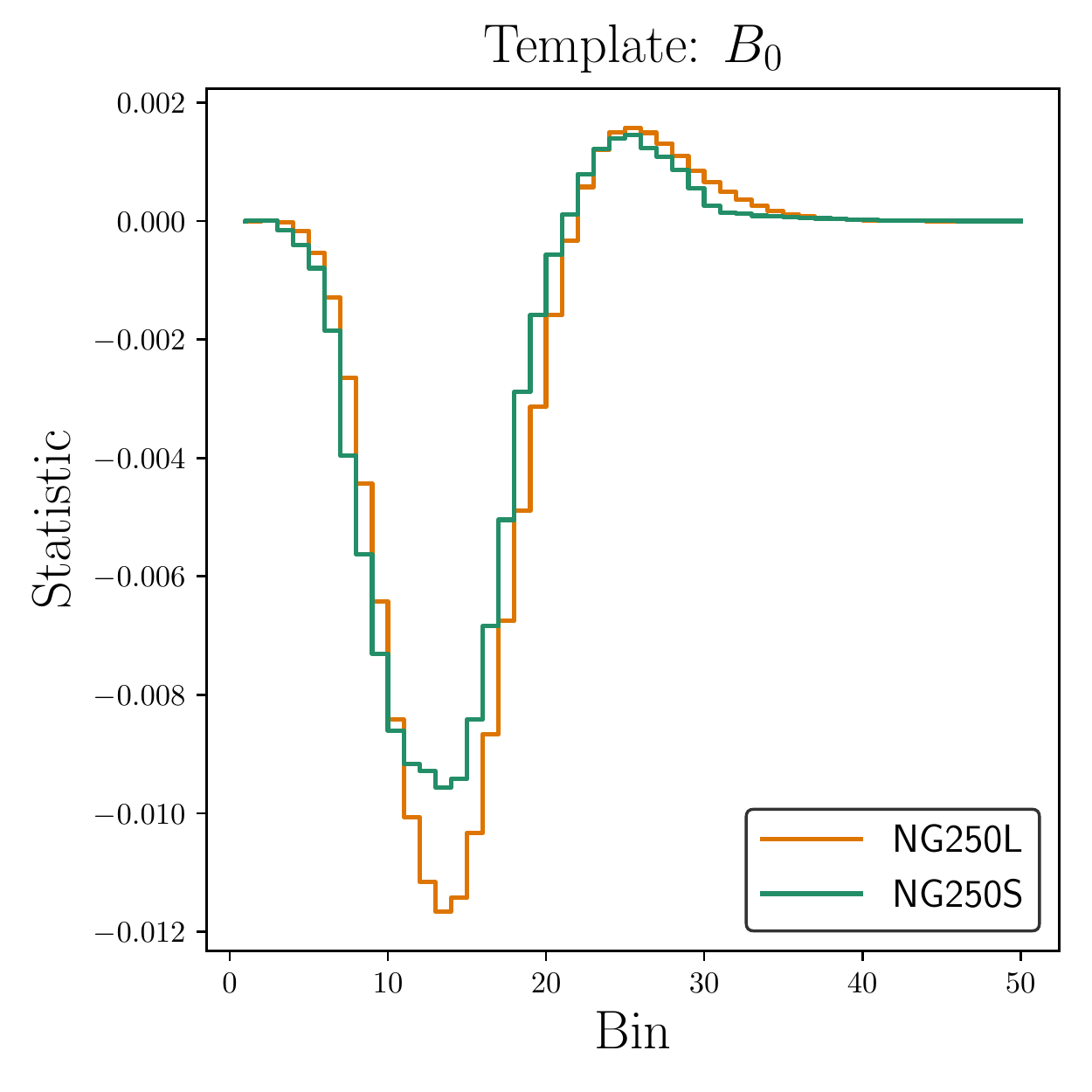}\includegraphics[width=0.33\textwidth]{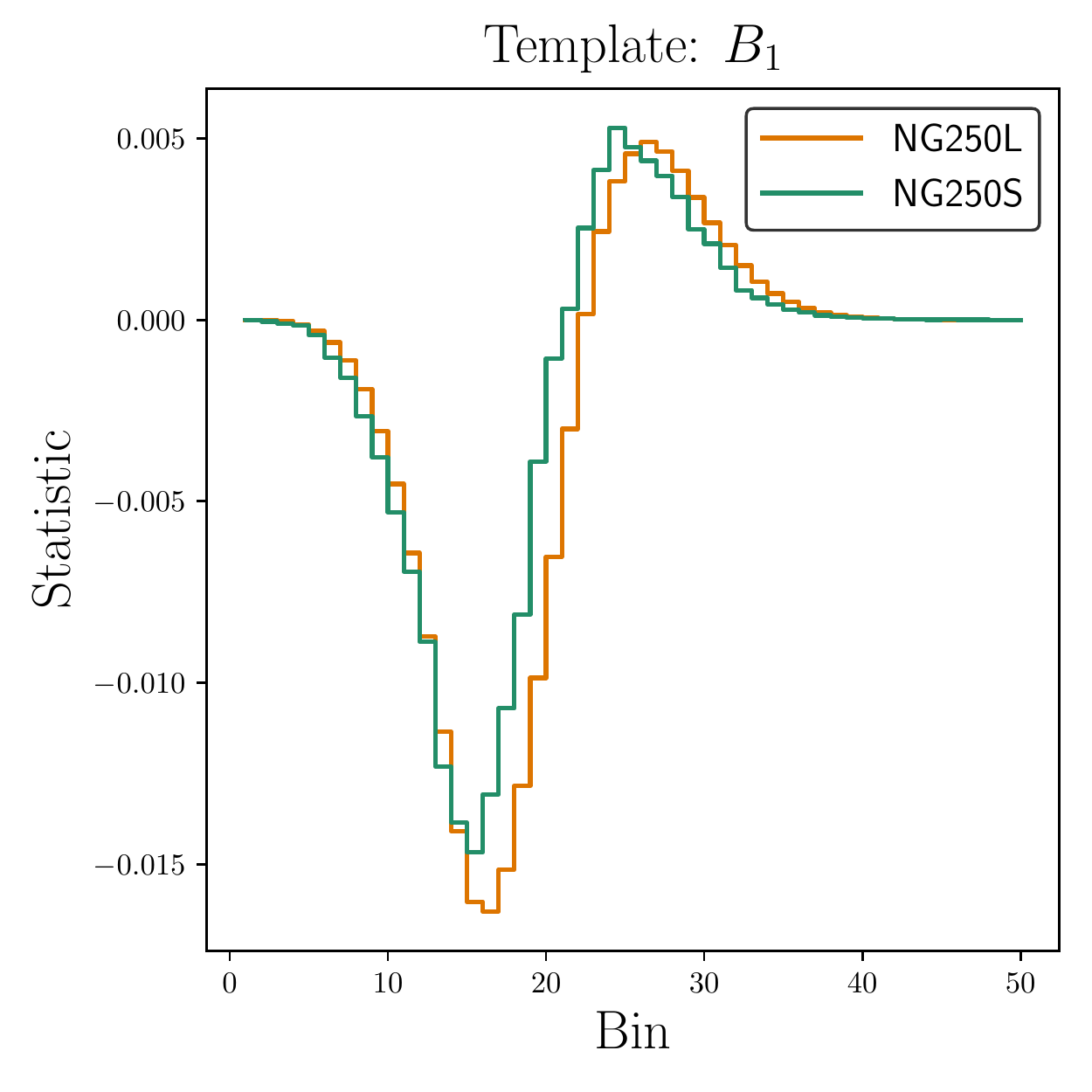}\includegraphics[width=0.33\textwidth]{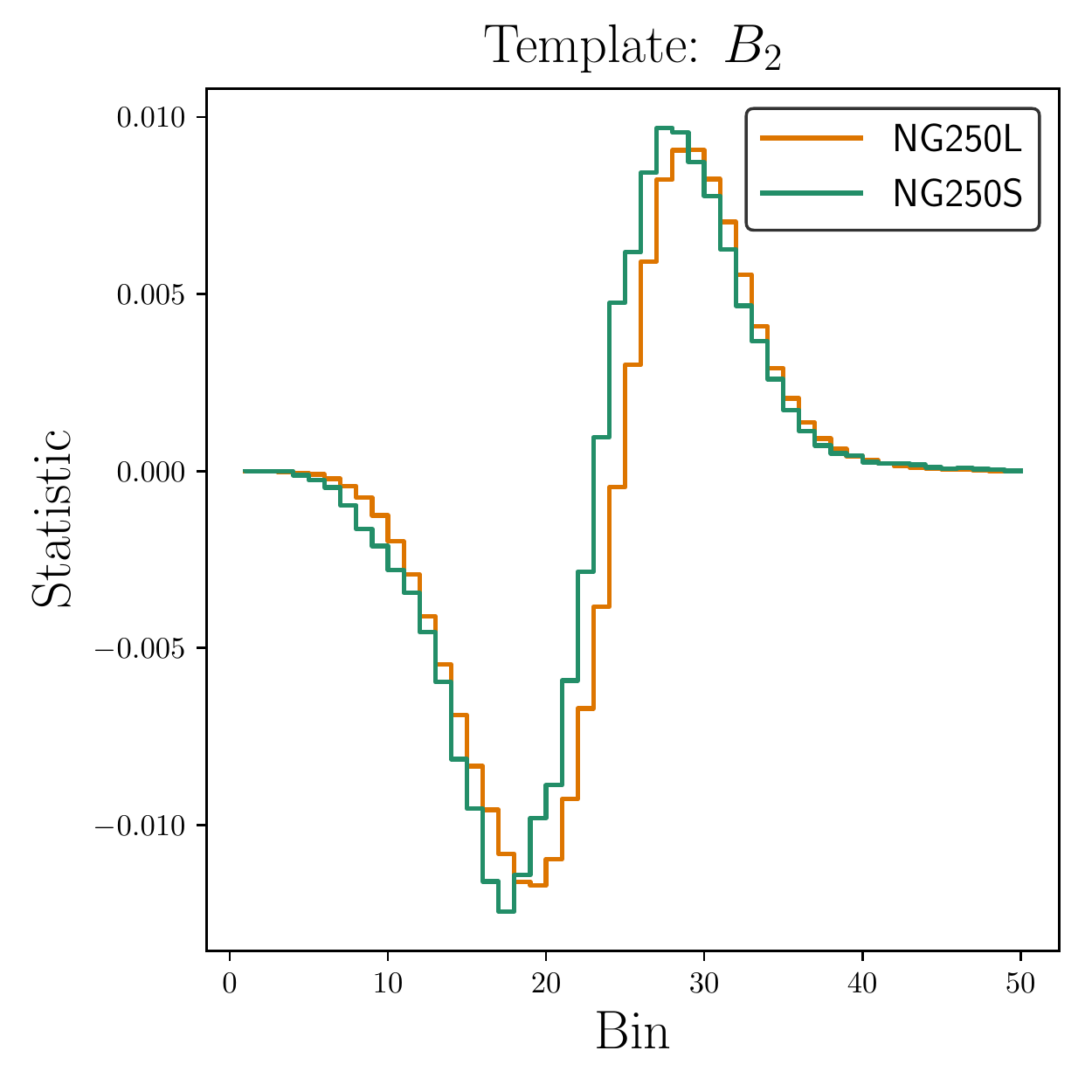}
            \includegraphics[width=0.33\textwidth]{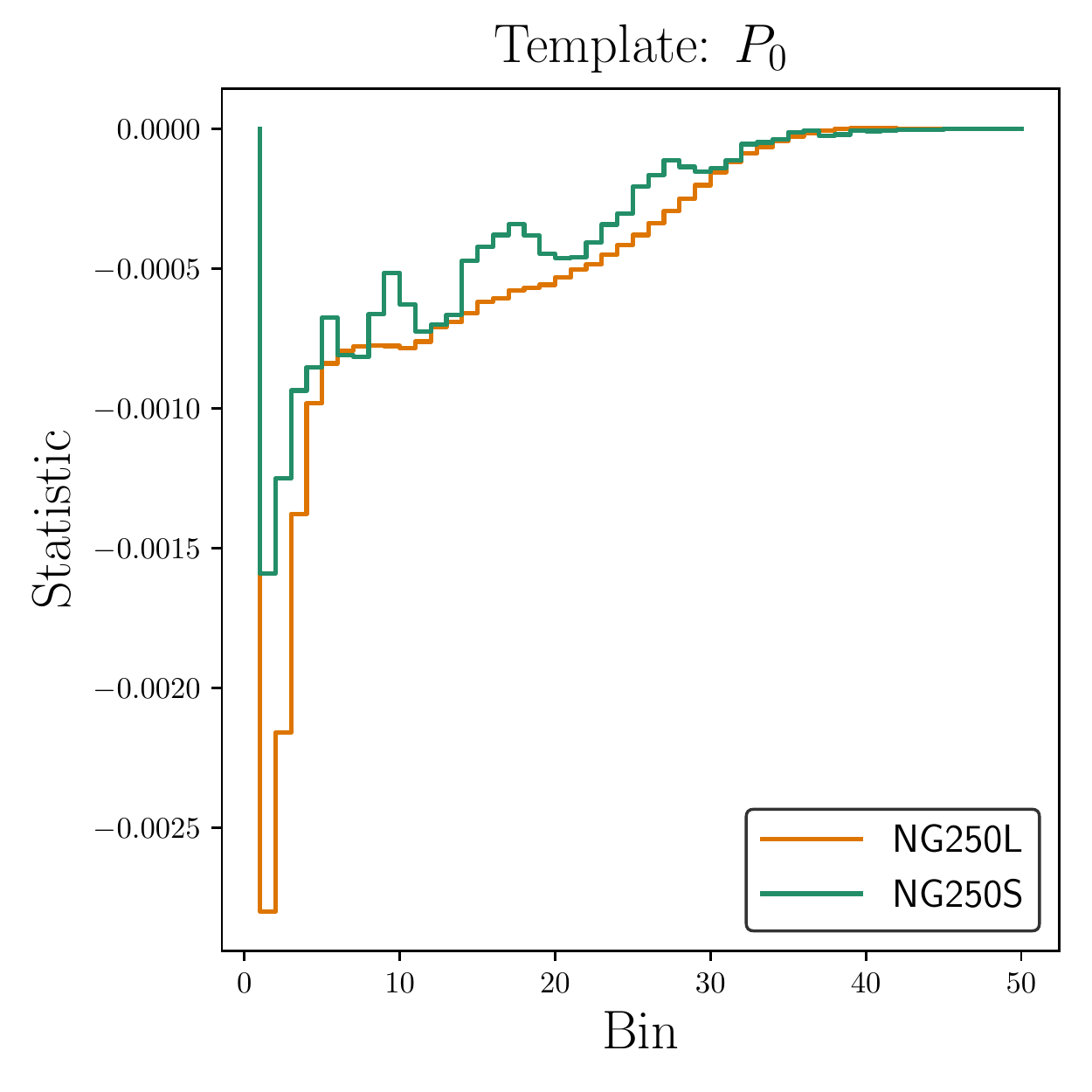}\includegraphics[width=0.33\textwidth]{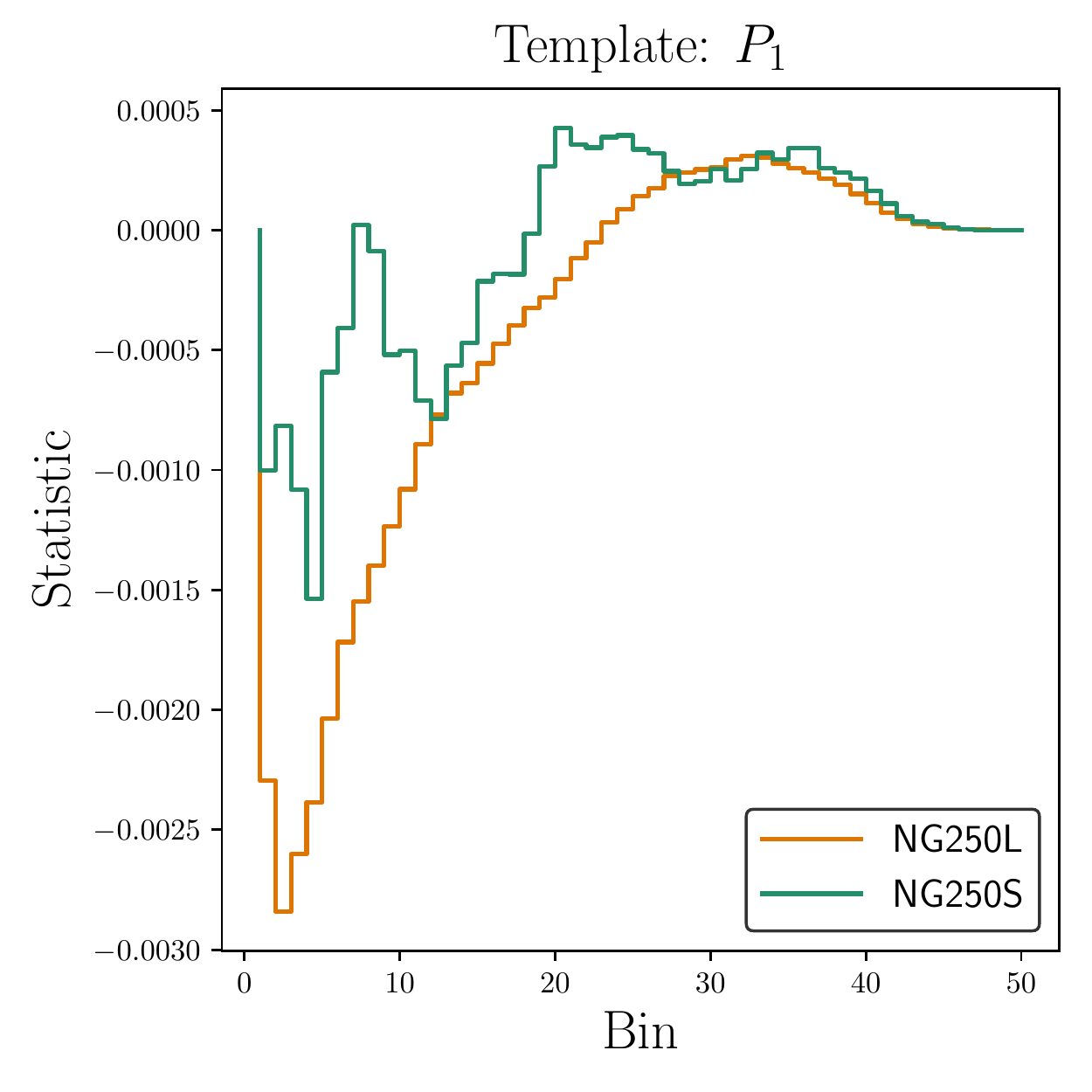}\includegraphics[width=0.33\textwidth]{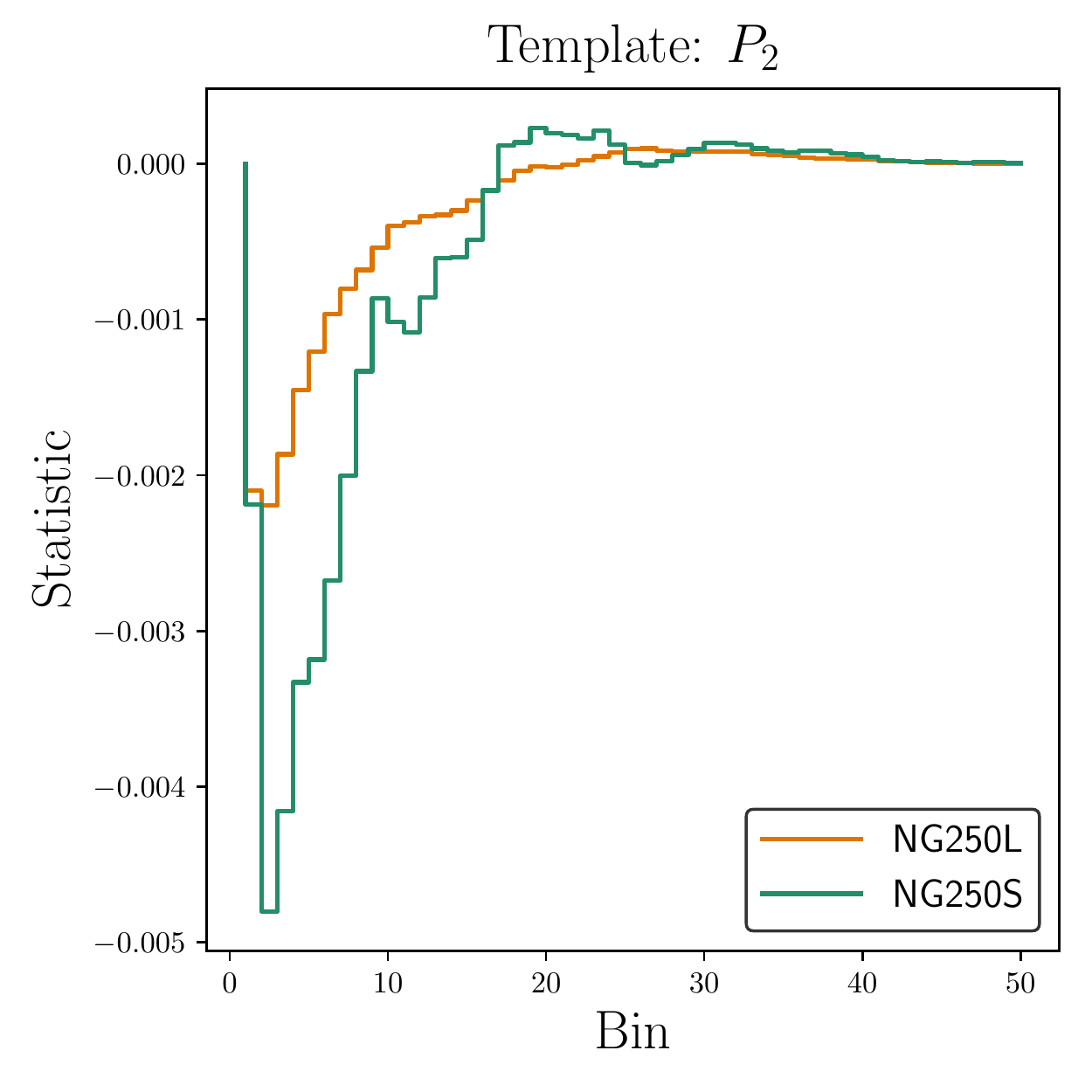}
        \includegraphics[width=0.33\textwidth]{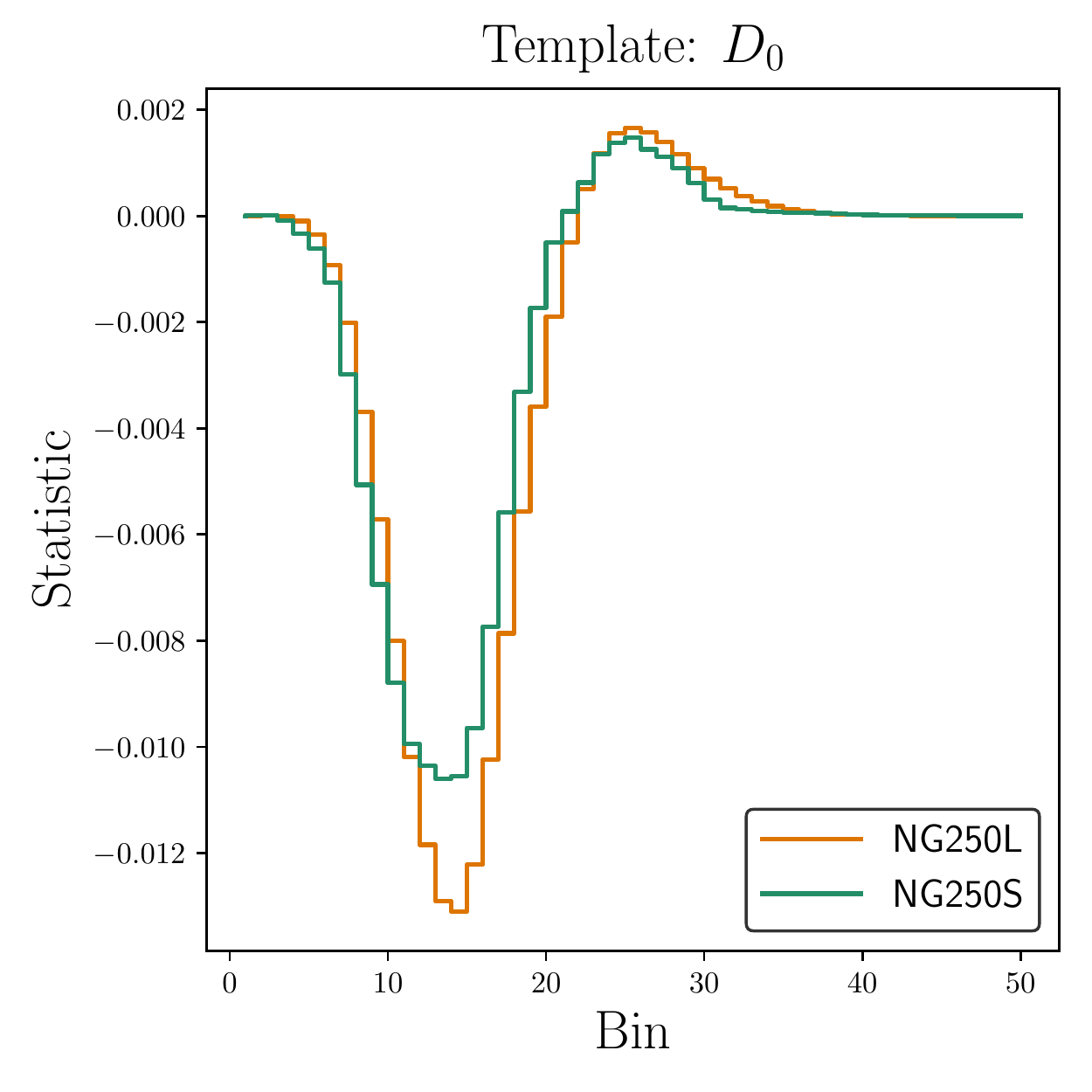}\includegraphics[width=0.33\textwidth]{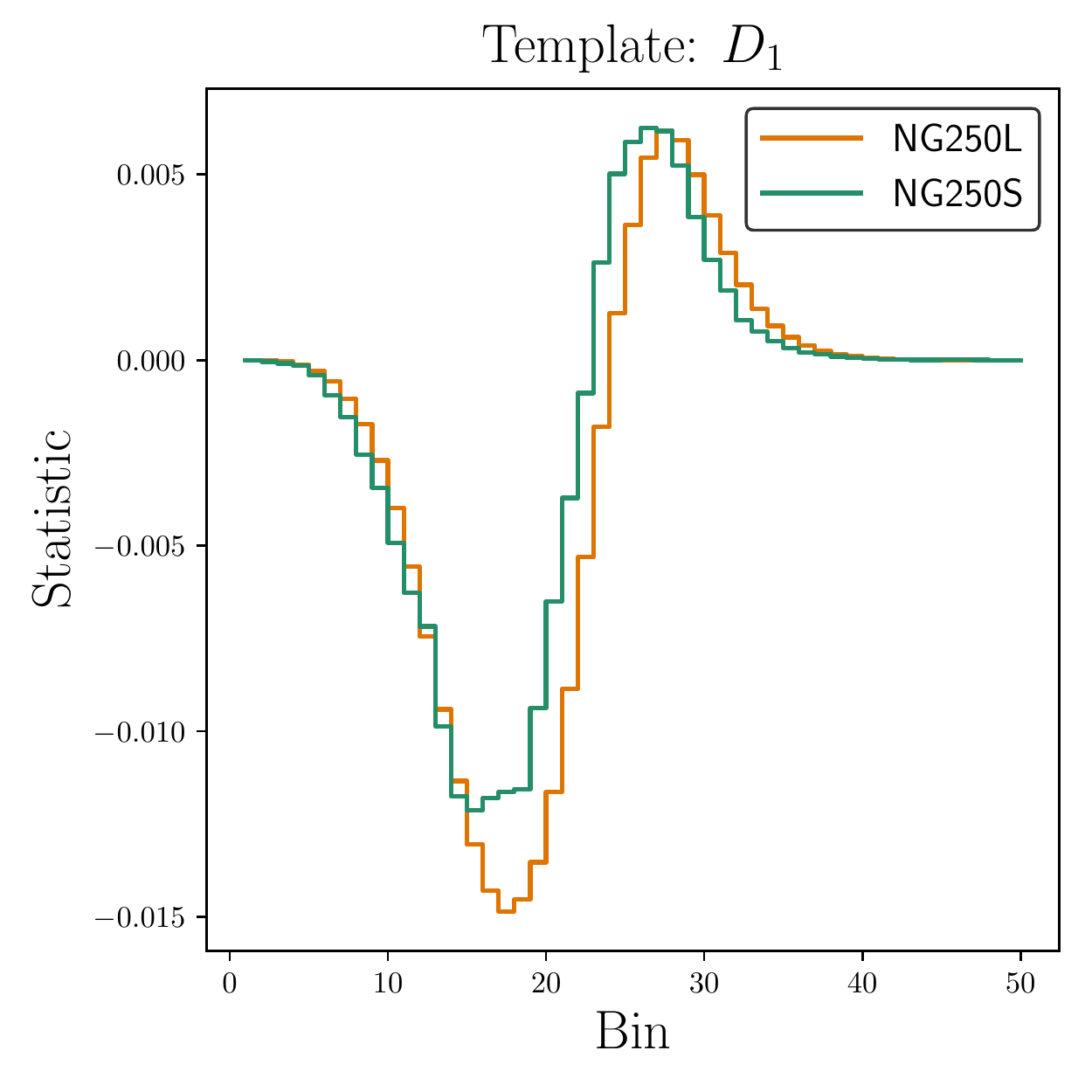}\includegraphics[width=0.33\textwidth]{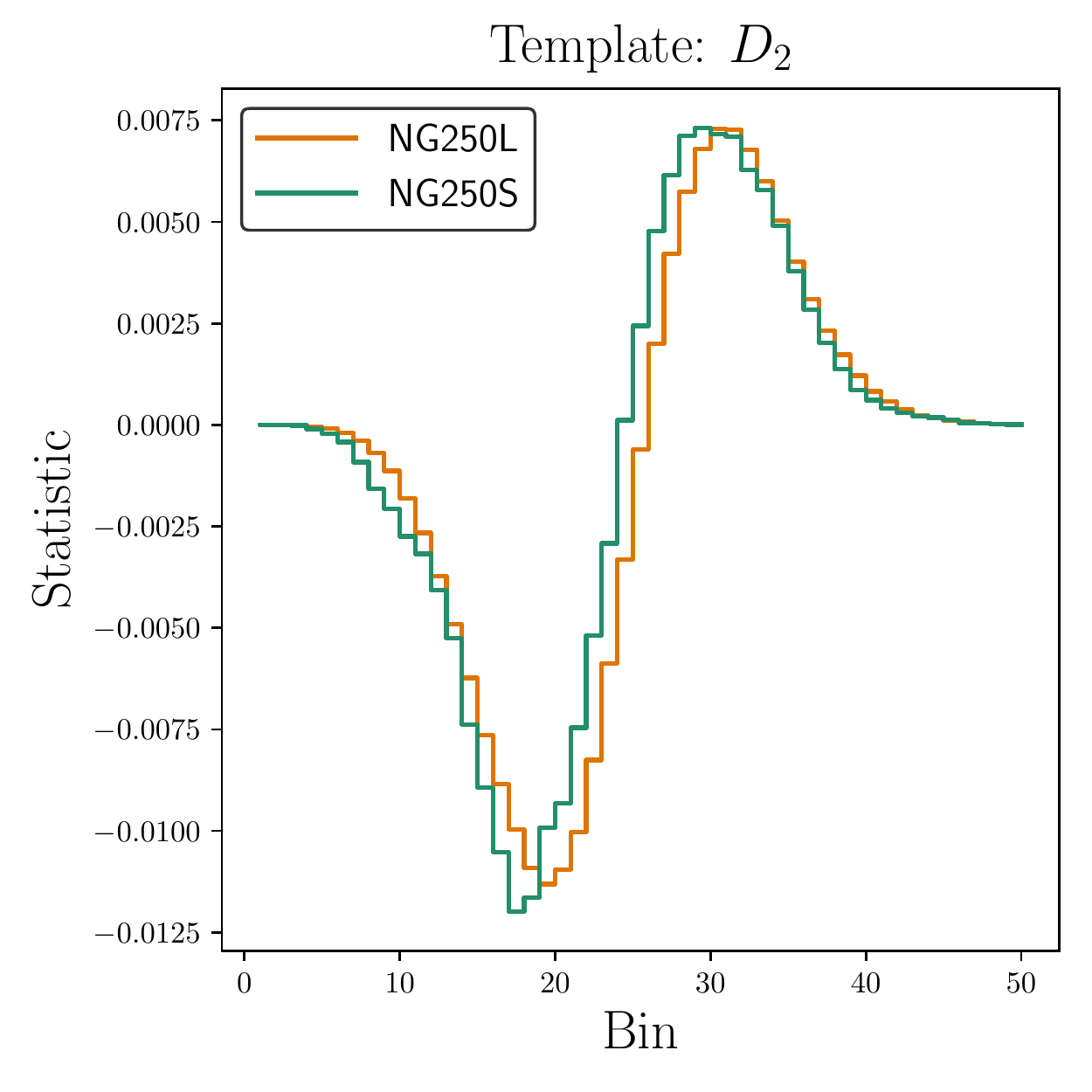}
    \caption{Non-Gaussian templates computed from large-box and small-box simulations. The excellent overlap between the templates confirms that the signatures of non-Gaussianity lives at scales complementary to the scale-dependent bias. This allows us to generate templates via less costly simulations.}
    \label{fig:templateLvsS}
\end{figure}
\afterpage{\FloatBarrier}
The combination of these results suggests we can safely use simulations with large non-Gaussianity, computed in small boxes, to detect more reasonable levels of non-Gaussianity in data with potentially different systematics than are modeled in a mock data set.
We observe excellent overlap between the two methods for generating simulations. This is promising not only from a computational perspective but also because we can imagine the difference in simulations as being replaced in an observational context by differences between survey systematics and their modeling in the mock data.  For the specific case of the search of primordial non-Gaussianity, Figure \ref{fig:templateLvsS} also confirms our expectation that the signature of $\fnl$ is important at small scales rather than large scales for the statistics we study (remember that features have scales $\sim \mathcal O(10)$ Mpc/h), in contrast with the distinctive signature at large scales provided by the scale dependent bias. Upon validating this pipeline in more detail, this would imply that when looking for primordial non-Gaussianity of the local type using our method in a galaxy survey, while still requiring a large volume, it would not be required to probe very large scales.

\newpage 

\bibliographystyle{utphys}
\bibliography{Halos-TDA}

\end{document}